\RequirePackage[2020-02-02]{latexrelease}
\documentclass{article} % One-column default
\usepackage[authoryear]{natbib}
\setcitestyle{numbers,square,comma}

\usepackage{preprint}

\usepackage{amsmath, amsthm, amssymb, amsfonts}

\usepackage{placeins}

\usepackage[utf8]{inputenc}	% allow utf-8 input
\usepackage[T1]{fontenc}	% use 8-bit T1 fonts
\usepackage{xcolor}		% colors for hyperlinks
\usepackage{hyperref}
\hypersetup{
    colorlinks = true,
    linkcolor = purple,
    urlcolor  = blue,
    citecolor = cyan,
    anchorcolor = black
}
\usepackage{cleveref}
            
\usepackage{booktabs} 		% professional-quality tables
\usepackage{float}			% Allows for figures within multicol
\usepackage{multicol}		% Multiple columns (Method B)

\usepackage{titlesec}
\titlespacing\section{0pt}{12pt plus 3pt minus 3pt}{1pt plus 1pt minus 1pt}
\titlespacing\subsection{0pt}{10pt plus 3pt minus 3pt}{1pt plus 1pt minus 1pt}
\titlespacing\subsubsection{0pt}{8pt plus 3pt minus 3pt}{1pt plus 1pt minus 1pt}

\usepackage{tikz,xcolor}

\definecolor{lime}{HTML}{A6CE39}
\DeclareRobustCommand{\orcidicon}{
	\begin{tikzpicture}
	\draw[lime, fill=lime] (0,0) 
	circle [radius=0.16] 
	node[white] {{\fontfamily{qag}\selectfont \tiny ID}};
	\draw[white, fill=white] (-0.0625,0.095) 
	circle [radius=0.007];
	\end{tikzpicture}
	\hspace{-2mm}
}
\foreach \x in {A, ..., Z}{\expandafter\xdef\csname orcid\x\endcsname{\noexpand\href{https://orcid.org/\csname orcidauthor\x\endcsname}
			{\noexpand\orcidicon}}
}
\title{Efficient Inverse-designed Structural Infill for Complex Engineering Structures}

\usepackage{xwatermark}
\usepackage{authblk}

\author[1\thanks{\tt{pdlj@dtu.dk}}]{Peter D\o rffler Ladegaard Jensen\orcidA{}}
\author[2]{Tim Felle Olsen\orcidB{}}
\author[2]{J. Andreas B\ae rentzen\orcidC{}}
\author[1]{Niels Aage\orcidD{}}
\author[1]{Ole Sigmund\orcidE{}}

\affil[1]{Department of Civil and Mechanical Engineering,  Solid Mechanics, Technical University of Denmark. Koppels All\'{e}, B.404, 2800 Kgs. Lyngby, Denmark.}

\affil[2]{Department of Applied Mathematics and Computer Science, Visual Computing, Technical University of Denmark. Richard Petersens Plads, B.321, 2800 Kgs. Lyngby, Denmark.}

\usepackage{graphicx}
\usepackage{array}
\usepackage{caption}
\usepackage{subcaption}
\usepackage{appendix}
\usepackage[flushleft]{threeparttable}
\usepackage{bm}

\newcommand{\unit}[1]{\,\mathrm{\left[#1\right]}}
\newcommand{\newparallel}{\mathbin{\!/\mkern-5mu/\!}}

\newcolumntype{L}{>{\raggedright\arraybackslash}X}
\begin{document}

% \twocolumn[ % Method A for two-column formatting
%   \begin{@twocolumnfalse} % Method A for two-column formatting
  
\maketitle

\begin{abstract}
Inverse design of high-resolution and fine-detailed 3D lightweight mechanical structures is notoriously expensive due to the need for vast computational resources and the use of very fine-scaled complex meshes. Furthermore, in designing for additive manufacturing, infill is often neglected as a component of the optimized structure. In this paper, both concerns are addressed using a so-called de-homogenization topology optimization procedure on complex engineering structures discretized by 3D unstructured hexahedrals.
Using a rectangular-hole microstructure (reminiscent to the stiffness optimal orthogonal rank-3 multi-scale) as a base material for the multi-scale optimization, a coarse-scale optimized geometry can be obtained using homogenization-based topology optimization. Due to the microstructure periodicity, this coarse-scale geometry can be up-sampled to a fine single-scale physical geometry with optimized infill, with only a minor loss in structural performance and at a fraction of the cost of a fine-scale solution. The upsampling on 3D unstructured grids is achieved through stream surface tracing which aligns with the optimized local orientation. The periodicity of the physical geometry can be tuned, such that the material serves as a structural component and also as an efficient infill for additive manufacturing designs.
The method is demonstrated through three examples of varying geometrical complexity. It achieves comparable structural performance to state-of-the-art methods but stands out for its significant computational time reduction, much faster than the base-line method. By allowing multiple active layers, the mapped solution becomes more mechanically stable, leading to an increased critical buckling load factor without additional computational expense. The proposed approach achieves promising results, benchmarking against large-scale SIMP models demonstrates computational efficiency improvements of up to 250 times. 
\end{abstract}
%\keywords{First keyword \and Second keyword \and More} % (optional)
% \vspace{-0.5cm}
%
%   \end{@twocolumnfalse} % Method A for two-column formatting
% ] % Method A for two-column formatting
%
% \begin{multicols}{2} % Method B for two-column formatting (doesn't play well with line numbers), comment out if using method A
%
%%%%%%%%%%%%%%%  Main text   %%%%%%%%%%%%%%%
% \linenumbers
\begin{multicols}{2}[\section{Introduction}\label{sec:intro}]
Inverse design approaches, such as topology optimization \cite{Bendsoe2004}, provide a systematic approach to obtain optimized design based on different objectives like stiffness, strength, weight, etc., with the evident advantage of designs with less resource consumption and longer lifetime. However, the procedure can be computationally expensive on fine-scale complex geometries. Moreover, systematic inverse design methods tailored for industrial-grade additive manufacturing are needed to exploit infill as an optimized structural component. Designing for stiffness maximization with an infill requirement results in a material distribution problem on two length scales: macro- and micro-scale. The material distribution at the macro-scale can be optimized to minimize the amount of material used while achieving desired mechanical properties and performance. Simultaneously, the material distribution at the micro-scale can be optimized to achieve specific mechanical properties such as stiffness or strength.

Using standard topology optimization methods, multi-scale material distribution on a single length scale requires high-resolution and large-scale modeling capabilities to obtain designs with a sufficient level of detail ~\cite{Sigmund2018}.
To address this problem, much research over the past two decades has been aimed at making this feasible by utilizing parallel programming paradigms; domain decomposition approaches~\cite{Mahdavi2006, Evgrafov2008, Aage2013}, and efficient multigrid preconditioning techniques~\cite{Amir2013, Aage2015}. With these advances, it is now possible to solve large-scale problems with billions of design variables~\cite{Aage2017,Baandrup2020}.
While large-scale topology optimization is an important tool to gain new insights into structural design, the point-wise material or void description results in length-scale dependent solutions. Furthermore, large-scale topology optimization comes at a significant computational cost, limiting commercial access to only the largest companies and hampering widespread access to high-resolution optimal design.

In order to overcome the need for high-resolution finite element analysis and tackle the length-scale problem, the homogenization theory can be used to compute the mechanical properties of optimized local periodic microstructure of the structural elements in the macroscopic structure. The optimization problem can thus be \textit{relaxed} by posing the material distribution problem as a macroscopic continuous description, independent of length scale. This multi-scale design problem, also known as homogenization-based topology optimization~\cite{BendsoeKikuchi1988} can significantly reduce the size of the numerical problem that needs to be solved.
\end{multicols}
\begin{figure}[t!]
    \centering
    \makebox[\textwidth][c]{\includegraphics[width=190mm]{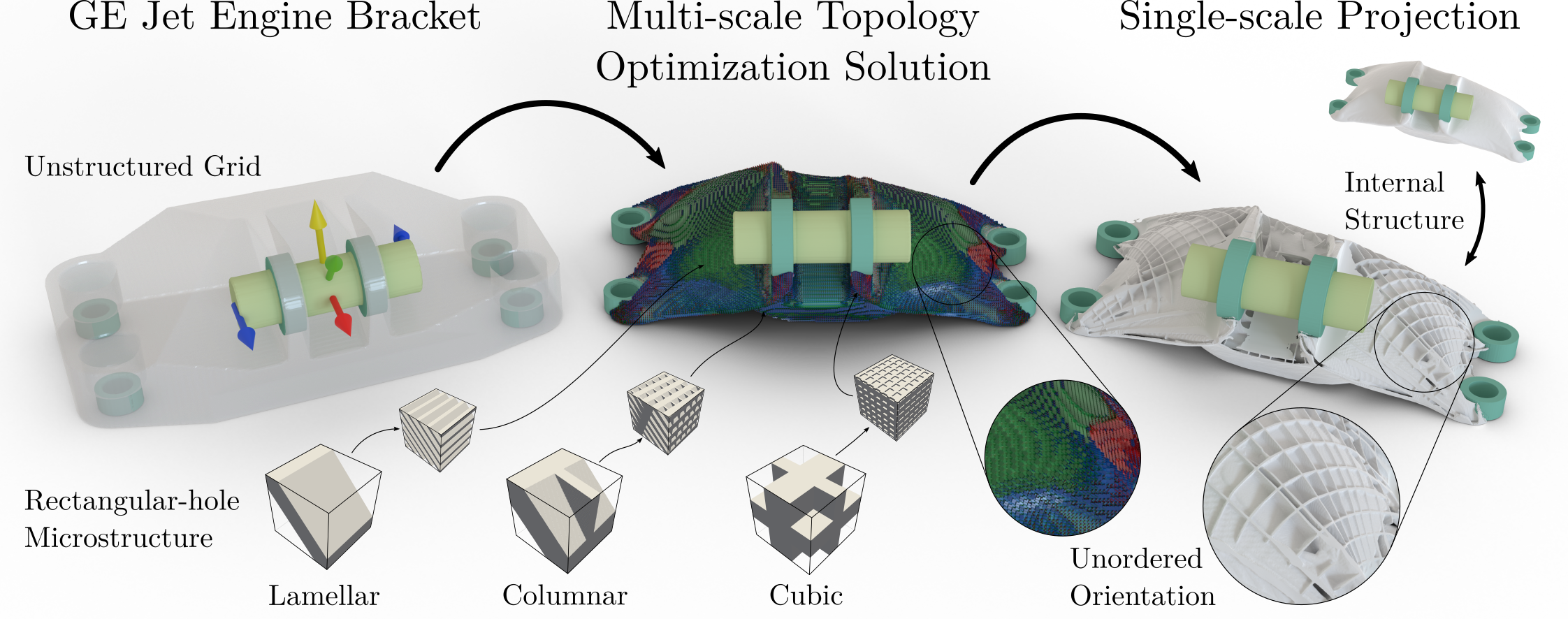}}%
    \caption{Graphical depiction of the de-homogenization procedure on complex geometries.}
    \label{fig:GraphicalIntroduction}
\end{figure}
\begin{multicols}{2}
It is well-known that multi-scale periodic laminated microstructures, so-called rank-$N$ microstructures (where $N$ indicates the number of laminations)~\cite{Francfort1986,Lurie1984,Milton1986,Norris1985}, achieve the optimal stiffness-to-density ratio under different domain settings and loading conditions. Hence these microstructures, which can be interpreted as infill structures, are the obvious choice for the multi-scale design problem as they  will result in optimal stiffness. In 3D under a single load case, an orthogonal rank-3 microstructure is stiffness optimal~\cite{Avellaneda1987,Francfort1995}, but its physical realization is challenging. The corresponding, easy-to-realize, single-length-scale, rectangular-hole microstructure approaches the rank-3 laminate in terms of stiffness; however, its mechanical properties cannot be calculated with a closed-form solution.

The multi-scale design problem requires a physically realizable interpretation to be applicable. The so-called de-homogenization approach introduced by~\citet{Pantz2008} and improved by~\citet{Groen2018}, maps the homogenization-based topology optimization solution to a single-scale solution with low computational cost by the construction of global mapping that approximates microstructure orientations (orientation mapping fields) and lamination widths. The procedure has been extended to multiple load case problems by \citet{Jensen2022} and to 3D, on structured grids by \cite{Groen2020}. However, the approach is challenged by microstructure non-uniqueness and ordering issues, particularly in 3D. While unordered microstructure orientations can be ordered in 2D, the problem is much more complex in 3D. Strict regularization and starting guess schemes have been proposed, but orientation singularities remain an issue.
Similar approaches have been made with open lattice structures, as shown in 3D by~\citet{Geoffroy-Donders2020,Wu2021b,Wang2022}, however, these methods produce open-walled structures which from a stiffness performance perspective, are inferior to closed-walled microstructures~\cite{Sigmund2016}.

\citet{Stutz2022} showed that a global mapping of the de-homogenization can be obtained from a number of partial solutions. A set of stream surfaces are obtained from random starting points and aligned with the microstructure orientations. Singularities are avoided in solid and void regions, and a subset of stream surfaces is selected through an optimization process to achieve uniform spacing for a target length scale. Selected surfaces are then converted to implicit solids, forming mapped single-scale, shell-like surfaces that align closely to microstructure orientations without implicit regularization. The resulting output is a union of surfaces, making the addition of other solids and structures easy, such as an exterior boundary shell.
A promising approach similar to \cite{Stutz2022} was published by~\citet{Garnier2022} where the de-homogenized structure is generated with a reaction/diffusion approach, however, this is again only shown for open lattice structures.

The de-homogenization procedure is a two-step multi-scale design problem. The first step of the procedure is to solve a multi-scale topology optimization problem using near-optimal periodic microstructures as base material. The topology optimization can be obtained on a relatively coarse grid allowing for extremely fast computations on single workstations. The second step of the procedure is mapping the coarse-scale solution to a fine-scale solution (the de-homogenization), which only should introduce a minor loss in performance.

De-homogenization procedures have so far only been developed for regular/structured grids, with limited application to practical engineering problems. To effectively capture the essential characteristics of such problems, unstructured grids must be employed. However, this introduces complexities for homogenization-based topology optimization, requiring careful consideration of factors such as obtaining a high-quality unstructured grid, regularization of microstructure material, and solver setup. Moreover, interpolations and upscaling become more challenging for the mapping procedure due to the unstructured coarse-scale solution. Therefore, this paper aims to address these concerns and extend the stream surface-based de-homogenization topology optimization procedure to complex geometries on unstructured grids in 3D. The proposed approach is applied to practical engineering design problems, including infill as a significant structural component. The proposed de-homogenization procedure is graphically depicted in \cref{fig:GraphicalIntroduction}. The proposed procedure advances the methodology closer to the ultimate goal of reproducing the giga-scale wing~\cite{Aage2017} and bridge~\cite{Baandrup2020} examples at an affordable computational cost without any loss in the level of structural details. 
\end{multicols}

\begin{figure}[tb]
    \centering
    \makebox[\textwidth][c]{
        \includegraphics[width=190mm]{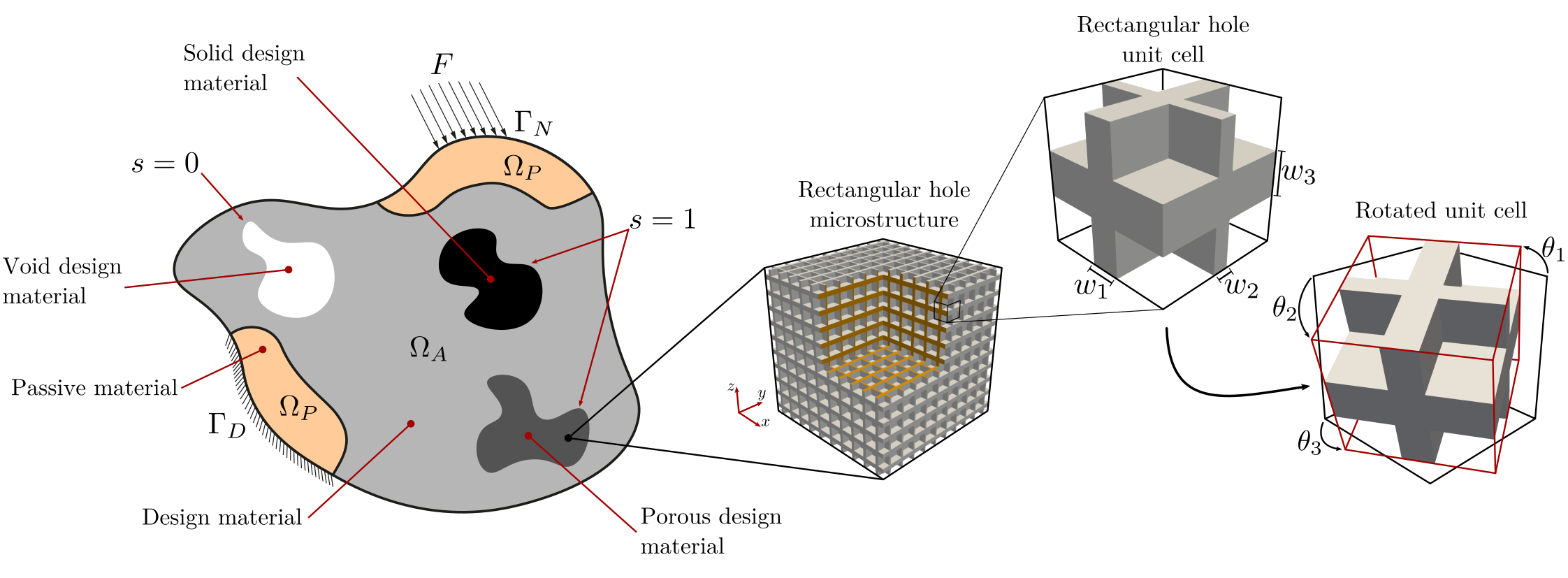}
    }
    \caption{Illustration of the design problem. The optimization problem is solved on $\Omega$, which is partitioned into an active design domain  $\Omega_A$ and a passive domain $\Omega_P$. Fixed displacements are applied to Dirichlet boundaries $\Gamma_D$, while tractions, $F$, are applied to Neumann boundaries $\Gamma_N$. The point-wise homogeneous microstructure design material has macroscopic varying design variables, thicknesses $w_i$, orientations $\theta_j$, and material indicator $s_k$.}
    \label{fig:rectHole}
\end{figure}

\begin{multicols}{2}
The paper is organized as follows: \cref{sec:homTO} presents the multi-scale optimization problem, including new microstructure geometry and infill regularization. In \cref{sec:DeHom} the frame field-aligned de-homogenization is reintroduced with a short description of previous work in addition to modifications and extensions included in this paper. \cref{sec:imp} includes the numerical implementation. In \cref{sec:ex} the method is demonstrated on different examples. The final \cref{sec:Discussion} includes conclusions and final remarks to the work.
\end{multicols}
\begin{multicols}{2}[\section{Multi-Scale Topology Optimization}\label{sec:homTO}]
The goal of the first step of the de-homogenization procedure is to obtain an optimized multi-scale solution that can be mapped and upscaled to a physical structure. The multi-scale optimization problem is illustrated in \cref{fig:rectHole} with the computational domain $\Omega \in \mathbb{R}^3$, which is partitioned into an active design domain  $\Omega_A \subseteq \Omega$ and a passive domain $\Omega_P \subseteq \Omega$. $\Omega$ is subjected to tractions $F$ applied to Neumann boundaries $\Gamma_N$, while zero prescribed displacements are applied to Dirichlet boundaries $\Gamma_D$. The objective of the optimization problem is to maximize stiffness, i.e., minimize compliance (external work) on $\Omega$. The domain $\Omega$ is discretized by $N_e$ finite elements on an unstructured grid, with $N_e^A$ elements in $\Omega_A$. The microstructure design material is assumed to be point-wise homogeneous on the macroscopic scale. Hence, the multi-scale design problem is cast as a homogenization-based topology optimization problem, i.e. the design domain is explicitly posed on the macroscopic scale and implicitly on the microscopic scale. The microstructure design material description is constant within each finite element. The following sections will formulate the microstructure parametrization, optimization problem, objective penalization and design regularization.
 
\subsection{Microstructure Design Parametrization}
The microstructure is parameterized on the macroscopic scale to control the mechanical properties of the homogenized microstructure. There are numerous ways to parameterize the microstructure. In this work, the microstructure is considered as a two-phase composite where one phase is stiff with a base Young's modulus of $E_0$ and the other phase is compliant with a base Young's modulus of $E_{\min} = 10^{-6}\,E_0$, mimicking void. Both phases have a base Poisson's ratio of $\nu_0 = 1/3$. The microstructure is parameterized by three different sets of design variables for a material point on the macroscale $e \in \Omega$,
\begin{equation}
    \bm{x}_e = \{w_i,\theta_j,s_k \}, \quad i\in\{1,...,N_w\}, \, j\in\{1,...,N_\theta\}, \, k\in\{1,...,N_s\},
\end{equation}
where $w_i \in [w_{\min},w_{\max}]$ is the $i$th relative thickness variable of $N_w$ thicknesses. $w_{\min}$ and $w_{\max}$ are upper and lower thickness bounds, $0 \leq w_{\min} \leq w_{\max} \leq 1$. The Euler angle $\theta_j\in [-4\pi,4\pi]$ refers to the $j$th Euler angle variable of $N_\theta$ total angles, used to determine the orientation of the microstructure. 
Remark that the bounds on $\theta_j$ are chosen to ensure enough slack such that the variable will not reach these bounds during optimization. The $k$th material indicator variable, $s_k \in [{0},{1}]$, introduced in ~\cite{Giele2021a} is included to control present and non-present material, and regulate material layout as discussed in \cref{sec:reg}. Regularization schemes are used on $w_i$ and $s_k$ to obtain $\hat{w}_{i}$ and $\hat{s}_k$, respectively, which are then used for computing the objective; see \cref{sec:reg} for detail. The relative density of the microstructure is found from,
\begin{equation}\label{eq:dens}
    \rho = 1 - \prod_{i=1}^{N_w}(1 - w_{i}).
\end{equation}
\end{multicols}
\begin{figure}[tb]
    \centering
    \makebox[\textwidth][c]{
    \begin{subfigure}{62mm}
     \includegraphics{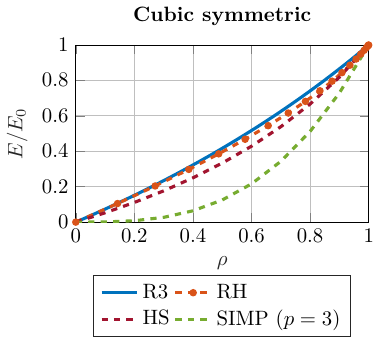} 
     \caption{}
     \label{fig:Emoduli_Cubic}
    \end{subfigure}%
    \begin{subfigure}{62mm}
     \includegraphics{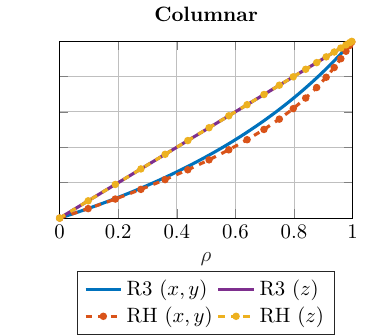}
     \caption{}
     \label{fig:Emoduli_Columnar}
    \end{subfigure}%
    \begin{subfigure}{62mm}
     \includegraphics{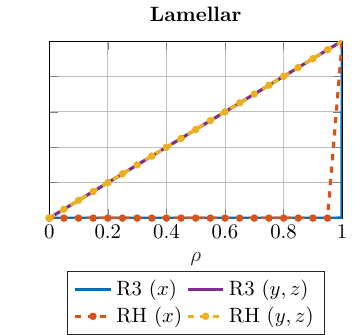}
     \caption{}
     \label{fig:Emoduli_Lamellar}
    \end{subfigure}
    }
    \caption{
    Specific modulus of (a) cubic symmetric, (b) columnar symmetric, and (c) lamellar microstructures, for the orthogonal rank-3 (R3) and rectangular-hole (RH) microstructure. In (a) the specific modulus of the isotropic Hashin–Shtrikman (HS) upper bound and the SIMP with $p=3$ are also present. Note that for the isotropic Hashin–Shtrikman the Poisson’s ratio is also dependent on the relative density, while it is constant for the other material models.
    }
    \label{fig:Emoduli}
\end{figure}
\begin{multicols}{2}
The near stiffness-optimal orthotropic rectangular-hole microstructure single-scale material model is considered for the microstructure due to its single-scale superior design freedom and further de-homogenization properties. The material is similar to the well-known orthogonal rank-3 multi-scale material (see formulation in~\cite{Groen2021}); however, consisting of a rectangular cuboid cavity with the relative dimensions $(1 - w_1) \times (1 - w_2) \times (1 - w_3)$, hence the surrounding walls, or plates (referred to as laminates), have the relative thickness $w_1$, $w_2$, and $w_3$. Therefore, $N_w=3$ and due to the orthogonality only one set of Euler angles are needed to realise the orientations such that $N_{\theta}=3$. The microstructure is illustrated in \cref{fig:rectHole}. A detailed description of the rectangular-hole microstructure is found in \cref{sec:microMat}.
The elastic properties are determined by numerical homogenization. The homogenized constitutive matrix $\tilde{\mathbf{C}}^H \in \mathbb{R}^{6\times 6}$ (Voigt notation) is obtained in the microscopic reference frame, only dependent on $w_i$. The microstructure design is expanded to include geometric rotations by,
\begin{equation}\label{eq:const}
    \tilde{\mathbf{C}}^H(w_i,\theta_j,s_k) = \mathbf{T}({\theta_j}) \, \mathbf{C}^H({w_i},s_k)\mathbf{T}^\top({\theta_j}).
\end{equation}
Here $\mathbf{T} \in \mathbb{R}^{6\times 6}$ is the orthogonal transformation matrix of the $\mathbf{R} \in \mathbb{R}^{3\times 3}$ proper orthogonal rotation matrix, determined by $\theta_j$, as detailed in \cref{sec:microMat}. 

Additionally, an isotropic Hashin-Shtrikman (HS) upper-bound microstructure~\cite{Hashin1963} is considered as a reference to the orthotropic microstructure. The HS microstructure is realized as a quasi-periodic microstructure with six uniform laminates~\cite{Allaire2002}, $N_w = 6$, and $N_{\theta} = 0$. Due to the uniform laminates, $w_i$ is an irrelevant design parameter from a multi-scale HS optimization perspective. The microstructure density, \cref{eq:dens}, is instead a much more reasonable design parameter and is thus used as the effective design parameter. A detailed description of the isotropic HS model is found in \cref{app:hs}.

As a representation of large-scale topology optimization, the well-known Solid Isotropic Microstructure with Penalization (SIMP)~\cite{Bendsoe1999} model is used for benchmarking. Like the isotropic HS microstructure, the density parameter is considered the design variable with $N_w = 1$ and $N_{\theta} = 0$. The auxiliary variable $s_k$ is not necessary for this model and hence excluded, whereas the design variable is regularized based on the same robust formulation as applied to $s_k$. A detailed description of the SIMP model is found in \cref{app:simp}.

A comparison in specific Young's modulus is seen in \cref{fig:Emoduli}. The specific Young's modulus is calculated by evaluating the compliance tensor obtained from \cref{eq:const} in relation to the relative density. A slight difference is seen between the rank-3 and rectangular-hole microstructures, which are contributed to the single-scale representation of the rectangular-hole microstructure. It is seen that the two isotropic materials are penalized, compared two the orthotropic materials. Note that isotropic Hashin-Shtrikman Poisson's ratio is also dependent on $\rho$, resulting in further penalization of the constitute tensor. The two orthotropic materials are seen to have a high range of design freedom, from the quasi-linear specific modulus, hence making them extremely efficient for stiffness-specific multi-scale topology optimization.

\subsection{Homogenization-based Topology Optimization Problem}
For the homogenization-based topology optimization Problem, the three different sets of design variables are defined as three design vectors for all elements in $\Omega_A$,
\begin{equation}
    \mathbf{x} = \{ \bm{w},\bm{\theta},\bm{s} \}.
\end{equation}
The compliance minimization problem is defined with the objective function $f(\mathbf{x})$, a weighted sum of the compliance, $\mathcal{J}(\bm{w},\bm{\theta},\bm{s})$, and two regularization penalty functions; orientation $\mathcal{P}^{\theta}(\bm{\theta})$, and relative thickness, $\mathcal{P}^{s}(\bm{s})$. The compliance is computed as,
\begin{equation}
    \mathcal{J} = \sum_i^M\mathbf{f}_i^{\top}  \mathbf{u}^{\,}_i,
\end{equation}
where index $i$ refers to one of the $M$ load cases, $\mathbf{f}_i$, and $\mathbf{u}_i$ is the finite element load and displacement vectors obtained from the solution of the linear finite element problems
\begin{equation}
    \mathbf{K}(\bm{w},\bm{\theta},\bm{s}) \mathbf{u}_i = \mathbf{f}_i,
\end{equation}
 where $\mathbf{K}$ is the finite element stiffness matrix based on \cref{eq:const}. Remark that the microstructure description is maintained for the multiple load case formulation. The optimization problem is defined as,
\begin{equation}\label{eq:TopOpt2}
\begin{aligned}
 & & \displaystyle \min_{\mathbf{x}} & :   f(\mathbf{x}) = \gamma_1 \mathcal{J}(\bm{w},\bm{\theta},\bm{s})  + \gamma_2 \mathcal{P}^{\theta}(\bm{\theta})  + \gamma_3  \mathcal{P}^{s}(\bm{s}),     \\
 & & \textrm{s.t.}             & :  \mathbf{K}(\bm{w},\bm{\theta},\bm{s}) \mathbf{u} = \mathbf{f},  \\% \text{   for } m = \{1,...,M\}, 								   \\
 & & 						     & :  g(\bm{w},\bm{s}) \leq 0, 	\\
 & &                           & :     \underline{\mathbf{x}} \leq  \mathbf{x}   \leq  \overline{\mathbf{x}}.			\\
\end{aligned}
\end{equation}
The compliance objective is scaled by the compliance of a solid design domain ($\bm{w} = \bm{1}$), by considering $\gamma_1 = 1/\mathcal{J}^{(0)}$. The two penalty functions are scaled by $\gamma_2$ and $\gamma_3$, respectively, and $\mathcal{P}^{\theta}(\bm{\theta})$ is computed as an aggregated sum of the dot products between neighboring element lamination normals, whereas $\mathcal{P}^{s}(\bm{s})$ is computed as an aggregation of all elements with a non-zero volume fraction. A detailed description of the penalty functions is found in \cref{sec:penal}. The optimization problem is subjected to an inequality constraints $g(\bm{w},\bm{s})$ associated with the macroscopic volume fractions, detailed in \cref{sec:lv}. Finally, the design variables are subjected to box constraints, $\underline{\mathbf{x}}$ and $\overline{\mathbf{x}}$. A set of outer move limits are imposed on the design variable changes, which limits the change between design iterations as $\mathbf{x}_\text{move} = \left\{\{0.1\}, \{0.05\}, \{0.2\}\right\}$, for $\bm{w}$, $\bm{\theta}$, and $\bm{{s}}$, respectively. Sensitivities with respect to the optimization problem are found using the discrete adjoint method.

\subsection{Regularization}\label{sec:reg}
Regularization schemes are often used to avoid unphysical behavior and control length scale. In this work, additional regularization is applied to the relative thicknesses to avoid low-density material and control the lower bound on the thicknesses. This approach is similar to \citet{Jensen2022} but extended both to 3D unstructured grids and capabilities to control the number of active layers. This regularization, as stated before, is based on the material indicator fields, which work both laminate-wise ($N_s=N_w$) i.e. $k=i$ or material-wise ($N_s=1$) i.e. $k=1$. $s_k$ is compounded with $w_i$. The variables $w_i$ and $s_k$ are filtered using the PDE-filter~\citep{Lazarov2011} with filter radii $R^{w}$ and $R^{s}$ to obtain $\tilde{w}_{i}$ and $\tilde{s}_{k}$, respectively. The SIMP density design variable is filtered with the filter radii $R^{\rho}$. The PDE-filter is implemented with Robin boundary conditions on $\partial\Omega_A$~\cite{Wallin2020} to avoid the use of padded design domains, unless $\partial\Omega_A = \partial\Omega_P$ for which Neumann boundary conditions are used. The indicator field, $\tilde{s}_{k}$, is transformed to a smooth unit step $\bar{{s}}_k^{m}$ using a smooth Heaviside projection following the modified robust approach~\citep{Wang2011} with step approximation parameter $\beta_s$ and threshold values $\eta_s^{m}$. Here $m\in\{e,i,d\}$ indicate \textit{eroded}, \textit{intermediate}, and \textit{dilated} fields. The steepness parameter, $\beta_s$, is continuated from $\beta_s = 0.1$ to $\beta_s = 64$ every 20th design iteration with an update exponent of $1.5$. The threshold values is defined as $\eta_s^{i} = 0.5$, $\eta_s^{d} = \eta_s^{i} - 0.01$ and $\eta_s^{e} = \eta_s^{i} + 0.01$, hence $\bar{{s}}_k^{m}$  is obtained by
\begin{equation}\label{Eq:HomoWidth.1}
    \bar{{s}}_k^{m}= \mathit{H}(\tilde{{s}}_{k},\,\beta_s,\,\eta_s^{m}),
\end{equation}
where $\mathit{H} \in [0,1]$ is a smooth Heaviside function,
\begin{equation}
    \mathit{H}(x,\beta,\eta) = \frac{\tanh\left(\beta \eta\right) + \tanh\left(\beta \left(x- \eta\right)\right)}{\tanh\left(\beta \eta\right) + \tanh\left(\beta \left(1- \eta\right)\right)}.
\end{equation}
In a previous 2D application~\cite{Jensen2022} of the material indicator field, the field was enforced as a material-wise indicator to advance microstructural stability as the critical member length is minimized this way. However, in 3D, enforcing all laminates to be active to advance stability is unnecessary. Hence it is desirable to let the number of active laminates be controlled by the user, which also enables more direct manipulation of the infill layout. To this end, let $\eta_a=N_a/N_s$ be a fraction of active laminates in the microstructure, where $N_a \in \{0,N_s\}$ is the required number of active laminates. Let $\mathit{A} \in [0,1]$ be a normalized $2\text{-}\mathrm{norm}$ function of the number of active laminates in an element microstructure,
\begin{equation}
   \mathit{A}(\bm{\bar{{s}}}^m) = N_s^{-1/2}\,||\bm{\bar{{s}}}^m||_2, \quad \bm{\bar{{s}}}^m = \{ \bar{{s}}_1^m,...,\bar{{s}}_k^m\},
\end{equation}
The smooth Heaviside function is now used to project $\mathit{A}$  at point $\eta_a$ to determine if the microstructure has $N_a$ active laminates,
\begin{equation}
    \mathit{\bar{A}}(\bm{\bar{{s}}}^m) = \mathit{H}\left( \mathit{A}(\bm{\bar{{s}}}^m) ,\,\beta_a,\,\eta_a \right),
\end{equation}
where $\mathit{\bar{A}} \in [0,1]$ is the indicator quantity and $\beta_a=24$ is the steepness parameter. Finally, $\mathit{\bar{A}}$ is used as an indicator design variable penalty scalar,
\begin{equation}\label{eq:indicatorPeanlty}
    \hat{s}_k^m = \bar{{s}}_k^m \, \mathit{\bar{A}}(\bm{\bar{{s}}}^m).
\end{equation}
Now, the material indicator field will be directly penalized if the active laminates requirement count is not satisfied. Note that setting $N_a = N_s$, corresponds to setting $N_s = 1$.

The computation of the constitutive matrix, in \cref{eq:const}, is now performed for $\hat{w}^{e}_{i}$, which is obtained from the eroded field $\hat{s}^{e}_k$ as follows
\begin{equation}\label{Eq:HomoWidth.2}
    \hat{w}^{e}_{i} = \hat{s}_k^{e} \, \tilde{w}^{\,}_{i}.
\end{equation}
\vspace{-5mm}
\subsection{Penalization}\label{sec:penal}
To ensure smooth orientations in the subsequent mapping procedure and prevent the presence of extensive regions with low material density, the two previously introduced penalty functions are employed in the objective. The penalty functions are based on those presented in ~\cite{Groen2021} and~\cite{Jensen2022}, for $\mathcal{P}^{\theta} \in  [0,1]$, and $\mathcal{P}^{s} \in [0,1]$, respectively, but are here extended to also allow for unstructured grids.

To ensure smooth orientations fields, an element penalization function $\mathcal{P}^{f}_i \in [0,1]^{N_{f}} $, where $N_{f}$ is the number of element pairs, is introduced to penalize the difference between the $i$th laminate normal of the element neighboring $f$th surface pair, $a$ and $b$, respectively. Let $d_i$ be the dot product between laminate normal $a$ and $b$,
\begin{equation}
    d_i(\{ \bm{\theta}_i \}^f) = \mathbf{n}_i(  \bm{\theta}_i  ^{a}) \cdot \mathbf{n}_i(\bm{\theta}_i^{b}), \quad \{\bm{\theta}_i\}^f = \{ \bm{\theta}_i^a,\bm{\theta}_i^b \}, \quad \bm{\theta}_i = \{\theta_1,...,\theta_j\},
\end{equation}
where $\mathbf{n}_i\in \mathbb{R}^3$ is the laminate surface normal, see \cref{sec:microMat} for definition.
$\mathcal{P}^{f}_i$ must take the minimum value when the neighboring laminates are orthogonal or parallel, i.e., $d = \{-1,0,1\}$, and the maximum value when the neighboring laminates are separated by $\pi/4$, i.e. $d = \pm \sqrt{2}/2$. Finally, $\mathcal{P}^{f}_i$ is aggregated and normalized with respect to all laminates and element pairs to $\mathcal{P}^{\theta}$,
\begin{equation}
    \mathcal{P}^{\theta}(\bm{\theta}) = \frac{1}{3\,N_f} \sum_{f=1}^{N_{f}}\sum_{i=1}^{N_w}  \mathcal{P}^{f}_i (\{\bm{\theta}_i\}^f), 
\end{equation}    
    with,
\begin{equation}    
    \mathcal{P}^{f}_i (\{\bm{\theta}_i\}^f)=  4 d_i(\{\bm{\theta}_i\}^f) ^2 - 4  d_i(\{\bm{\theta}_i\}^f)^4.
\end{equation}
In order to avoid large regions of the design domain where the microstructure wall thickness obtains the minimum thickness, i.e., $w_i = w_{\min}$, the sum of volumes of the material indicator fields is minimized~\cite{Giele2021a}. Let $\mathcal{P}^{s} \in [0,1]$ be defined as,
\begin{equation}\label{Eq:HomoWidth.4}
    \mathcal{P}^{s} (\bm{s}) = \frac{1}{V_{\Omega_A} N_s} \sum_{k=1}^{N_s}  \int_{\Omega_A} \hat{{{s}}}_k^{e} \, \mathrm{d}\Omega_A,
\end{equation}
where $V_{\Omega_A}$ is the total volume of $\Omega_A$. Hence $\mathcal{P}^{s} = 1$ if all elements have material, and if $\mathcal{P}^{s} = 0$ none have.

\subsection{Volume Constraint}\label{sec:lv}
The optimization problem is subjected to a macroscopic volume fraction constraint that is imposed on the dilated design, 
\begin{equation}\label{Eq:HomoWidth.3}
g(\bm{w},\bm{s}) = \frac{V^d}{V_{\Omega_A} f^d} - 1, \quad \text{where,} \quad f^{d} = \frac{V^{d}}{V^{i}} f^{i},
\end{equation}
% where, 
% \begin{equation}
% f^{d} = \frac{V^{d}}{V^{i}} f^{i},
% \end{equation}
with,
\begin{equation}
V^{m} = \int_{\Omega_A} \rho(\hat{{w}}_i^m) \, \text{d}\Omega_A.
\end{equation}
where $f^{i}$ is the specified intermediate volume fraction, and $f^{d}$ is the dilated volume fraction which is updated every 20 design iterations to ensure that the volume of the intermediate design meets $f^{i}$.
\end{multicols}
\begin{multicols}{2}[\section{Frame Field-Aligned De-homogenization}\label{sec:DeHom}]
The microstructure design can be approximated on a single scale using a de-homogenization approach. In general, any de-homogenization process operates by employing a set of vector fields to represent microstructure orientations, $\mathcal{F} = \left\{\mathbf{n}(\bm{\theta}_1), ... , \mathbf{n}(\bm{\theta}_{N_w}) \right\}$, and a set of scalar fields representing laminate thicknesses, $\mathcal{W} = \{w_1, ..., w_{N_w} \}$. The goal of the process is first to find a geometric description that aligns with $\mathcal{F}$ at every point in $\Omega$. Then, create a volumetric solid from the geometric description, representing $\mathcal{W}$.
In the case of the rectangular-hole microstructure, the set of orientations $\mathcal{F}$ is often referred to as a Frame-Field, since the orthonormal microstructure orientations locally lead to a set of basis vectors, or a frame.
In this work, the stream surface de-homogenization procedure~\cite{Stutz2020} is used and extended in the following ways: Improvements have been made to the thickness of each surface in the final structure; the outer hull briefly mentioned by~\cite{Stutz2020} is now default, and used to define the outer shape; surface precision now adapts to embedded error metrics and along structural boundaries; frame field implementation have been extended to accept unstructured meshes.
The following sections briefly summarize the original procedure and highlight modifications.

\begin{figure}[b!]
    \centering
    \makebox[\textwidth][c]{
    \begin{subfigure}{47.5mm}
    \includegraphics[width=\linewidth]{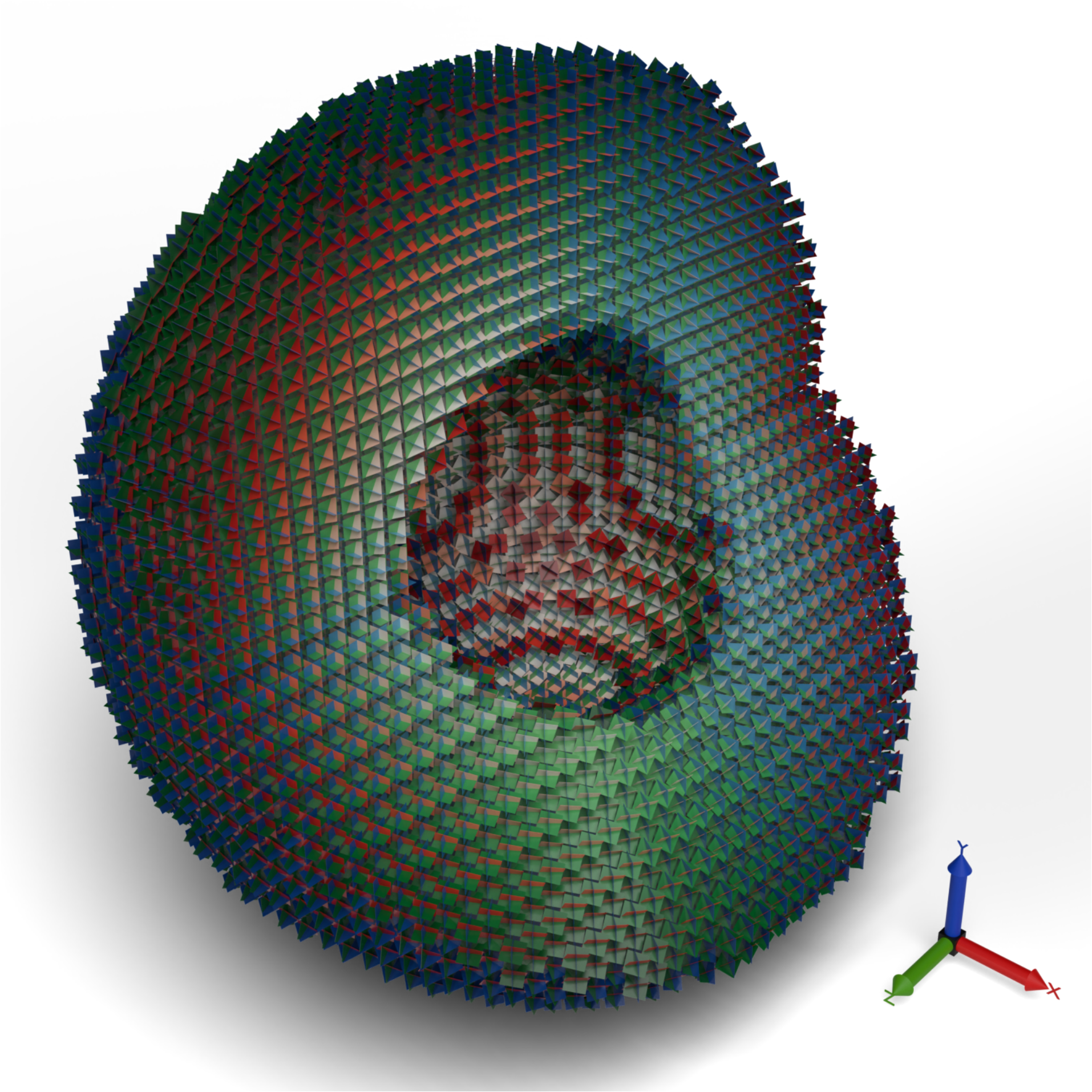}
    \caption{}
    \label{fig:deHom:ss_T}
    \end{subfigure}%
    \begin{subfigure}{47.5mm}
    \includegraphics[width=\linewidth]{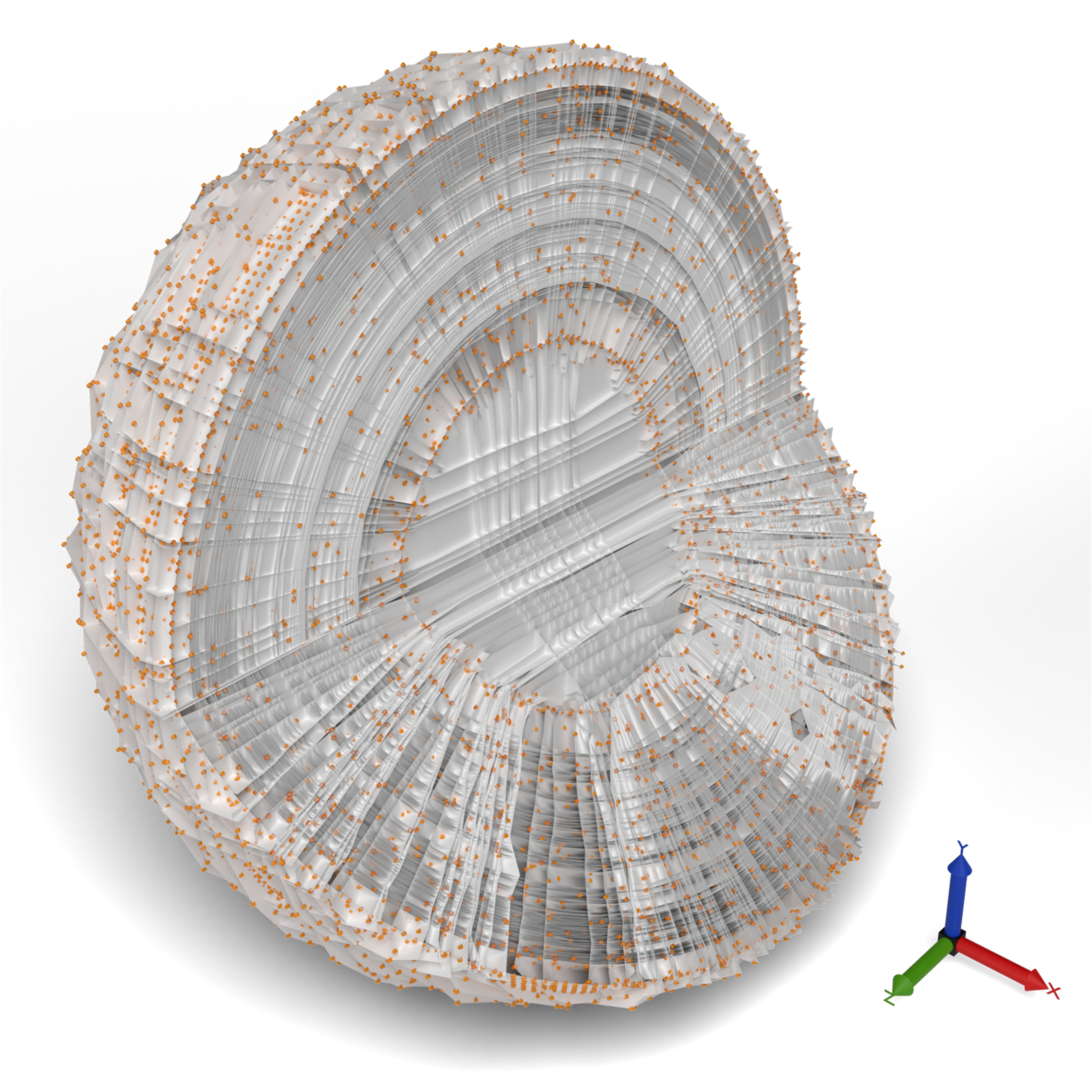}
    \caption{}
    \label{fig:deHom:ss_F}
    \end{subfigure}%
    \begin{subfigure}{47.5mm}
    \includegraphics[width=\linewidth]{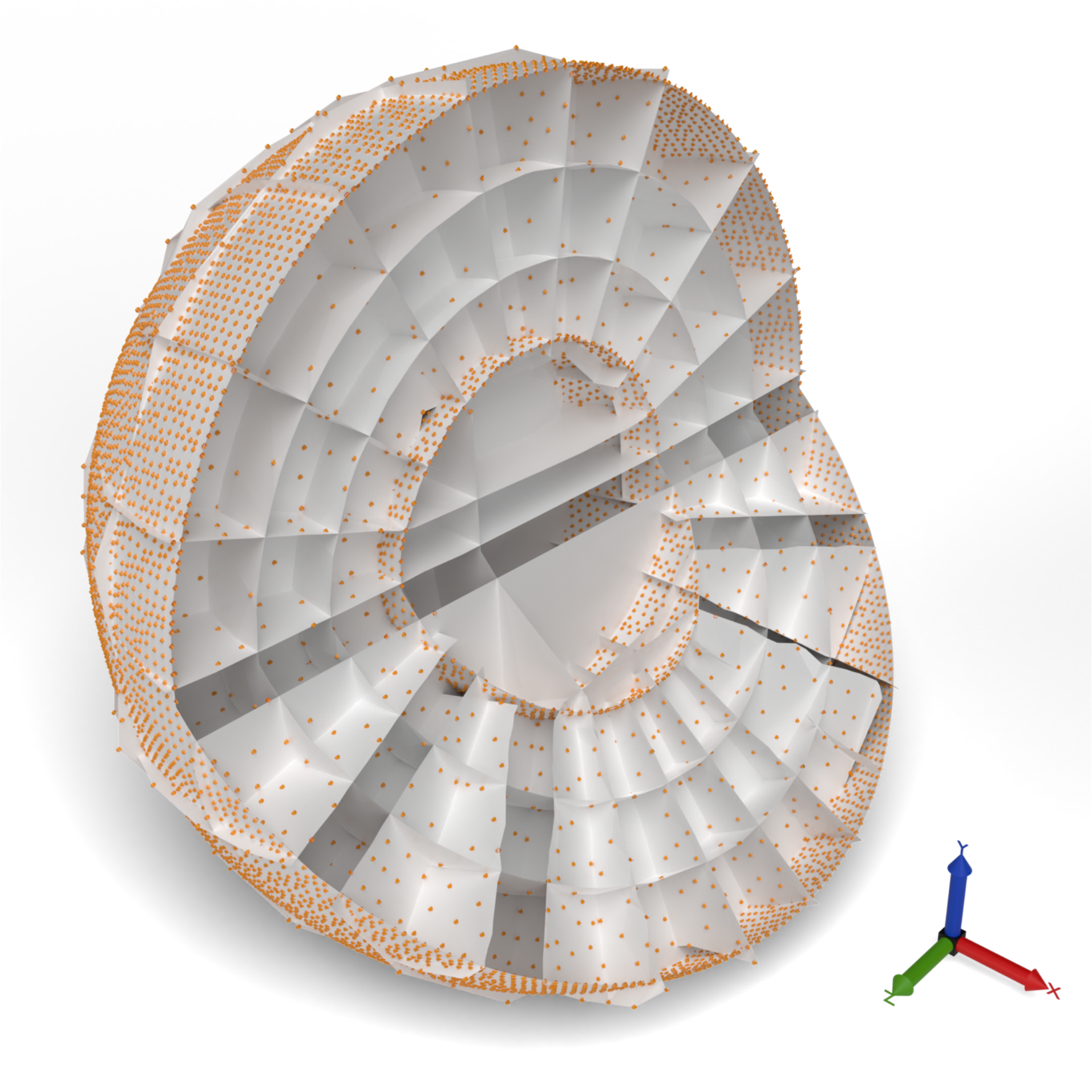}
    \caption{}
    \label{fig:deHom:ss_O}
    \end{subfigure}%
    \begin{subfigure}{47.5mm}
    \includegraphics[width=\linewidth]{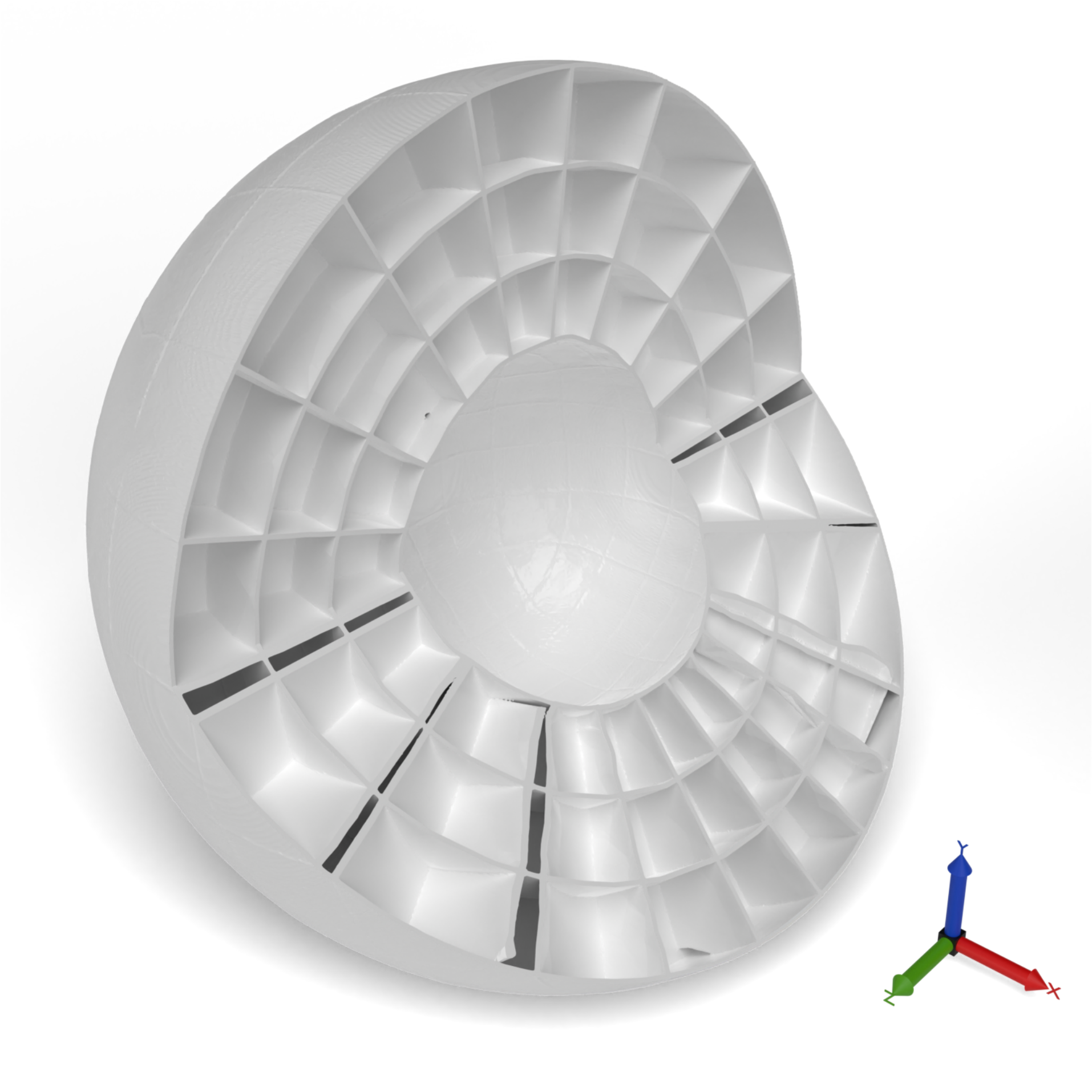}
    \caption{}
    \label{fig:deHom:ss_S}
    \end{subfigure}
    }%
    \caption{Illustration of a hollow sphere example's stream surface de-homogenization procedure.
    (a) naive illustration of the frame field $\mathcal{F}$ in the form of point-wise tangent planes to the laminate frame normals. The first laminate is normal to the surface of the sphere, while the two other laminates are tangential to the sphere surface and a fixed plane. The red, blue, and green colors of the tangent planes represent the thickness of $\mathcal{W}$.
    (b) shows the full set of surfaces, $\mathcal{S}$, generated from $\mathcal{F}$, which densely contains the whole domain.
    (c) shows a selected set of surfaces $\mathcal{S}_{\text{opt}}$ with uniform spacing.
    (d) shows how a selected set of surfaces $\mathcal{S}_\text{opt}$ is merged together into a volumetric solid based on the relative thicknesses $\mathcal{W}$. }
    \label{fig:deHom:ss}
\end{figure}

The stream surface de-homogenization procedure is a three-stage process illustrated in \cref{fig:deHom:ss}. The element-wise constant $\mathcal{F}$ and $\mathcal{W}$ from \cref{eq:TopOpt2} are passed as volume-weighted average node data from the original finite element grid to start the de-homogenization procedure illustrated in \cref{fig:deHom:ss_T}. This input data is used to generate a super-set of surfaces, $\mathcal{S}$, perpendicular to $\mathcal{F}$, densely covering $\Omega$ (\cref{fig:deHom:ss_F}). From the super-set, a set of uniformly distributed surfaces, $\mathcal{S}_\text{opt}$, can be selected (\cref{fig:deHom:ss_O}). Finally, a volumetric solid domain $\Omega_S \subseteq \Omega$ is created from $\mathcal{S}_\text{opt}$, and the corresponding $\mathcal{W}$ (\cref{fig:deHom:ss_S}). $\Omega_S$ can then be meshed for mechanical analysis and processed for manufacturing.

\subsection{Generation of Member Super-set}\label{sec:surfaceGeneration}
A stream surfaces approach will only require local separation of the frame field. Each stream surface can be constructed from the well-established theory of surface tracing in a vector field. As was the case of~\citet{Stutz2022}, a Runge-Kutta method of fourth order (RK4)~\cite{Chapra2011Applied} is used to construct the stream surfaces. An outline of the method and any modifications is given below, however, see \cite{Stutz2022} for the full details.

A surface, defined by points $\mathbf{p}$, is started at a random seed point $p_S$, where a random laminate normal is chosen to represent the normal of the surface, $\mathbf{n}_S$. The surface is generated as an expanding front, where all points added to the surface will be used to create new points. This allows the surfaces to be densely generated in $\Omega$, stopping when the laminate thickness falls under the void threshold ($w_{\min}$) or the local density exceeds the solid threshold ($\rho_{\max}$). The solid and void thresholds are defined by the parameter $\eta_S$, such that $\eta_S = w_{\min}$ and $\rho_{\max} = 1 - \eta_S$. Thus, a new point $p_i$ with associated normal $\mathbf{n}_i$ is only added to the surface, if the evaluated layer thickness, $w_i$, and local density satisfy the following,
\end{multicols}
\newpage
\begin{multicols}{2}
\begin{equation}
    \begin{split}
        &w_i = \mathcal{W}(p_i, \mathbf{n}_i) \geq w_{\min}, \\
        &\rho(p_i) \leq \rho_{\max}.
    \end{split}
\end{equation}

Solid regions are afterward appended to the volumetric solid as $w_i = 1$. The void region is explicitly defined in the solution of \cref{eq:TopOpt2}; however, the solid region is not explicitly defined and is thus an estimate.
Furthermore, if the surface differs significantly from a plane locally, the surface tracing is also discontinued to prevent surface splitting and to maintain a smooth, continuous surfaces. New points are added to the surface by selecting a random direction orthogonal to the surface normal at the parent point and the RK4 method is used to determine the location of the new point, such that the surface follows the frame field locally. In order to improve the spatial consistency of points, all nearby points are used to estimate the position of the new point.

In the work of~\citet{Stutz2022} the surface generation was based on a uniform step size. In this work, an adaptive step size is used instead. The surface's point density can adapt to various domain changes using the Dormand-Prince method of 5th order with a 4th order embedded error measure~\cite{Dormand1980}. This is particularly useful for complex unstructured geometries, where domain features can easily disturb a smooth strain field.
As described by \citet{Stutz2022}, a vast super-set of surfaces should be constructed. Estimates of how many surfaces should be created based on the expected complexity of the final structure can be found in that paper.

\subsection{Selection of Surfaces}\label{sec:surfaceSelection}
Given the generated super-set of $N_S$ surfaces, $\mathcal{S}$, constructed as explained above, \citet{Stutz2022} provide the specifications for selecting a uniformly distributed sub-set $\mathcal{S}_\text{opt} \subseteq \mathcal{S}$. For completeness, a short overview of the selection method as described in \cite{Stutz2022} is given below.

Given a desired minimal feature thickness, $\delta_{\min}$, each surface will activate some region of $\Omega$. Hence, $\mathcal{S}_\text{opt}$ must be selected such that the surfaces are uniformly distributed. At the same time, the volume fraction is satisfied in order to comply with the solution of \cref{eq:TopOpt2}. The set $\mathcal{S}_\text{opt}$ is obtained by minimizing an energy $\mathcal{E}$ over binary design variables $\bm{\omega}\in\{0,1\}$ that will be assigned to the stream surfaces. Note if the surface selection strictly abides by the solution of \cref{eq:TopOpt2}, $N_\mathcal{S} \rightarrow \infty$ and $\delta_{\min} \rightarrow w_{\min}\lambda$, where the local surface spacing $\lambda \rightarrow 0$, however, this is not practical. Thus selecting $0 < \delta_{\min}$ will dissociate the solution of  \cref{eq:TopOpt2} with the de-homogenization solution.

Let $\mathbf{I} \in \{0,1\}^{N_\mathcal{S}}$ be an indicator function used to determine if a point is activated for a given surface $S$ can be described by,
\begin{align} \label{eq:dehom:activation}
    \mathbf{I}_{S,\mathbf{n}}(\mathbf{p}) & = \begin{cases}
                            1, & \text{if } \mathbf{n} \newparallel \mathbf{n}_\mathcal{S} \wedge \left\| \mathbf{p} - S \right\| < \frac{\lambda}{2}, \\
                            0, & \text{otherwise}.
                        \end{cases}
\end{align}
where ``$\newparallel$'' indicates parallelism.

\citet{Stutz2022} defined a binary optimization problem that is solved in 2 steps. First, a relaxed continuous problem is solved to find a set of continuous weights $\tilde{\bm{\omega}}\in[0,1]$ used to restrict the solution space.
\begin{equation}\label{eq:dehom:select:relax}
\begin{aligned}
 & & \displaystyle \min_{ \tilde{\bm{\omega}} } & :  \mathcal{E}(\tilde{\bm{\omega}}) = \int_\Omega \left\| \sum_{S \in \mathcal{S}} \tilde{\omega}_S \mathbf{I}(\mathbf{p}) - \mathbf{1} \right\|_{L_1} \mathrm{d} \mathbf{p},     \\
 & & \textrm{s.t.}             & :   \tilde{\omega}_S = 1, \quad S \in \left\{ \mathcal{S}_f \right\}.
\end{aligned}
\end{equation}
Here $\mathcal{S}_f$ is a set of fixed surfaces that can manually be added to $\Omega$.
Based on the restricted solution space, the binary problem could be solved to find the selected list of surfaces $\bm{\omega}$.
\begin{equation}\label{eq:dehom:select:full}
\begin{aligned}
 & & \displaystyle \min_{ {\bm{\omega}} } & :   \mathcal{E}(\bm{\omega}) =\int_\Omega \left\|\sum_{S\in\mathcal{S}} \omega_S \mathbf{I}(\mathbf{p})-\mathbf{1} \right\|_{L_1} \mathrm{d} \mathbf{p},     \\
 & & \textrm{s.t.}             & :   \omega_S = 0, \quad S \in \left\{ \mathcal{S} \mid \tilde{\omega}_S = 0 \right\}, \\
 & &                           & :   \omega_S = 1, \quad S \in \left\{ \mathcal{S} \mid \tilde{\omega}_S = 1 \right\},\\
 & &                           & :   \omega_S = 1, \quad S \in \left\{ \mathcal{S}_f \right\}.
\end{aligned}
\end{equation}
The addition of $\mathcal{S}_f$ has been used to add an enveloping membrane around the material in $\Omega$. This membrane will ensure the connection between internal members while providing a high-quality definition of the outer shape.

\subsection{Volumetric synthesis}
The optimized selected set of surfaces $\mathcal{S}_\text{opt}$ is defined as point clouds. However, the surfaces must be merged together into a volumetric structure in $\Omega_S$ for practical applications of the procedure. The process for constructing the volumetric model was described in detail~\cite{Stutz2022}; however, in this work, the construction is notably different in relation to the thickness of each surface. The original approach assumed that the final structure adhered very closely to the target spacing. That was not always true, especially close to boundary conditions where the structural members often converge. If the spacing locally is larger than the target spacing, then surfaces are too thin, which can be a problem for manufacturing, stability, etc. If the spacing is locally smaller than the target spacing, then too much material is allocated to the region. In order to correct this, the distances are computed between neighboring surfaces with approximately the same orientation and scale the thicknesses of the surfaces according to the computed distances.
\end{multicols}
\begin{multicols}{2}[\section{Implementation on Complex Unstructured Geometries}\label{sec:imp}]
The topology optimization problems is implemented in a PETSc-based 3D unstructured topology optimization framework \cite{Traff2021,Jensen2021,Hoeghoej2022}. The finite element system is solved using a geometric multigrid preconditioned Flexible Generalized Minimal RESidual method~\cite{petsc-web-page,petsc-user-ref,petsc-efficient}.
The optimization problem is solved in a nested form, using an implementation of the Method of Moving Asymptotes (MMA)~\cite{Svanberg1987,Aage2013}. The finite element mesh is generated with isoparametric hexahedral tri-linear elements for advantageous accuracy and performance. The hierarchy of meshes necessary for the geometric multigrid preconditioner is all generated using the meshing software Coreform Trelis Pro~\cite{cubit}.
In order to get good multigrid performance, the geometric multigrid levels are obtained on the same Cartesian grid alignment. The weights for the prolongations are computed using the trilinear finite element shape functions for fine grid mesh nodes lying within the coarse mesh and an inverse weighting scheme for fine grid nodes lying outside the bounding box of the coarse grid. The multigrid preconditioner utilizes four steps of either SOR or Chebyshev smoothening, and the coarse grid correction is obtained using a parallel LU solver.
The optimization results are obtained on the DTU Sophia cluster with two AMD EPYC 7351 16-Core processors and 128 GB memory per node.
The stream-surface de-homogenization is implemented primarily as a C++ framework. The framework is written with explicit support for the Message Passing Interface (MPI). However, the selection problem is solved through a Branch-and-Bound algorithm, implemented in a Matlab-based library for convex optimization CVX, using the commercial Mosek solver \cite{cvx,Mosek}. Combining surfaces together into a volumetric structure is done in a mostly single-threaded C++ program.

In order to run a mechanical analysis of the post-processed volumetric structures the geometry is meshed with Coreform Trelis Pro again using isoparametric hexahedral tri-linear elements. Due to the geometric complexity of the volumetric structures, an algebraic multigrid solver is used to solve the linear elasticity problem for post-analysis.
\end{multicols}

\begin{multicols}{2}[\section{Numerical Examples}\label{sec:ex}
\begin{figure}[b!]
    \centering
    \makebox[\textwidth][c]{
        \begin{subfigure}{82.728mm}
            \includegraphics[width=\linewidth]{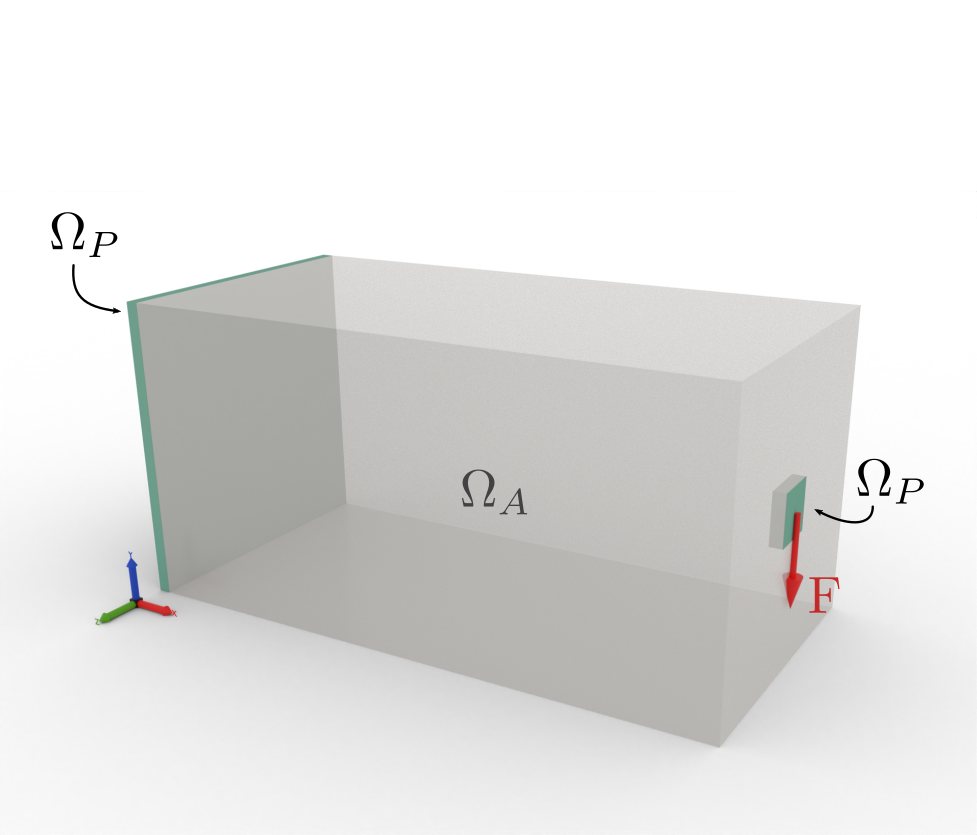}
            \caption{}
            \label{fig:MichellModel}
        \end{subfigure}%
        \begin{subfigure}{36.887mm}
            \includegraphics[width=\linewidth]{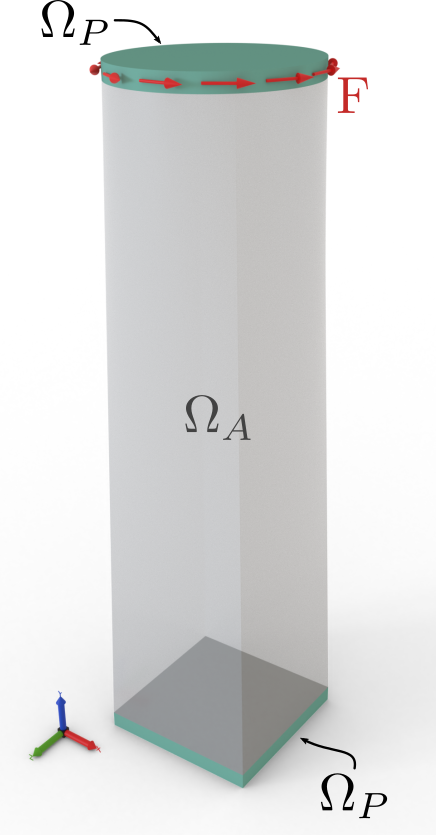}
            \caption{}
            \label{fig:lotteModel}
        \end{subfigure}%
        \begin{subfigure}{70.386mm}
            \includegraphics[width=\linewidth]{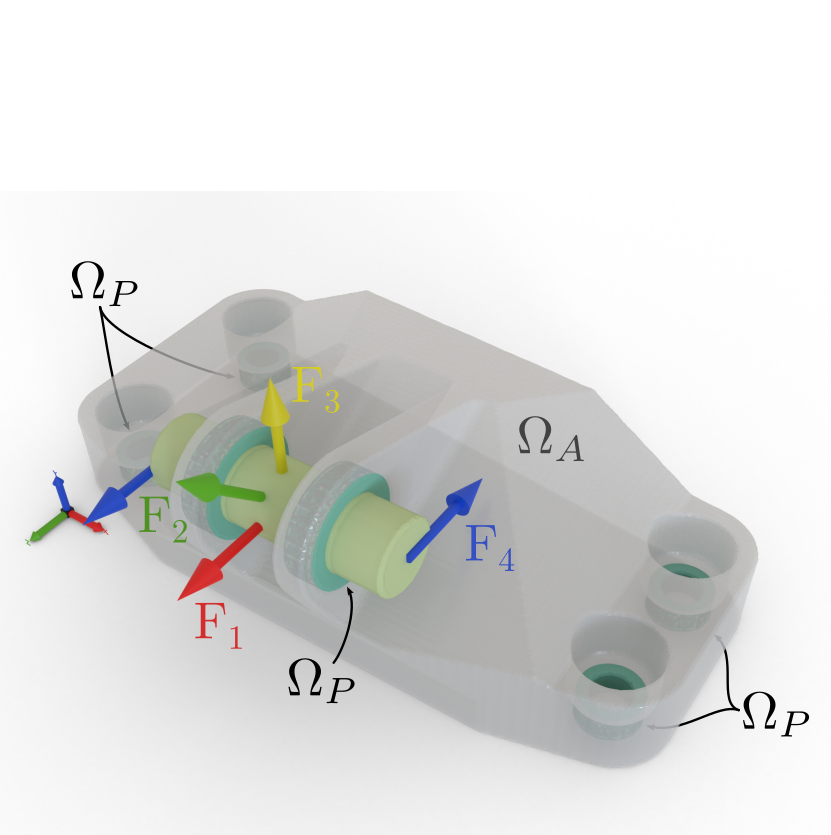}
            \caption{}
            \label{fig:bracketModel}
        \end{subfigure}
    }
    \caption{
        Numerical examples: The gray semi-transparent regions represent the active design domains, $\Omega_A$, while the passive domains, $\Omega_P$, are represented in green. Boundary conditions are exclusively applied to $\Omega_P$, with surface tractions indicated by arrows.
        (a) The Michell cantilever: The domain is extended with a passive back plate where the cantilever is clamped at the end. The front of the cantilever is loaded with a surface traction.
        (b) The Lotte tower: The domain is designed as a square that morphs into a circle from the bottom to the top. Both the square and the circle have the same surface area. The tower includes passive domains at the top and bottom. The bottom surface is fixed, while a tangential surface traction is applied to the side of the top passive domain.
        (c) The GE Jet Engine Bracket: The domain is modelled after a jet engine bracket, secured by four passive bolt connections represented as solid fixed rings. Four load cases are applied to a passive shaft with artificially higher stiffness (light green color). The shaft is connected to the design domain through passive solid rings.
    }
    \label{fig:examples}
\end{figure}

]
Three examples are considered to verify the procedure: a standard Michell cantilever, the Lotte Tower, and a GE Jet Engine Bracket. All three examples are illustrated in \cref{fig:examples} including the load cases, boundary conditions, passive domain, and design domain. The active design domain, $\Omega_A$, is shown in transparent gray color, while the passive domain, $\Omega_P$, is shown in green color. The load cases and boundary cases are only applied to $\partial\Omega_P$. The loads are applied as surface tractions shown as arrows on the respective surfaces. The boundary conditions are also only applied over surfaces.

$\Omega$ is discretized with the mesh $\mathcal{T}$ and $\mathcal{T}^\rho$ for the multiscale optimization and large-scale SIMP optimization, with average element sizes $h$ and $h^\rho$, respectively. $\Omega_S$ is discretized with the mesh $\mathcal{T}^S$ for the post-analysis, with an average element size $h^S = 5/4 \delta_{\min}$. The meshes and following results of the examples are available for download as Exodus II formats with side sets for applied load cases and boundary conditions. The files can be found at~\cite{uTopDeHom}.

The filter radii are defined as $R^w = 2h$ and $R^s = 6h$, and $R^\rho = 2.5h^\rho$ for the multiscale optimization and large-scale SIMP optimization problems, respectively.

The design variables are initialized uniformly with $w_i = 1 - (1 - f^i)^{1/N_w}$, $s_k = 1.0$, and $\theta_j = 0$ (Cartesian aligned normals) if nothing else is specified for the examples. The lower bound for $w_i$ is defined as $w_{\min} = 0.05$ to ensure manufacturability, and the upper bound is $w_{\max} = 1.0$. The number of indicator variables is set to the number of physical thickness variables for all examples, $N_s = N_w$.
\end{multicols}
\vspace{-0.5cm}
\newpage
\begin{table}[t]
    \centering
    %\begin{threeparttable}
        \caption{Michell Cantilever topology optimization results. For each model, the following is reported;
            Optimized compliance of full and active domain ($\mathcal{J}^*$, $\mathcal{J}^*_A$),
            Weighted compliance ($\varsigma^*$),
            Final penalization on angles and relative thickness ($\mathcal{P}^\theta$, $\mathcal{P}^s$),
            Time spent ($T_\text{wall}$, $T_\text{CPU}$).      
        }
        \begin{tabular}{@{}llllllll@{}}
            \toprule
            Model     & $\mathcal{J}^*$ & $\mathcal{J}^*_A$ & $\varsigma^*$ & $\mathcal{P}^\theta$   & $\mathcal{P}^s$ & $T_\text{wall}$                     & $T_\text{CPU}$                        \\ \midrule
            $N_a = 0$ & $237.964$       & $229.915$         & $23.796$      & $4.060 \times 10^{-3}$ & $0.178$         & $00 \mathord{:} 25 \mathord{:} 43 $ & $13 \mathord{:} 39 \mathord{:} 18 $   \\
            $N_a = 2$ & $247.003$       & $238.107$         & $24.700$      & $5.021 \times 10^{-3}$ & $0.242$         & $00 \mathord{:} 23 \mathord{:} 48 $ & $12 \mathord{:} 38 \mathord{:} 20 $   \\
            $N_a = 3$ & $259.141$       & $249.940$         & $25.914$      & $3.791 \times 10^{-3}$ & $0.234$         & $00 \mathord{:} 24 \mathord{:} 52 $ & $13 \mathord{:} 12 \mathord{:} 21 $   \\ \midrule
            HS Iso.   & $251.588$       & $241.396$         & $25.159$      & -                      & $0.104$         & $00 \mathord{:} 23 \mathord{:} 57 $ & $12 \mathord{:} 41 \mathord{:} 33 $   \\
            SIMP      & $247.875$       & $237.643$         & $24.788$      & -                      & -               & $03 \mathord{:} 55 \mathord{:} 10 $ & $4510 \mathord{:} 58 \mathord{:} 55 $ \\
            \bottomrule
        \end{tabular}
        \label{tab:TOcantilever}
    %\end{threeparttable}
\end{table}
\begin{multicols}{2}
As described in \cref{sec:surfaceGeneration}, point densities on the stream-surfaces vary, however, a minimal and maximal point spacing is set to $0.5h$ and $2h$ respectively. It should be noted that the Lotte Tower required a significantly higher point density, due to the level of detail combined with the aspect ratio, here the minimal point spacing was set to $0.2h$.

The optimization allows a maximum of 400 iterations. The multiscale optimization is conducted on a single compute node (32 cores). In contrast, the large-scale SIMP optimization problems for the Michell cantilever, Lotte Tower, and GE Jet Engine Bracket are executed on 36 nodes (1152 cores), 9 nodes (288 cores), and 22 nodes (704 cores), respectively. The variation in the number of nodes corresponds approximately to grid sizes employed for each problem. For the parallel computations of the stream-surface de-homogenization, a single compute node comprising 32 cores is employed.

The two penalty function weights are $\gamma_2 = 1.0$ and $\gamma_3 = 0.05$ for all examples, which are chosen based on extensive numerical experiments. The optimized compliance is denoted as $\mathcal{J}^*$, while the compliance found from de-homogenization is denoted as $\mathcal{J}^S$. The wall clock time, $T_{\text{wall}}$, and the total CPU time $T_{\text{CPU}}$, measured as [hh:mm:ss], are reported with the results, and a complete overview is seen in \cref{tab:overview1,tab:overview2}. Furthermore, since the volume cannot be guaranteed to be conserved for the de-homogenization designs, a weighted compliance measure, $\varsigma$, is used to evaluate the designs. $\varsigma$ is found by multiplying the compliance with the volume fraction:
\begin{equation}
    \varsigma = \mathcal{J}f
\end{equation}
It should be noted that this measurement quantity is nonlinear. Therefore, when comparing de-homogenization results with optimization results, the greater the deviation in volume fraction, the less reliable $\varsigma$ is. Additionally, the comparison is not fully "fair" as the quantities being compared are not obtained on the exact same mesh discretization, since there is no straightforward way to refine the optimized multiscale result on $\mathcal{T}$ to $\mathcal{T}^S$ without introducing interpolation errors. Moreover, it should be noted that $\mathcal{T}^S$ has a lower mesh quality compared to $\mathcal{T}$ and $\mathcal{T}^\rho$ due to the high level of mesh conformality with hexahedral elements.

The computational time for de-homogenization only accounts for the de-homogenization procedure presented here and does not include the time required for post-evaluation.

\subsection{Michell Cantilever}
The Michell cantilever is used as a benchmark model to compare to previous works of \citet{Groen2021}, and \citet{Stutz2022}. The cantilever has dimensions of $(2\times 1 \times 1) \unit{m}$. The cantilever domain is extended with a passive back plate with dimensions of $(1/24 \times 1 \times 1) \unit{m}$, where the cantilever is clamped at the end. The front of the cantilever is loaded by a surface-traction of $F = \{0,-36,0\}^\top \unit{N/m^2}$ on the passive domain with a size of $(1 \times 2 \times 2)/24 \unit{m}$.
The domain is discretized with $N_e = 225,792$ elements ($713,097$ degrees of freedom (dof)), corresponding to an element side length of $h=1/48 \unit{m}$. The mesh is similar to the one used in \cite{Groen2021, Stutz2022}, except for the passive back plate. Similarly, the SIMP benchmark model is obtained with $N_e = 115,605,504$ elements ($349,069,875$ degrees of freedom), corresponding to an element size of $h^\rho = 1/384 \unit{m}$. The volume fraction of the design domain is set as $f^i = 0.1$. The cantilever is de-homogenized with a minimal feature thickness of $\delta_{\min} = 5/778 \unit{m}$.

The optimization results are presented in \cref{tab:TOcantilever}, while the de-homogenized results are shown in \cref{tab:DEcantilever}. The relative differences in volume fraction, denoted as $\delta(f^*)$, and weighted compliance, denoted as $\delta(\varsigma^*)$, are provided with respect to the multi-scale solution. Additionally, the relative difference in weighted compliance, $\delta\left(\varsigma^\rho\right)$, compared to the large-scale SIMP solution, is also reported.

\begin{figure}[tb!]
    \centering
    \makebox[\textwidth][c]{
        \begin{subfigure}{62mm}
            \includegraphics[width=\linewidth]{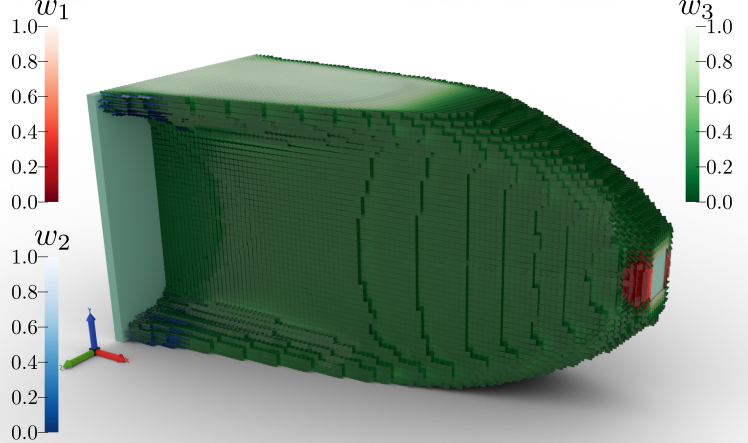}
            \caption{}
            \label{fig:michell_A0_TO}
        \end{subfigure}%
        \begin{subfigure}{62mm}
            \includegraphics[width=\linewidth]{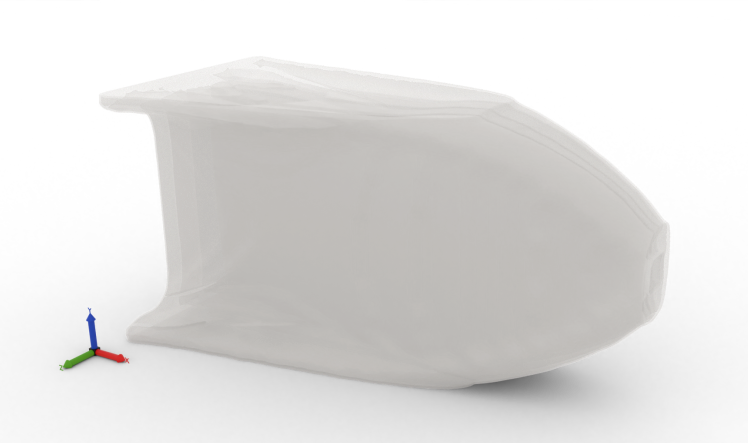}
            \caption{}
            \label{fig:michell_A0_deHom_o}
        \end{subfigure}%
        \begin{subfigure}{62mm}
            \includegraphics[width=\linewidth]{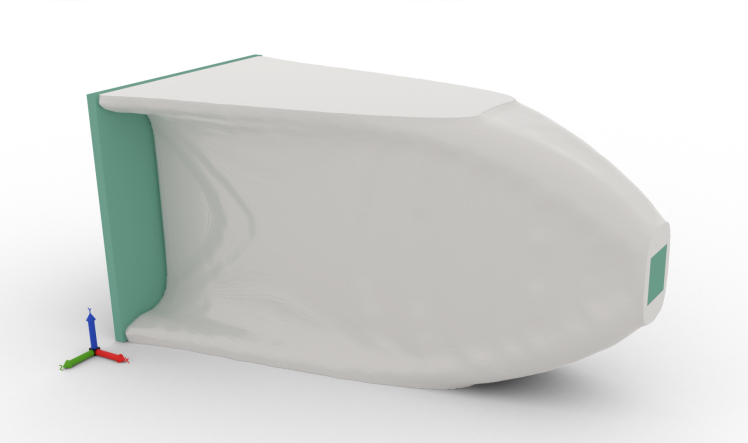}
            \caption{}
            \label{fig:michell_A0_deHom_mesh}
        \end{subfigure}
    }

    \makebox[\textwidth][c]{
        \begin{subfigure}{62mm}
            \includegraphics[width=\linewidth]{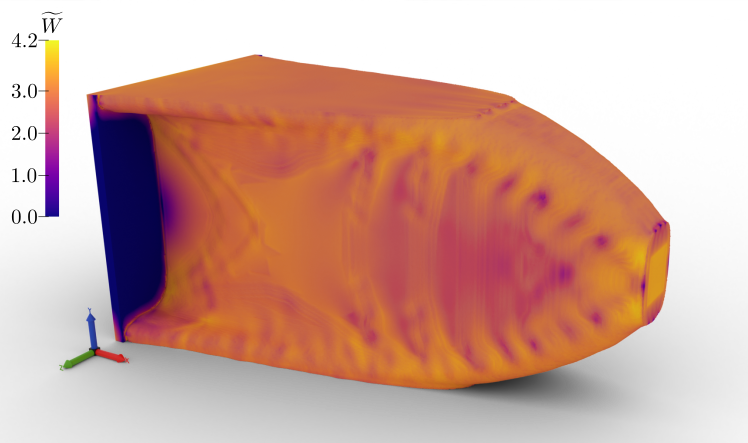}
            \caption{}
            \label{fig:michell_A0_deHom}
        \end{subfigure}%
        \begin{subfigure}{62mm}
            \includegraphics[width=\linewidth]{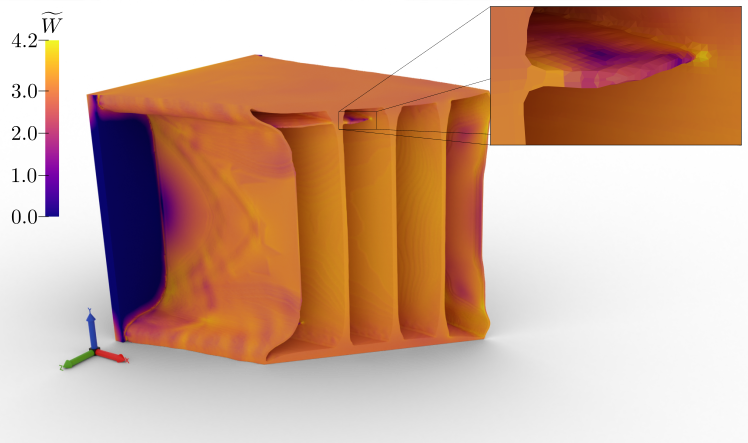}
            \caption{}
            \label{fig:michell_A0_deHom_Cut}
        \end{subfigure}%
        \begin{subfigure}{62mm}
            \includegraphics[width=\linewidth]{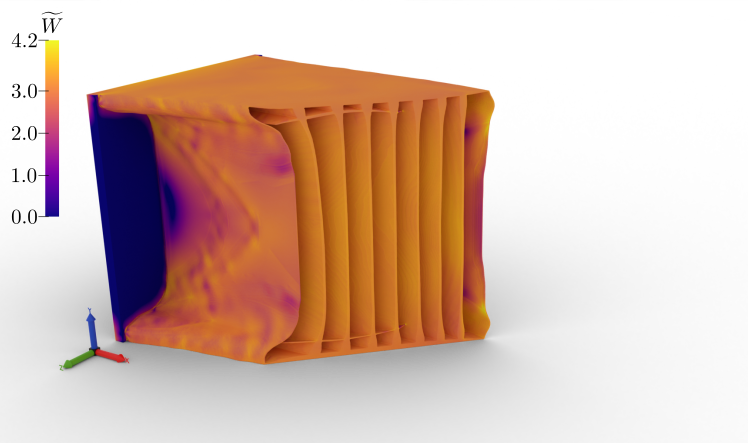}
            \caption{}
            \label{fig:michell_A0_deHom_fine_Cut}
        \end{subfigure}%
    }

    \makebox[\textwidth][c]{
        \begin{subfigure}{62mm}
            \includegraphics[width=\linewidth]{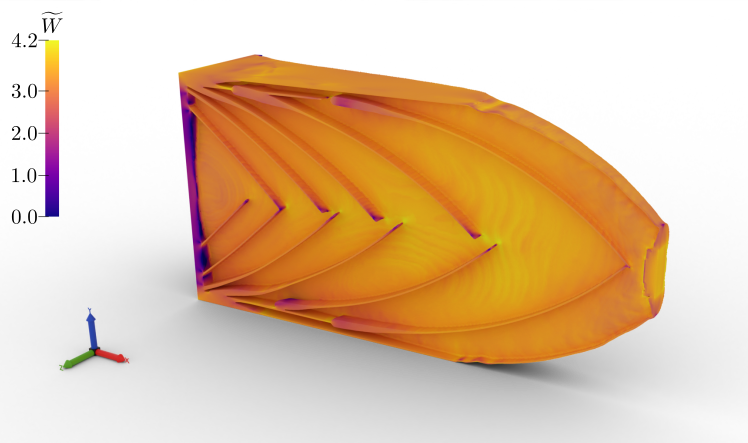}
            \caption{}
            \label{fig:michell_A12_deHom_cut}
        \end{subfigure}%
        \begin{subfigure}{62mm}
            \includegraphics[width=\linewidth]{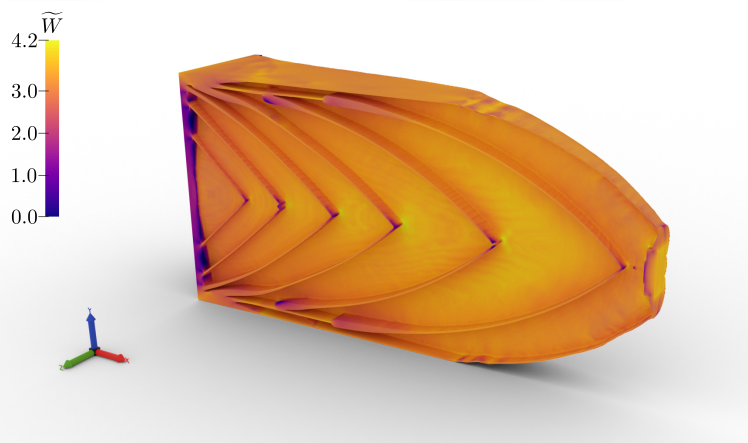}
            \caption{}
            \label{fig:michell_A12_sym_deHom_cut}
        \end{subfigure}%
        \begin{subfigure}{62mm}
            \includegraphics[width=\linewidth]{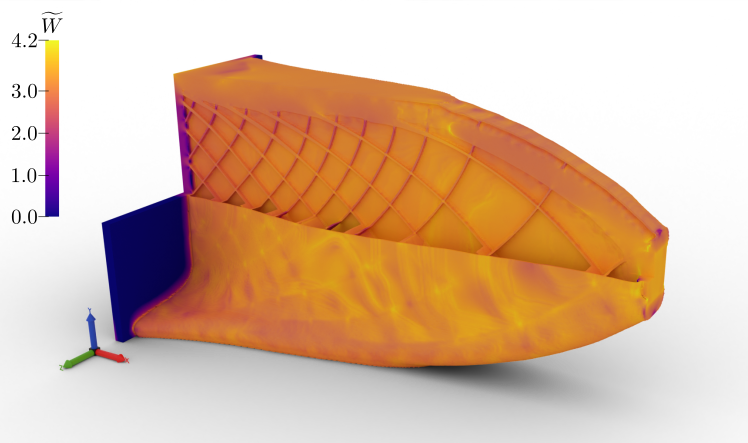}
            \caption{}
            \label{fig:michell_A13_deHom_cut}
        \end{subfigure}%
    }
    \caption{
        Michell cantilever results.
        (a) A simplified representation of the multi-scale solution, where the three differently colored plates represent $w_1$, $w_2$, and $w_3$.
        (b) The semi-transparent stream surfaces from the de-homogenization approach.
        (c) Finite element mesh of the de-homogenization result with $\Omega_P$ inserted.
        (d) Strain energy density plot of the de-homogenization result. Blue indicates no loading, while yellow indicates maximum loading.
        (e) Half-clip of the $N_a = 0$ de-homogenization result. The internal open surface is highlighted.
        (f) Half-clip of the $N_a = 0$ $(0.5,\delta_{\min})$ de-homogenization result.
        (g) and (h) Half-clip of the $N_a = 2$ de-homogenization result, where (h) exploits symmetry after de-homogenization. 
        (i) Quarter-clip of the $N_a = 2$ de-homogenization result. 
    }
    \label{fig:cant}
\end{figure}

The first example provides a direct comparison to the solution presented in \cite{Groen2021, Stutz2022}, where there is no restriction on the required number of active laminates, i.e., $N_a = 0$. A simplified representation of the optimized result is shown in \cref{fig:michell_A0_TO}, where the element-wise microstructure is illustrated using plates. The plates are oriented based on the microstructure's normal direction and are colored based on their relative thickness. Shades of red indicate $\hat{w}_1$, shades of blue indicate $\hat{w}_2$, and shades of green indicate $\hat{w}_3$. If $\hat{w}_i = 0$, the plate is removed. This illustration provides a basic visualization of the de-homogenization result.
It can be observed that high-relative thickness horizontal plates form at the top and bottom, supported by low-relative thickness vertical plates in the middle region, which is consistent with the solution presented in \cite{Groen2021, Stutz2022}.
In \cref{fig:michell_A0_deHom_o}, the volume-synthesized de-homogenized result is shown before $\Omega_P$ is inserted. The volume is semi-transparent to indicate the internal structure. After inserting $\Omega_P$ and discretizing the geometry, the mesh $\mathcal{T}^S$ is shown in \cref{fig:michell_A0_deHom_mesh}, with a plot of the strain energy density, $\widetilde{W} = \log_{10}\sum_i^M  W_i $, of the structure cut in half displayed in \cref{fig:michell_A0_deHom_Cut}. The strain energy density, ${W}$, is uniform within the internal structure, except for regions of interactions and boundaries, which is an expected result of the de-homogenization process. The relatively uniform distribution of ${W}$ indicates that the optimized solution is well-recovered in these regions since the optimality condition demands an even strain energy density distribution. At intersections, the smooth geometry becomes increasingly disrupted, displaying irregularities and discontinuities, resulting in stress raisers and increased values of ${W}$, as highlighted in \cref{fig:michell_A0_deHom_Cut}.
\end{multicols}

\begin{multicols}{2}
A significant amount of low-energy material is observed at the boundaries, such as the hull or surface edge regions. This is a direct result of the underlying microstructure model, which assumes infinite periodicity, perfect bonding, and separation of length scale. Hence, if $\delta_{\min} \rightarrow \infty$, this issue should be alleviated.
It should be noted that, as shown in \cref{tab:DEcantilever}, excluding the hull, i.e., having open walled structures, results in very poor weighted compliance.

The compliance is measured as the global compliance on $\Omega$, but for a comparison with \cite{Groen2021, Stutz2022}, it is necessary to compare the compliance of $\Omega_A$, denoted as $\mathcal{J}_A$. From the optimization problem, $\mathcal{J}_A$ is found to be very similar to \cite{Groen2021}, with only a 1\% difference, which is acceptable considering that the problem formulation is not exactly the same. The obtained result was achieved almost 23 times faster in terms of wall clock time; however, the CPU time is 23.5\% slower than \cite{Groen2021}, which could be attributed to an insufficient parallel setup and the added complexity with respect tto the unstructured mesh setup.

Comparing $\varsigma^S_A$, it is observed to be 2\%-3\% larger than in \cite{Groen2021, Stutz2022}. However, the volume fraction derived in \cite{Stutz2022} is violated by almost 20\%, and in \cite{Groen2021} by 7\%, whereas in this work, the volume fraction is violated by less than 2\%. The wall time for de-homogenization matches that of \cite{Groen2021}, but it is almost four times faster than in \cite{Stutz2022}. This increase in computational efficiency is despite changes made to achieve a more robust de-homogenization procedure, such as the addition of an outer hull, which clearly has a computational cost as can be seen in \cref{tab:DEcantilever}. The unsorted and unstructured nature of $\mathcal{F}$ and $\mathcal{W}$ also contributes to an increase in computational complexity. Furthermore, it should be noted that computational time in the de-homogenization procedure from \cite{Groen2021} does not depend on $\delta_{\min}$, whereas in \cite{Stutz2022} and this method, it has an impact.

It is remarkable that the de-homogenized structures perform well despite the extreme violation of scale separation. When using a half-minimal feature thickness, the performance increases significantly, as evident from the results. The structure can be seen in \cref{fig:michell_A0_deHom_fine_Cut}, where the finer periodicity is highlighted. However, it is important to note that achieving this finer resolution comes at a higher computational cost. Additionally, the requirements for post-processing also increase drastically, which should be taken into consideration.
\end{multicols}
\begin{table}[tb!]
    \centering
    %\begin{threeparttable}
        \caption{Michell Cantilever de-Homogenization results.
        For each model, the following is reported;
        Compliance of full and active domain ($\mathcal{J}^S$, $\mathcal{J}^S_A$),
        Volume fraction of design ($f^S$),
        Weighted compliance of full and active domain ($\varsigma^S$, $\mathcal{J}^S_A$),
        Relative difference to optimized design in volume and weighted compliance ($\delta(f^*)$, $\delta(\varsigma^*)$),
        Relative difference to SIMP design in weighted compliance ($\delta(\varsigma^\rho)$),
        Time spent ($T^S_\text{wall}$, $T^S_\text{CPU}$).}

        \begin{tabular}{@{}llllllllllll@{}}
            \toprule
            Model                            & $\mathcal{J}^S$ & $\mathcal{J}^S_A$ & $f^S$ & $\varsigma^S$ & $\delta(f^*)$ & $\delta(\varsigma^*)$ & $\delta\left(\varsigma^\rho\right)$ & $T_\text{wall}^S$ & $T_\text{CPU}^S$ \\ \midrule
            $N_a = 0$                        & 266.283         & 257.514           & 0.098 & 26.132        & -1.86         & 9.82                  & 5.43                                & 00:26:50          & 06:06:11         \\
            $N_a = 0$  (No hull)             & 301.036         & 290.556           & 0.096 & 28.834        & -4.22         & 21.17                 & 16.32                               & 00:13:45          & 03:59:40         \\
            $N_a = 0$ $(0.5\,\delta_{\min})$ & 272.510         & 263.323           & 0.093 & 25.303        & -7.15         & 6.33                  & 2.08                                & 01:55:54          & 13:40:47         \\
            $N_a = 0$ (Sym.)                 & 245.523         & 237.544           & 0.099 & 24.255        & -1.21         & 1.93                  & -2.15                               & -                 & -                \\ \midrule
            $N_a = 2$                        & 339.711         & 330.644           & 0.081 & 27.514        & -19.01        & 11.39                 & 11.00                               & 00:26:40          & 06:41:55         \\
            $N_a = 2$  (Sym.)                & 314.064         & 305.584           & 0.082 & 25.810        & -17.82        & 4.50                  & 4.13                                & -                 & -                \\ \midrule
            $N_a = 3$                        & 333.497         & 323.839           & 0.086 & 28.556        & -14.37        & 10.20                 & 15.20                               & 00:26:14          & 07:35:46         \\
            $N_a = 3$  (Sym.)                & 303.572         & 294.735           & 0.088 & 26.841        & -11.58        & 3.58                  & 8.29                                & -                 & -                \\
            \bottomrule
        \end{tabular}
        \label{tab:DEcantilever}
    %\end{threeparttable}
\end{table}
\begin{multicols}{2}
By forcing more active microstructure layers with $N_a = 2$ and $N_a = 3$, it is evident that the stiffness performance of the structure decreases due to the added layer requirement. However, it is also observed that the additional surfaces contribute to an infill that causes a more stable structure. This can be seen in \cref{fig:michell_A12_deHom_cut} and \cref{fig:michell_A13_deHom_cut}, where critical members are reduced compared to the previous case. This indicates that the added layers help to distribute the loads and improve the overall stability of the structure. Furthermore, this also demonstrates how the infill functions as a direct structural member, contributing to a manufacturable design.

In \cref{fig:michell_A12_deHom_cut}, it can be observed that many surfaces have open ends, which can lead to stress raisers and reduced structural performance. Ideally, the surfaces should meet to form a more continuous structure. However, the stream surface-based de-homogenization method in this study does not allow for such surface connections, as the surfaces are generated from random seeding and selected based on their spacing.
To address this issue, symmetry can be exploited in the case of the Michell Cantilever. By reflecting the upper part of the de-homogenized structure in the middle $xz$-plane, a more connected and continuous structure can be obtained, as shown in \cref{fig:michell_A12_sym_deHom_cut}. This symmetry exploration improves the overall performance of the structure, leading to a reduction in weighted compliance, as shown in \cref{tab:DEcantilever}.
This symmetry exploration approach can also be applied to the $N_a = 0$ and $N_a = 3$ models, resulting in a similar reduction in weighted compliance. By choosing the "better" half of the structure to reflect, i.e. the half with the best local weighted compliance, the overall performance can be further improved, leading to a reduction in weighted compliance of up to 7\%.

According to \cref{tab:TOcantilever,tab:DEcantilever}, it is evident that the de-homogenized result outperforms the HS optimization result. This outcome aligns with expectations when comparing an isotropic material model with a laminated orthotropic material model. Similar findings were also observed in a previous study by~\citet{Jensen2022}. Furthermore, by imposing a minimum wall thickness, a lower material density limit of $\underline{\rho} = 0.265$ is achieved, resulting in a design that approaches a nearly binary configuration, with only filtering artifacts causing porous design areas. Considering these factors, a direct comparison between de-homogenization and post-evaluation of HS results is not attempted.
\end{multicols}
\vspace{-1cm}
\begin{figure}[b!]
    \centering
    \makebox[\textwidth][c]{\includegraphics[width=137.500mm]{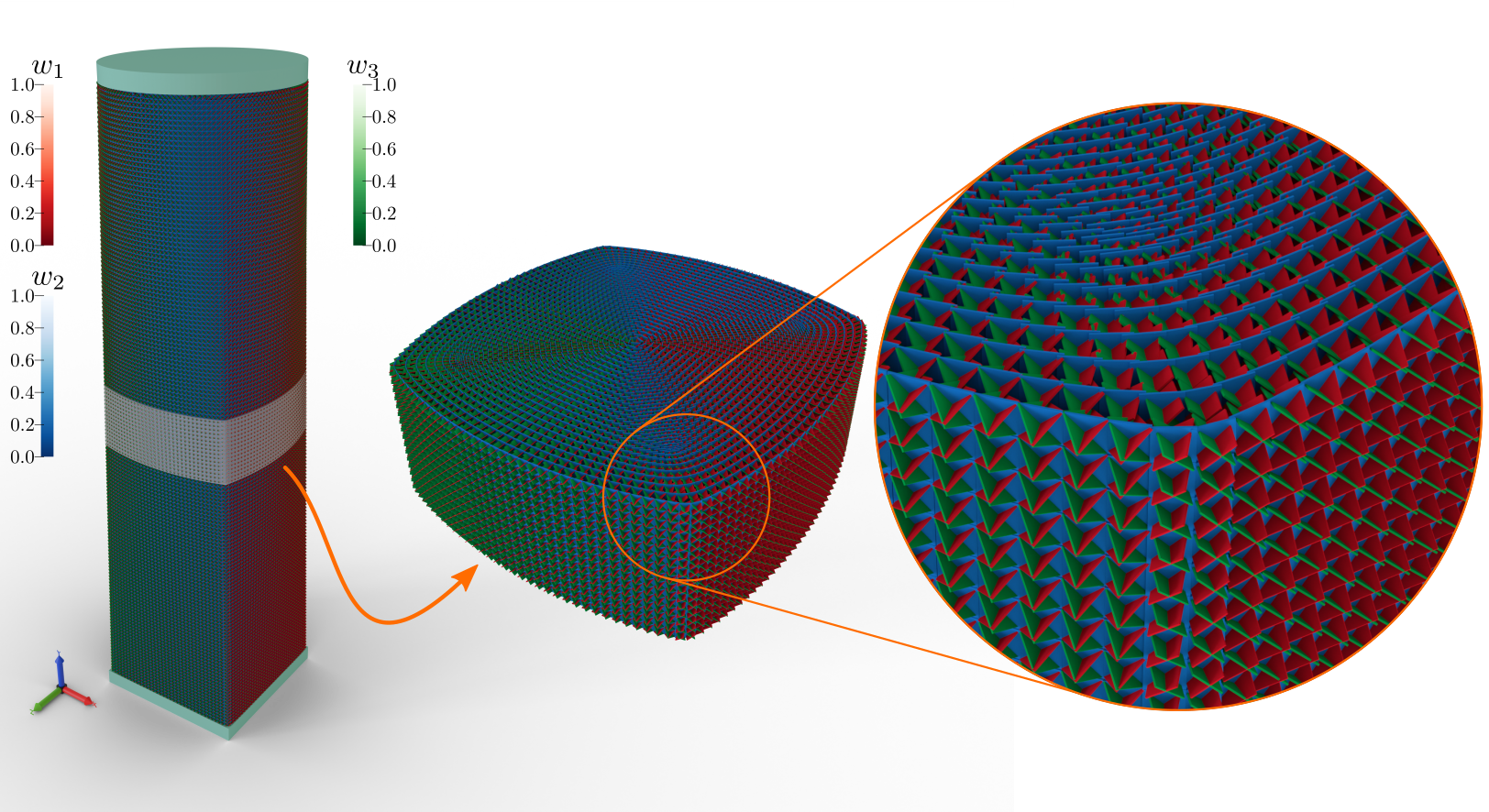}}%
    \caption{
        The initial guess for the optimization problem is based on the principal stress directions from a solid isotropic state. Due to the global torsion stress state, the normal direction related to $w_2$ is radial to the tower's center axis, while the normals related to $w_1$ and $w_3$ are tangential and form a helix with an angle of $\pi/4$ to the tower's center axis.}
    \label{fig:lotteTowerSG}
\end{figure}
\begin{figure}[t!]
    \centering
    \makebox[\textwidth][c]{
        \begin{subfigure}{47.5mm}
            \includegraphics[width=\linewidth]{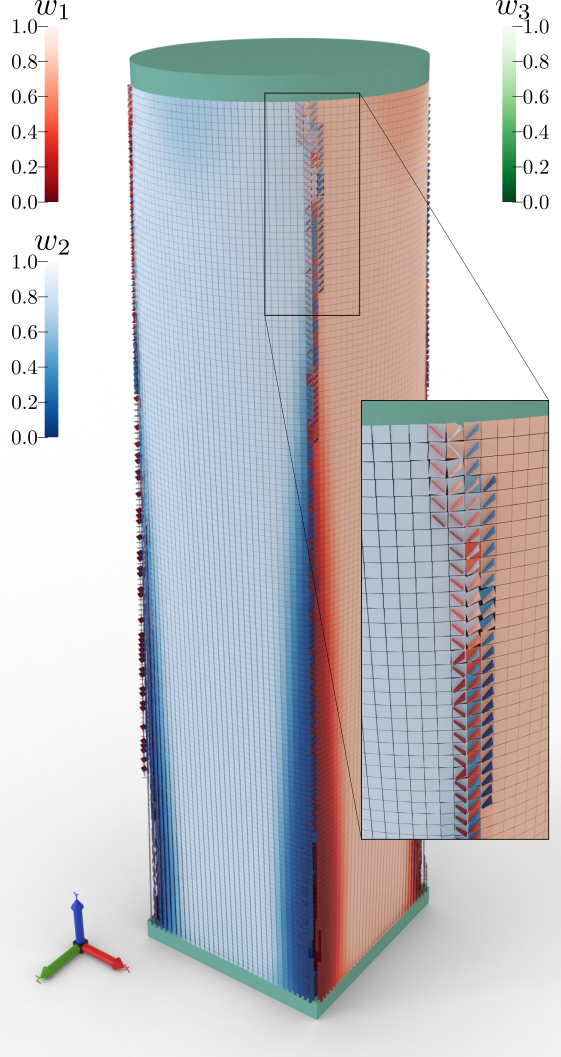}
            \caption{}
            \label{fig:lotteTower_Hom_A0_init0}
        \end{subfigure}%
        \begin{subfigure}{47.5mm}
            \includegraphics[width=\linewidth]{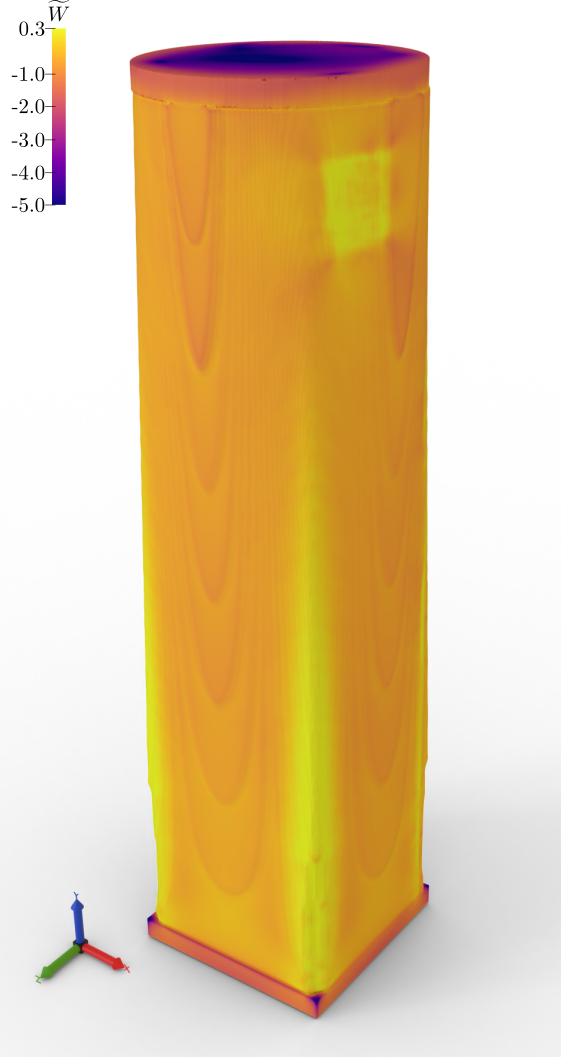}
            \caption{}
            \label{fig:lotteTower_deHom_A0_init0}
        \end{subfigure}%
        \begin{subfigure}{47.5mm}
            \includegraphics[width=\linewidth]{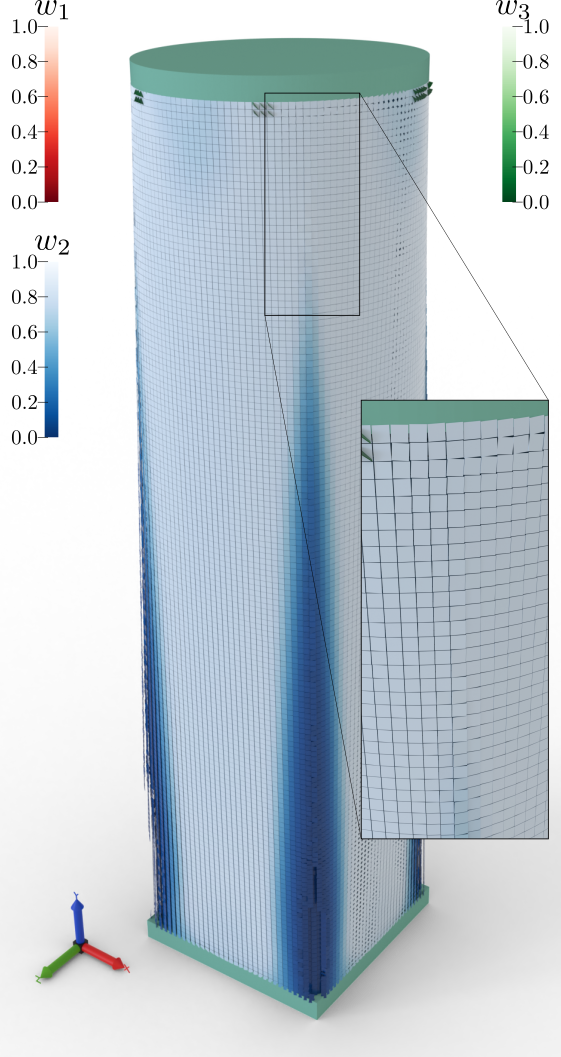}
            \caption{}
            \label{fig:lotteTower_Hom_A0_init1}
        \end{subfigure}%
        \begin{subfigure}{47.5mm}
            \includegraphics[width=\linewidth]{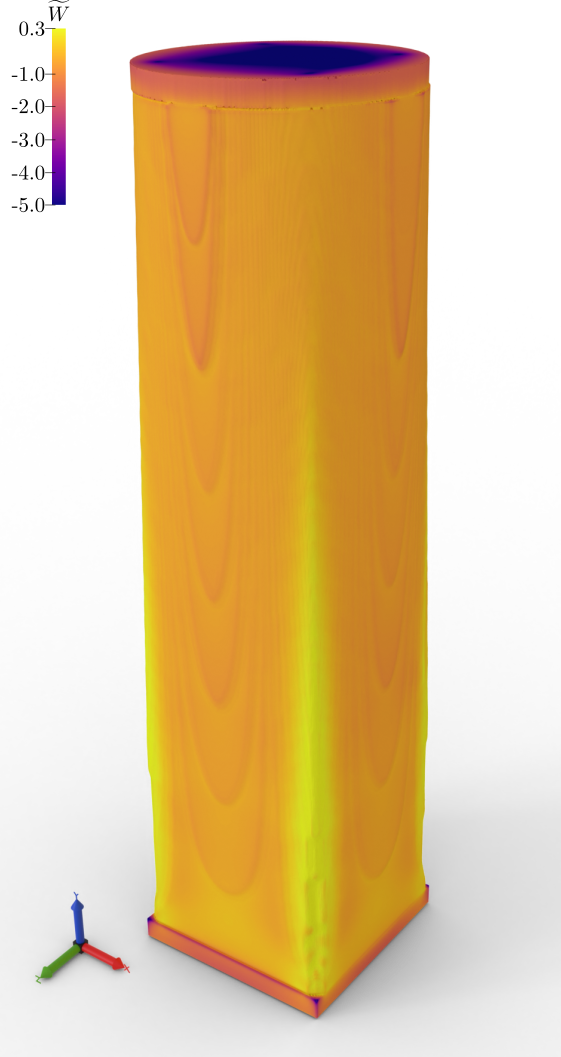}
            \caption{}
            \label{fig:lotteTower_deHom_A0_init1}
        \end{subfigure}
    }
    \caption{
        Lotte Tower $N_a = 0$ results.
        (a) $N_a = 0$ multi-scale optimization result from a uniform initial guess, resulting in a connecting seam where layers $w_1$ and $w_2$ meet.
        (c) $N_a = 0$ SG multi-scale optimization result from a uniform initial guess.
        (b) and (d) de-homogenized result of (a) and (c), respectively.
    }
    \label{fig:lotte0}
\end{figure}
\begin{multicols}{2}
When comparing the large-scale SIMP result to the de-homogenized result, it is found that they achieve similar weighted compliance. However, the de-homogenized result is obtained approximately 200 times faster than the large-scale SIMP result. \cref{fig:michell_SIMP} illustrates the SIMP result, and it is noted that the result bears similarities to the $N_a = 2$ de-homogenized result, indicating that the de-homogenized structure captures essential features of the SIMP optimization result.
\subsection{Lotte Tower}
The second example is a Lotte Tower model resembling the one found in \cite{Stromberg2011}. Howevere; in this work, the tower is only half the height with the approximate box dimensions of $(1 \times 1 \times 4) \unit{m}$. It is designed as a square that morphs into a circle from the bottom to the top. The square and the circle have the same surface area of $1 \unit{m^2}$. The tower includes passive domains at the top and bottom with a height of $0.1 \unit{m}$. The passive bottom surface is fixed, while a tangential surface-traction of $F = 0.1 \unit{N/m^2}$ is applied to the side of the top passive domain. Hence the tower is twisted. The volume fraction is set to $f^i = 0.1$.
In contrast to \cite{Stromberg2011}, this tower is discretized as a solid, resulting in an unstructured mesh. The domain is discretized with $N_e = 416,232$ elements ($1,291,950$ dof), corresponding to approximately an element side length of $h = 1/40 \unit{m}$. The reference large-scale SIMP solution is discretized with $N_e = 26,641,472$ elements ($80,607,987$ dof), corresponding to approximately an element side length of $h^\rho = 1/160  \unit{m}$.
The de-homogenized Lotte Tower has a minimal feature thickness of $\delta_{\min} = 1/128  \unit{m}$.

Four modeling cases of the Lotte Tower are considered, including two cases with a $N_a = 0$ layer restriction, where one is obtained by an initial guess (denoted SG) for the Euler angles $\bm{\theta}$. The other two cases have layer restrictions of $N_a = 2$ and $N_a = 3$.

The initial guess, SG, is obtained by recovering the principal stress directions from a solid isotropic state. Obtaining a good initial guess this way in 3D is intrinsically difficult since the principal stress directions are not unique. However, for this torsion case, there is only one globally dominant stress state. As a result, the normal directions related to $w_2$ are radial to the tower's center axis, while the normals related to $w_1$ and $w_3$ are tangential and form a helix with an angle of $\pi/4$ to the tower's center axis. Consequently, the initial guess has a singularity line in the center axis of the tower. The initial guess is depicted in \cref{fig:lotteTowerSG}.

The natural solution to this torsion problem is a closed circular-like hollow tube; however, due to the bounding geometry, a perfect straight cylinder is not possible. For a perfect hollow cylinder (with solid walls) with the given volume fraction, the theoretical compliance is found to be $J = 141.096 \times 10^{-3}$ (see~\cite{Sigmund2016, Sundstroem2010}). 
However, to achieve this theoretical compliance, a minimum length-scale of approximately $1.6h$ is required, in addition to the actual cylinder shape.
Therefore, for the homogenization-based topology optimization problem, it is not possible to create an encapsulating solid structure with the given minimum length-scale. 
However, for the large-scale SIMP problem, a minimum required length-scale of approximately $6.5h^\rho$ exists, making it feasible to construct a solid closed-walled structure.

The optimization result are stated in \cref{tab:lotteTowerTO} where no significant difference in compliance values between the two cases with $N_a = 0$ is observed. However, the layup differs, as depicted in \cref{fig:lotteTower_Hom_A0_init0} and \cref{fig:lotteTower_Hom_A0_init1}, for $N_a = 0$ without and with an orientation initial guess, respectively.
\end{multicols}
\begin{table}[t!]
    \centering
    %\begin{threeparttable}
        \caption{Lotte Tower topology optimization results.
        For each model, the following is reported;
        Compliance of the design ($\mathcal{J}^*$),
        Weighted compliance ($\varsigma^*$),
        Final penalization on angles and relative thickness ($\mathcal{P}^\theta$, $\mathcal{P}^s$),
        Time spent ($T_\text{wall}$, $T_\text{CPU}$).}
        \begin{tabular}{@{}llllllll@{}}
            \toprule
            Model          & $\mathcal{J}^*$            & $\varsigma^*$             & $\mathcal{P}^\theta$       & $\mathcal{P}^s$           & $T_\text{wall}$                      & $T_\text{CPU}$                        \\ \midrule
            $N_a = 0$      & $ 150.491 \times 10^{-3} $ & $ 15.049 \times 10^{-3} $ & $ 105.970 \times 10^{-4} $ & $86.454 \times 10^{-3} $  & $ 00 \mathord{:} 55 \mathord{:} 18 $ & $ 29 \mathord{:} 24 \mathord{:} 33 $  \\
            $N_a = 0$ (SG) & $ 150.435 \times 10^{-3} $ & $ 15.044 \times 10^{-3} $ & $ 94.670 \times 10^{-4} $  & $86.578 \times 10^{-3} $  & $ 00 \mathord{:} 56 \mathord{:} 38 $ & $ 30 \mathord{:} 09 \mathord{:} 05 $  \\ \midrule
            $N_a = 2$      & $ 151.922 \times 10^{-3} $ & $ 15.192 \times 10^{-3} $ & $ 114.380 \times 10^{-4} $ & $123.133 \times 10^{-3} $ & $ 00 \mathord{:} 48 \mathord{:} 01 $ & $ 25 \mathord{:} 32 \mathord{:} 06 $  \\ \midrule
            $N_a = 3$      & $ 154.611 \times 10^{-3} $ & $ 15.461 \times 10^{-3} $ & $ 106.330 \times 10^{-4} $ & $201.049 \times 10^{-3} $ & $ 00 \mathord{:} 47 \mathord{:} 35 $ & $ 25 \mathord{:} 18 \mathord{:} 26 $  \\ \midrule
            HS Iso.        & $ 158.725 \times 10^{-3} $ & $ 15.873 \times 10^{-3} $ & -                          & $104.260 \times 10^{-3} $                  & $ 00 \mathord{:} 43 \mathord{:} 47 $ & $ 23 \mathord{:} 16 \mathord{:} 59 $  \\
            SIMP           & $ 150.447 \times 10^{-3} $ & $ 15.045 \times 10^{-3} $ & -                          & -                         & $ 02 \mathord{:} 28 \mathord{:} 09 $ & $ 710 \mathord{:} 26 \mathord{:} 04 $ \\
            \bottomrule
        \end{tabular}
        \label{tab:lotteTowerTO}
    %\end{threeparttable}
\end{table}
\begin{multicols}{2}
Both solutions result in a circular-like hollow tube with a thickness corresponding to the minimum length-scale. Consequently, the material density is relatively high at the circumference of the tower. However, as the tower transitions from a circular to a square shape, the corners have lower density to alleviate the effects of the coarse mesh discretization, since maintaining a perfect circular shape becomes unfeasible. In fact, as observed in the SIMP solution (see \cref{fig:lotteTower_SIMP}), the final topology undergoes a transformation from a circle to a superellipse, achieving a compliance value very close to that of the multiscale solution (and to the theoretical). Therefore, with a finer mesh discretization for the multiscale optimization solution, a similar final shape would be expected, considering the comparable compliance attained by the SIMP approach.

The main difference between the two $N_a = 0$ cases is that the SG case primarily consists of the $w_2$ layer, whose orientation remains nearly unchanged from the initial guess, which is expected. However, for the uniform initial guess, the layup consists of the $w_1$ and $w_2$ layers connected by a seam, as highlighted in \cref{fig:lotteTower_Hom_A0_init0}. The presence of the seam is a result of the microstructure orthogonality constraint when transitioning between the $w_1$ and $w_2$ layers for a smooth connection. In the cases of $N_a = 2$ and $N_a = 3$, a helix pattern is formed (similar to the orientation initial guess), with one or two additional layers, respectively. These additional layers have a minimum thickness of $w_{\min}$, resulting in only a slight increase in compliance. This also indicates a more non-unique solution.

The de-homogenization problem is similar to the torsion sphere problem discussed in previous works such as~\cite{Groen2021, Stutz2022}. This problem is inherently challenging, as the stream surface, when tracing the cylinder shape, needs to meet itself, resulting in a semi-closed stream surface. However, due to interpolation errors and numerical integration, the shape may undergo slight morphing.
\end{multicols}
\begin{figure}[b!]
    \centering
    \makebox[\textwidth][c]{
        \begin{subfigure}{47.5mm}
            \includegraphics[width=\linewidth]{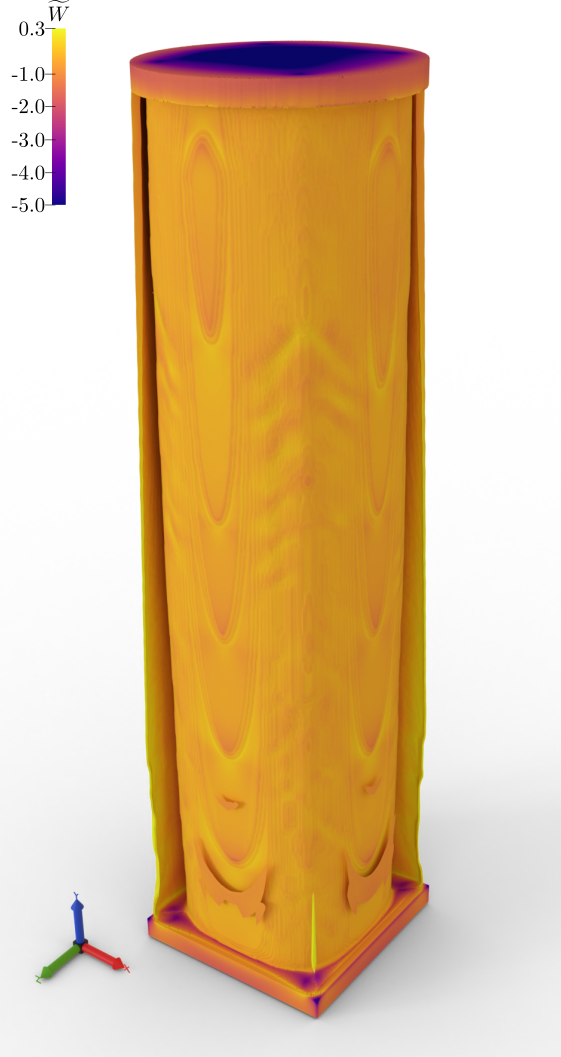}
            \caption{}
            \label{fig:lotteTower_deHom_A0_init0_cut2}
        \end{subfigure}%
        \begin{subfigure}{47.5mm}
            \includegraphics[width=\linewidth]{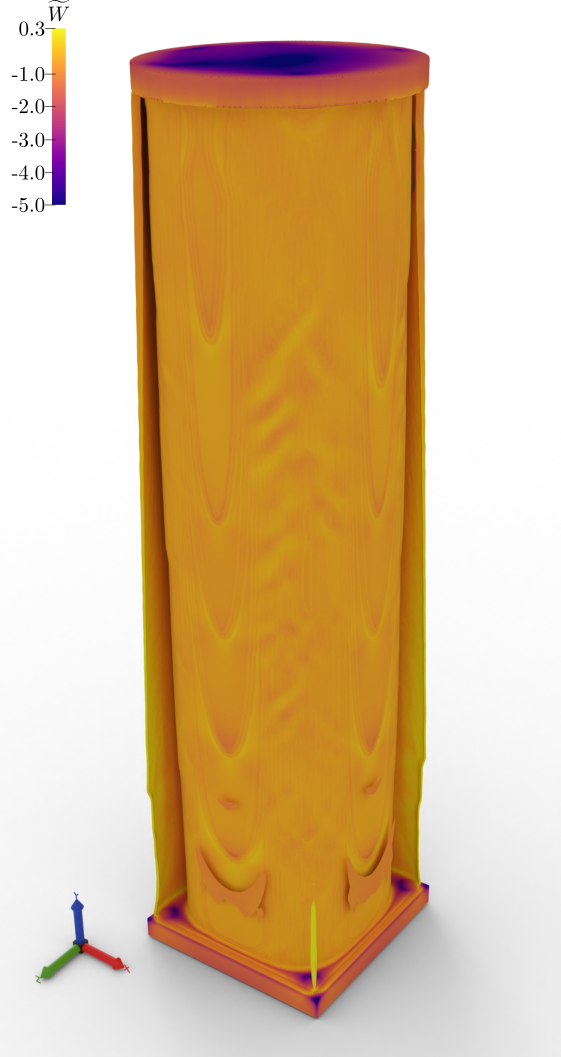}
            \caption{}
            \label{fig:lotteTower_deHom_A0_init1_cut2}
        \end{subfigure}%
        \begin{subfigure}{47.5mm}
            \includegraphics[width=\linewidth]{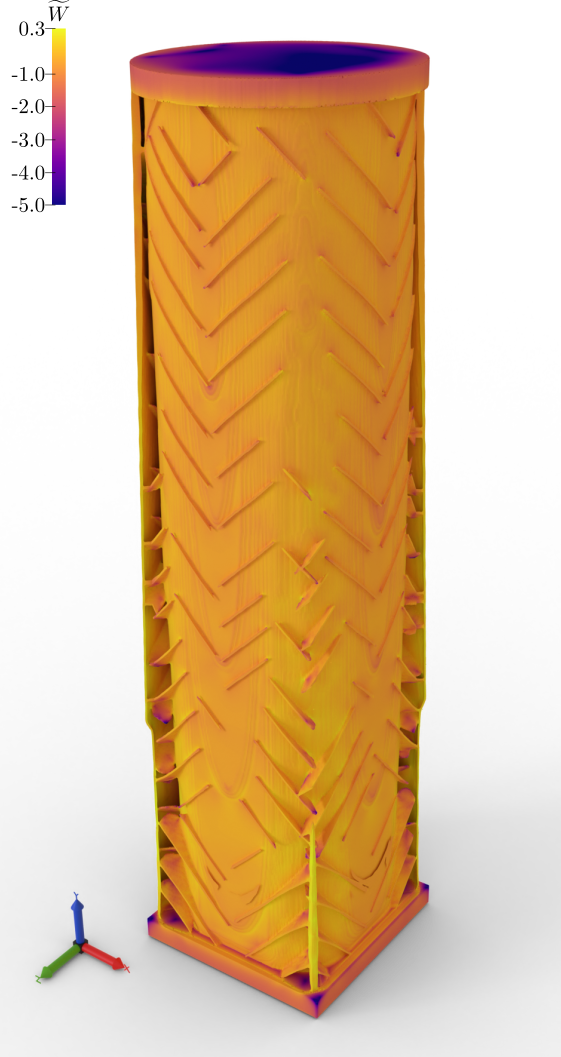}
            \caption{}
            \label{fig:lotteTower_deHom_A12_cut}
        \end{subfigure}%
        \begin{subfigure}{47.5mm}
            \includegraphics[width=\linewidth]{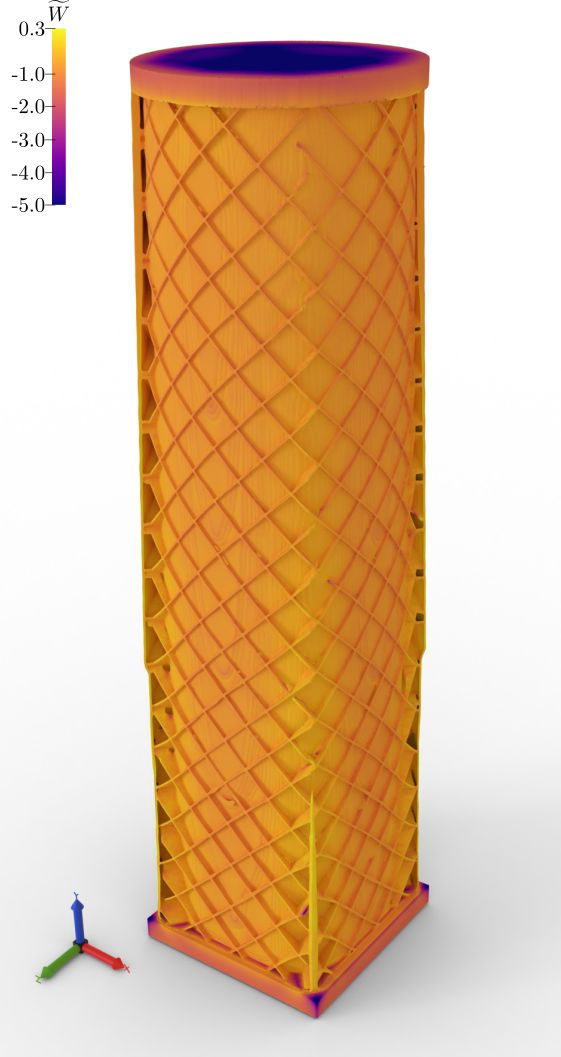}
            \caption{}
            \label{fig:lotteTower_deHom_A13_cut}
        \end{subfigure}
    }%
    \caption{
        Lotte Tower results.
        (a) and (b) Half-clip of the outer surface of $N_a = 0$ and $N_a = 0$ SG, respectively. The relatively small radial macroscopic feature size makes it challenging to achieve multiple radial surfaces without an extremely small feature size.
        (c) and (d) Half-clip of the outer surface of the de-homogenized result for $N_a = 2$ and $N_a = 3$, respectively, where the internal surfaces are evident.}
    \label{fig:lotte23}
\end{figure}
\begin{table}[t!]
    \centering
    %\begin{threeparttable}
        \caption{Lotte Tower de-homogenization results.
        For each model, the following is reported;
        Compliance of the design ($\mathcal{J}^S$),
        Volume fraction of design ($f^S$),
        Weighted compliance ($\varsigma^S$),
        Relative difference to optimized design in volume and weighted compliance ($\delta(f^*)$, $\delta(\varsigma^*)$),
        Relative difference to SIMP design in weighted compliance ($\delta(\varsigma^\rho)$),
        Time spent ($T^S_\text{wall}$, $T^S_\text{CPU}$).}
        \begin{tabular}{@{}lllllllllll@{}}
            \toprule
            Model        & $\mathcal{J}^S$          & $f^S$    & $\varsigma^S$           & $\delta(f^*)$ & $\delta(\varsigma^*)$ & $\delta\left(\varsigma^\rho\right)$ & $T_\text{wall}^S$ & $T_\text{CPU}^S$ \\ \midrule
            $N_a = 0$    & $138.971\times 10^{-3} $ & 0.129195 & $17.954\times 10^{-3} $ & 29.20         & 19.30                 & 19.34                               & 01:03:59          & 14:24:40         \\
            $N_a = 0$ SG & $140.189\times 10^{-3} $ & 0.127029 & $17.808\times 10^{-3} $ & 27.03         & 18.37                 & 18.37                               & 00:52:25          & 14:55:54         \\ \midrule
            $N_a = 2$    & $131.937\times 10^{-3} $ & 0.126765 & $16.725\times 10^{-3} $ & 26.77         & 10.09                 & 11.17                               & 01:09:18          & 24:18:40         \\ \midrule
            $N_a = 3$    & $116.642\times 10^{-3} $ & 0.137771 & $16.070\times 10^{-3} $ & 37.77         & 3.94                  & 6.81                                & 02:20:44          & 61:25:10         \\
            \bottomrule
        \end{tabular}
        \label{tab:lotteTowerDeHom}
    %\end{threeparttable}
\end{table}
\begin{multicols}{2}
The de-homogenized result can be seen in \cref{tab:lotteTowerDeHom}, while the structure of the two cases with $N_a = 0$ can be observed in \cref{fig:lotteTower_deHom_A0_init0} and \cref{fig:lotteTower_deHom_A0_init1}, respectively, along with their corresponding cross-sections in \cref{fig:lotteTower_deHom_A0_init0_cut2} and \cref{fig:lotteTower_deHom_A0_init1_cut2}. The structures for $N_a = 2$ and $N_a = 3$ are shown in \cref{fig:lotteTower_deHom_A12_cut} and \cref{fig:lotteTower_deHom_A13_cut}, respectively.

From \cref{tab:lotteTowerDeHom}, it is seen that the volume fraction is violated by more than $25\%$, which makes the structural performance comparison somewhat ambiguous. The volume violation is a result of a combination of several factors, including the presence of high-density material, the enforcement of a closed hull, and a relatively small overall feature size on $\mathcal{T}$ compared to the minimum feature size on $\mathcal{T}^S$, leading to excessive local surface spacing (periodicity) in the radial direction of the tower.
To capture the radial local surface spacing, a much smaller $\delta_{\min}$ is required, which becomes impractical. This observation is also supported by the fact that $\varsigma^S$ approaches $\varsigma^*$ for $N_a = 2$ and $N_a = 3$, as the overall feature size for the additional layers allows for a suitable local surface spacing, as seen in \cref{fig:lotteTower_deHom_A12_cut} and \cref{fig:lotteTower_deHom_A13_cut}.

The main difference between the two $N_a = 0$ cases lies in computational time, with SG being twice as fast, which is a direct result of the smoother surface generation in the SG case. This difference can be attributed to the fact that $\mathcal{F}$ is smoother for SG, making the surface tracing process more straightforward. Additionally, all four cases exhibit higher strain energy density near the corners of the tower, resulting from the macroscopic length scale that does not allow for perfectly rounded corners, leading to less efficient designs. Furthermore, \cref{fig:lotteTower_deHom_A0_init0} shows an overloaded patch on the cylinder wall where the wall segment is thinner. This anomaly is not found in the optimized solution but can be attributed to imperfections that can arise due to the heuristic nature of the stream surface de-homogenization procedure.

The enhanced mechanical stability of the results obtained for $N_a>1$ becomes visible when considering the structures with $N_a = 2$ and $N_a = 3$. However, to evaluate the actual stability performance of the structure, a linear buckling analysis (LBA) is performed on $\mathcal{T}^S$. In this analysis, the following generalized eigenvalue problem is solved;
\end{multicols}
\begin{figure}[b!]
    \centering
    \makebox[\textwidth][c]{
        \begin{subfigure}{47.5mm}
            \includegraphics[width=\linewidth]{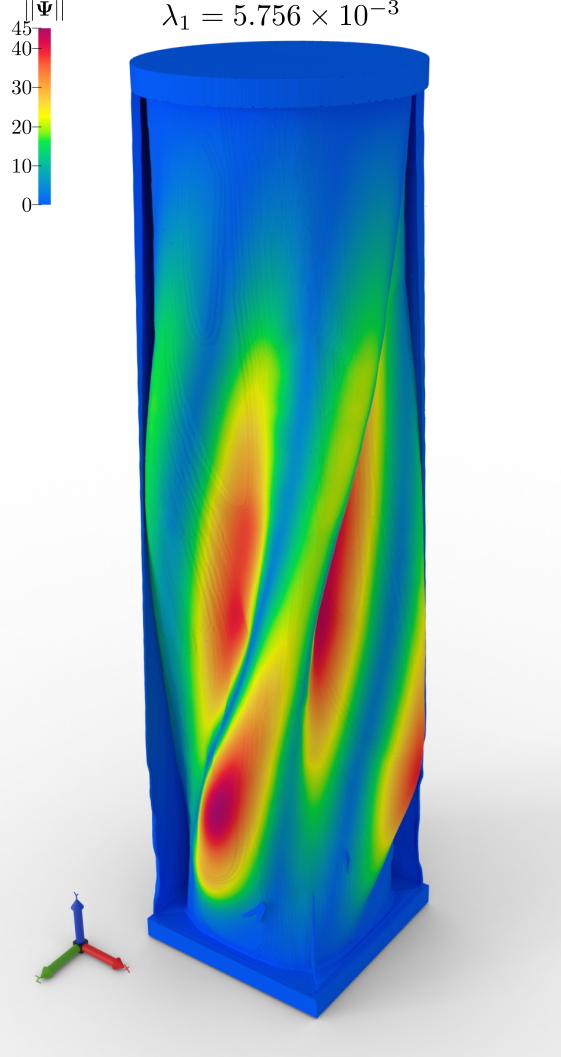}
            \caption{}
            \label{fig:lotteTower_deHom_A0_init0_buckling}
        \end{subfigure}%
        \begin{subfigure}{47.5mm}
            \includegraphics[width=\linewidth]{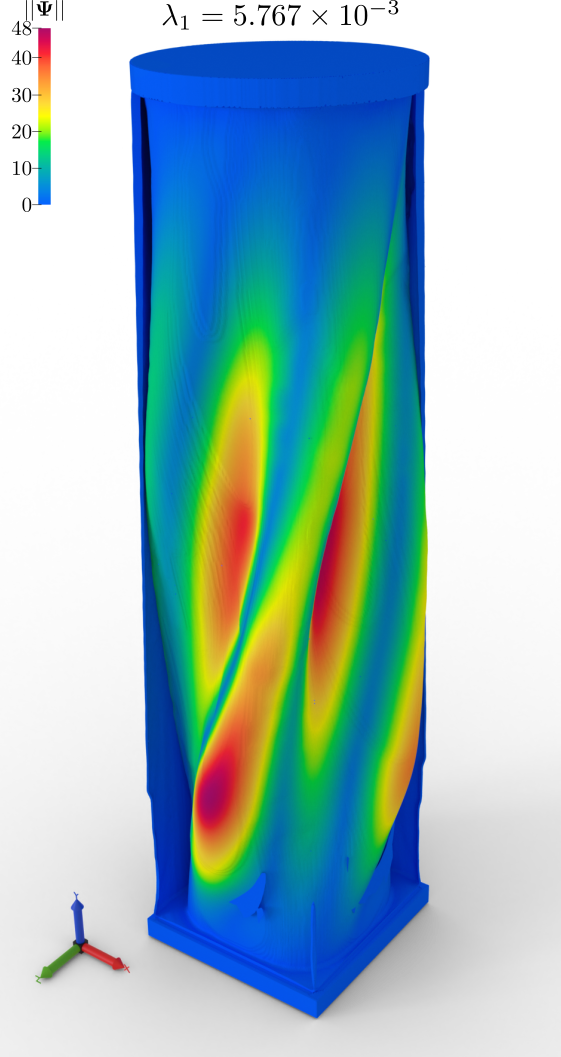}
            \caption{}
            \label{fig:lotteTower_deHom_A0_init1_buckling}
        \end{subfigure}%
        \begin{subfigure}{47.5mm}
            \includegraphics[width=\linewidth]{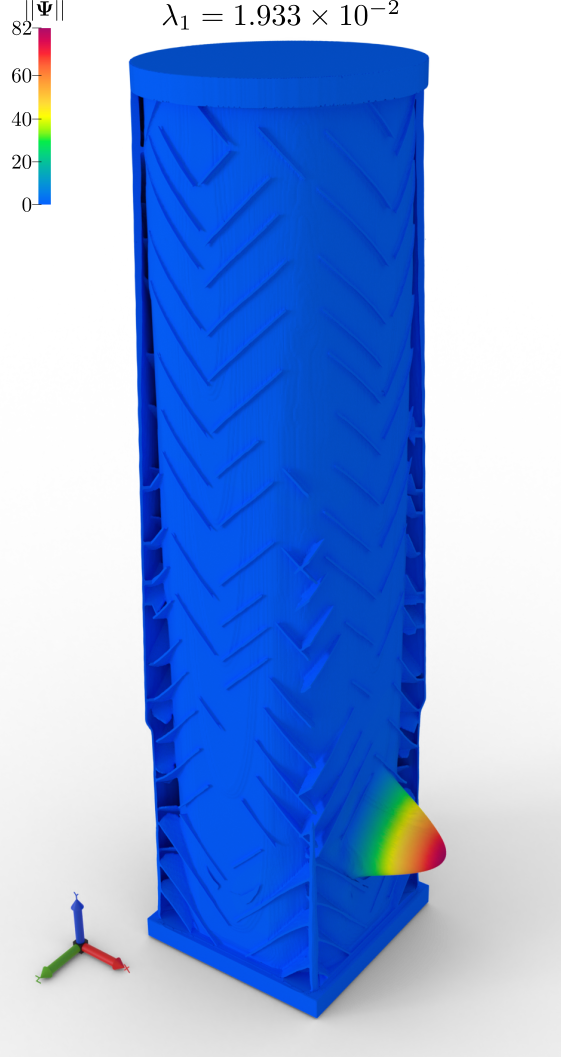}
            \caption{}
            \label{fig:lotteTower_deHom_A12_buckling}
        \end{subfigure}%
        \begin{subfigure}{47.5mm}
            \includegraphics[width=\linewidth]{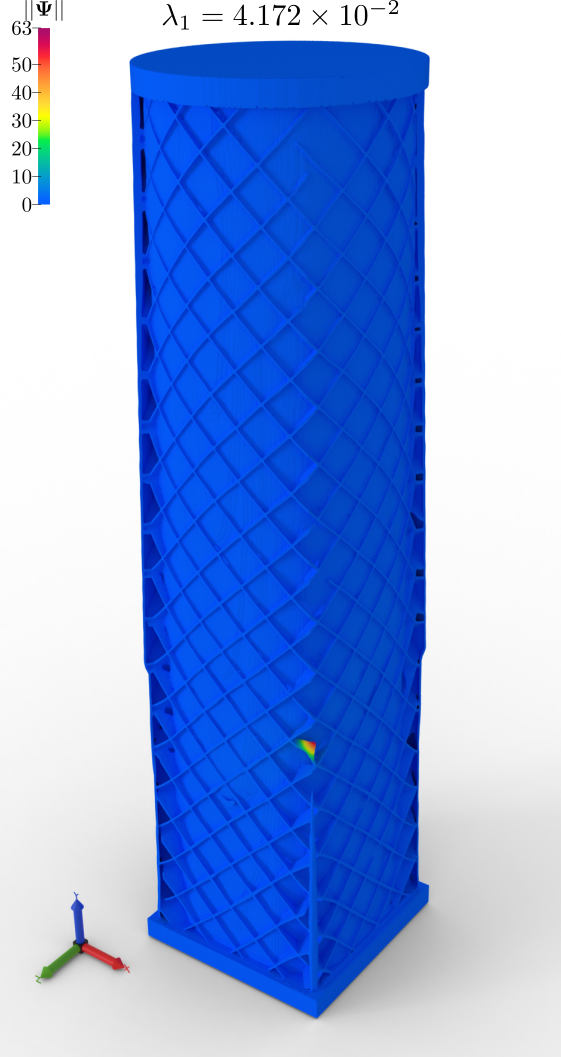}
            \caption{}
            \label{fig:lotteTower_deHom_A13_buckling}
        \end{subfigure}
    }%
    \caption{
        First BLF and buckling mode of the four Lotte Tower cases:
        (a) and (b) The $N_a = 0$ and $N_a = 0$ SG cases, respectively. Almost similar global modes are obtained. The inner surface is thinner than the outer; hence the buckling mode is located there.
        (c) and (d) The $N_a = 2$ and $N_a = 3$ cases, respectively. The additional surface makes critical members shorter, resulting in localized modes.}
    \label{fig:lotteTower_deHom_buckling}
\end{figure}

\begin{multicols}{2}
\begin{equation}
    \left(\mathbf{K} + \lambda_{h}\mathbf{G}(\mathbf{u})\right)\bm{\Psi}_{h} = 0, \quad h\in\{1,...,N_{\lambda}\}.
\end{equation}
Here, $\mathbf{G}$ is the stress stiffness matrix dependent on the displacement field $\mathbf{u}$, $\lambda_1$ is the smallest eigenvalue (i.e., the critical buckling load factor or BLF), and $\bm{\Psi}_1$ is the associated eigenvector, representing the buckling mode. The LBA is implemented within the aforementioned topology optimization framework using the Scalable Library for Eigenvalue Problem Computations (SLEPc)\cite{Roman2010} framework, with an implementation inspiration form~\cite{Wang2021,Ferrari2021}.

The LBA problem is solved for $N_\lambda = 24$. The first BLF and the corresponding first eigenvector $\bm{\Psi}_1$ is shown in \cref{fig:lotteTower_deHom_buckling}. All BLFs are plotted in \cref{fig:lotteTowerBuckling}.

It is observed that the BLFs increase by an order of magnitude when considering $N_a = 2$ and $N_a = 3$, compared to the single layer case $N_a = 0$, this is also to be expected based on the work of \cite{Clausen2016}, where isotropic infill studied exploited for stability. \cref{fig:lotteTower_deHom_buckling} illustrates the buckling behavior, where for $N_a = 0$, the inner part of the tower experiences a global buckling mode, whereas for $N_a = 2$ and $N_a = 3$, the modes are localized. In fact, the appearance of semi-global modes only occurs after the first 8 and 14 modes for $N_a = 2$ and $N_a = 3$, respectively. Many of these localized modes seem to be influenced by imperfections in the surfaces, leading to local stress raisers. Therefore, a post-cleanup of the surfaces could potentially further increase the BLFs.

This study clearly demonstrates the added benefits of incorporating layer restrictions in the optimization problem, as it enables infill with stability considerations in the de-homogenized results. This approach is thus more computationally efficient compared to explicitly including buckling constraints in the optimization problem, as buckling constraints can be computationally expensive to include~\cite{Ferrari2021}. However, it should be noted that the layer restrictions themselves are not considered as direct buckling constraints.

\subsection{GE Jet Engine Bracket}
To demonstrate the capabilities of the proposed de-homogenization methodology on a more complex problem, the GE Jet engine bracket~\cite{GEJETBRACKET} is considered as an engineering case study. The original bracket description can be found in~\cite{Carter2014}, with some modifications made to the model used in this study, which is similar to the one described in~\cite{Hoeghoej2022}. The bracket has approximate dimensions of $178 \times 106 \times 62 \unit{mm}$. It is subjected to four load cases ($F_1$ - $F_4$). Additionally, for this study, two single load cases with $F_2$ and $F_4$ are considered. These cases are denoted as LCA, LC2, and LC4, representing the multiple load case and the single load cases, respectively.
The loads are applied to a passive shaft (light green colored in \cref{fig:examples}) with artificially higher stiffness (10 times higher) to transfer the loads accurately. The bracket is fixed at the four bolt connections. The shaft and bolt connections are modeled with passive rings connected to the design domain. The allowed total volume fraction is $f^i = 0.137$, corresponding to a total weight of $300 \unit{g}$ of $\Omega$.

The bracket model is discretized using $N_e = 1,185,879$ elements ($3,718,200$ degrees of freedom), resulting in an approximate element side length of $h=0.8 \unit{mm}$. The corresponding large-scale SIMP model is discretized using $N_e = 63,940,348$ elements ($194,309,049$ degrees of freedom), with an approximate element side length of $h=0.2 \unit{mm}$.
The bracket is analyzed both without and with different layer restrictions. The layer restrictions configurations and resulting optimization results are presented in \cref{tab:bracketTO}.

In \cref{tab:bracketTO}, it can be observed that the computational cost for the bracket optimization is higher compared to the two previous examples, mainly due to the increased number of finite elements. The multiple load case example (LCA) has a higher computational cost than the others due to the four right-hand side solves required each iteration. The slower computation time can be attributed, in part, to the challenges associated with generating a high-quality mesh for complex geometries with increased detail. It is difficult to obtain a conforming coarse hexahedral mesh of high quality, which adds complexity to the geometric multigrid levels as well. The lower-quality mesh also contributes to fluctuating computational times.

The introduction of layer restrictions has a minor impact on the compliance, indicating a high degree of non-uniqueness in the solutions. This non-uniqueness is particularly evident in LC4.
\end{multicols}

\begin{table}[b!]
    \centering
    %\begin{threeparttable}
        \caption{Bracket topology optimization results. 
        For each model, the following is reported;
        Compliance of the design ($\mathcal{J}^*$),
        Weighted compliance ($\varsigma^*$),
        Final penalization on angles and relative thickness ($\mathcal{P}^\theta$, $\mathcal{P}^s$),
        Time spent ($T_\text{wall}$, $T_\text{CPU}$).}
        \begin{tabular}{@{}lllllll@{}}
            \toprule
            Model         & $\mathcal{J}^*$           & $\varsigma^*$             & $\mathcal{P}^\theta$       & $\mathcal{P}^s$           & $T_\text{wall}$                      & $T_\text{CPU}$                          \\
            \midrule
            LCA $N_a = 0$ & $ 359.379 \times 10^{2} $ & $ 49.283 \times 10^{2} $  & $ 134.600 \times 10^{-4} $ & $121.584 \times 10^{-3} $ & $ 08 \mathord{:} 04 \mathord{:} 38 $ & $ 5685 \mathord{:} 10 \mathord{:} 53 $  \\
            %LCA $N_a = 2$ & $ 358.210 \times 10^{2} $ & $ 49.123 \times 10^{2} $  & $ 114.470 \times 10^{-4} $ & $190.790 \times 10^{-3} $ & $ 07 \mathord{:} 24 \mathord{:} 42 $ & $ 237 \mathord{:} 05 \mathord{:} 48 $   \\
            LCA $N_a = 3$ & $ 366.333 \times 10^{2} $ & $ 50.237 \times 10^{2} $  & $ 105.620 \times 10^{-4} $ & $203.994 \times 10^{-3} $ & $ 08 \mathord{:} 00 \mathord{:} 20 $ & $ 5634 \mathord{:} 44 \mathord{:} 03 $  \\
            \midrule
            LCA HS Iso.   & $ 371.319 \times 10^{2} $ & $ 50.921 \times 10^{2} $  & -                          & $142.520 \times 10^{-3}$                   & $ 09 \mathord{:} 30 \mathord{:} 50 $ & $ 304 \mathord{:} 22 \mathord{:} 40 $   \\
            LCA SIMP      & $381.955 \times 10^{2} $  & $52.379 \times 10^{2} $   & -                          & -                         & $ 14 \mathord{:} 34 \mathord{:} 41 $ & $ 10260 \mathord{:} 52 \mathord{:} 01 $ \\
            \midrule
            LC2 $N_a = 0$ & $ 807.740 \times 10^{1} $ & $ 110.769 \times 10^{1} $ & $ 68.660 \times 10^{-4} $  & $81.673 \times 10^{-3} $  & $ 03 \mathord{:} 10 \mathord{:} 25 $ & $ 101 \mathord{:} 29 \mathord{:} 58 $   \\
            LC2 $N_a = 3$ & $ 820.471 \times 10^{1} $ & $ 112.515 \times 10^{1} $ & $ 68.890 \times 10^{-4} $  & $157.640 \times 10^{-3} $ & $ 03 \mathord{:} 05 \mathord{:} 05 $ & $ 98 \mathord{:} 39 \mathord{:} 20 $    \\
            \midrule
            LC2 HS Iso.   & $ 815.085 \times 10^{1} $ & $ 111.777 \times 10^{1} $ & -                          & $139.690 \times 10^{-3} $                   & $ 02 \mathord{:} 46 \mathord{:} 30 $ & $ 88 \mathord{:} 44 \mathord{:} 19 $    \\
            LC2 SIMP      & $845.269 \times 10^{1} $  & $115.916 \times 10^{1} $  & -                          & -                         & $ 07 \mathord{:} 02 \mathord{:} 43 $ & $ 4941 \mathord{:} 11 \mathord{:} 27 $  \\
            \midrule
            LC4 $N_a = 0$ & $ 196.507 \times 10^{1} $ & $ 26.948 \times 10^{1} $  & $ 54.400 \times 10^{-4} $  & $76.255 \times 10^{-3} $  & $ 03 \mathord{:} 01 \mathord{:} 32 $ & $ 96 \mathord{:} 46 \mathord{:} 08 $    \\
            LC4 $N_a = 2$ & $ 196.870 \times 10^{1} $ & $ 26.998 \times 10^{1} $  & $ 42.330 \times 10^{-4} $  & $112.847 \times 10^{-3} $ & $ 04 \mathord{:} 49 \mathord{:} 54 $ & $ 154 \mathord{:} 33 \mathord{:} 39 $   \\
            \midrule
            LC4 HS Iso.   & $ 195.473 \times 10^{1} $ & $ 26.806 \times 10^{1} $  & -                          & $139.82 \times 10^{-3}$                   & $ 02 \mathord{:} 34 \mathord{:} 30 $ & $ 82 \mathord{:} 20 \mathord{:} 03 $    \\
            LC4 SIMP      & $202.571  \times 10^{1} $ & $27.780 \times 10^{1}$    & -                          & -                         & $ 05 \mathord{:} 57 \mathord{:} 16 $ & $ 4189 \mathord{:} 58 \mathord{:} 03 $  \\
            \bottomrule
        \end{tabular}
        \label{tab:bracketTO}

    %\end{threeparttable}
\end{table}
\newpage
\begin{figure}[h!]
    \centering
    \makebox[\textwidth][c]{
        \begin{subfigure}{63.333mm}
            \includegraphics[width=\linewidth]{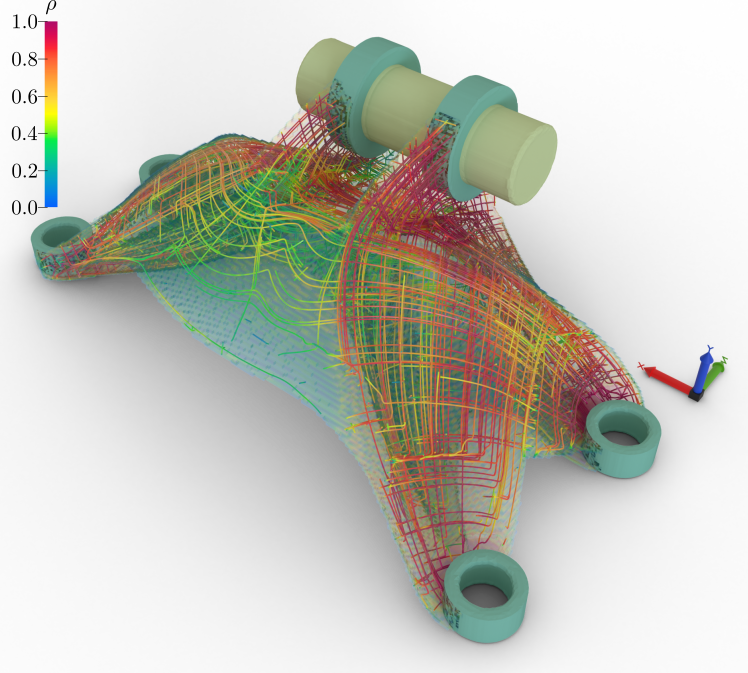}
            \caption{}
            \label{fig:Bracket_LC0_HOM_STREAM}
        \end{subfigure}%
        \begin{subfigure}{63.333mm}
            \includegraphics[width=\linewidth]{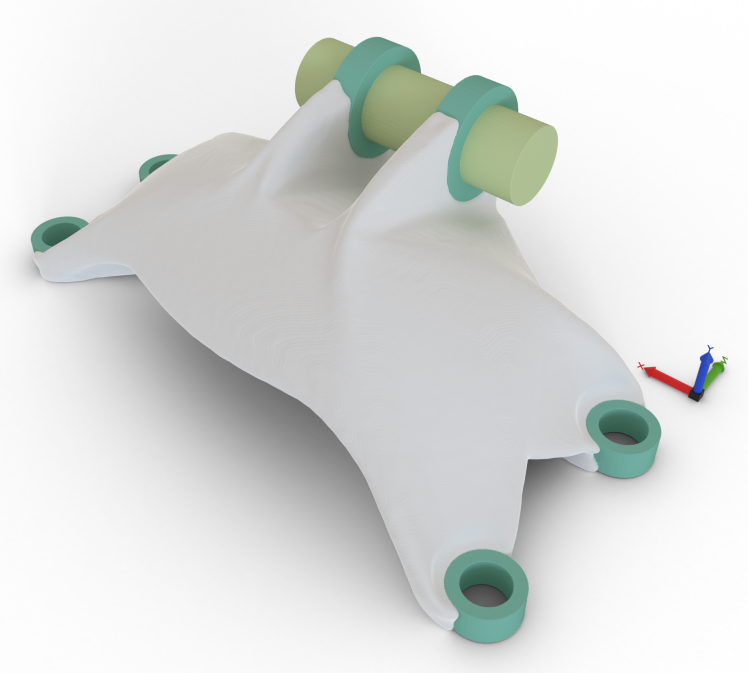}
            \caption{}
            \label{fig:Bracket_LC0_DEHOM}
        \end{subfigure}%
        \begin{subfigure}{63.333mm}
            \includegraphics[width=\linewidth]{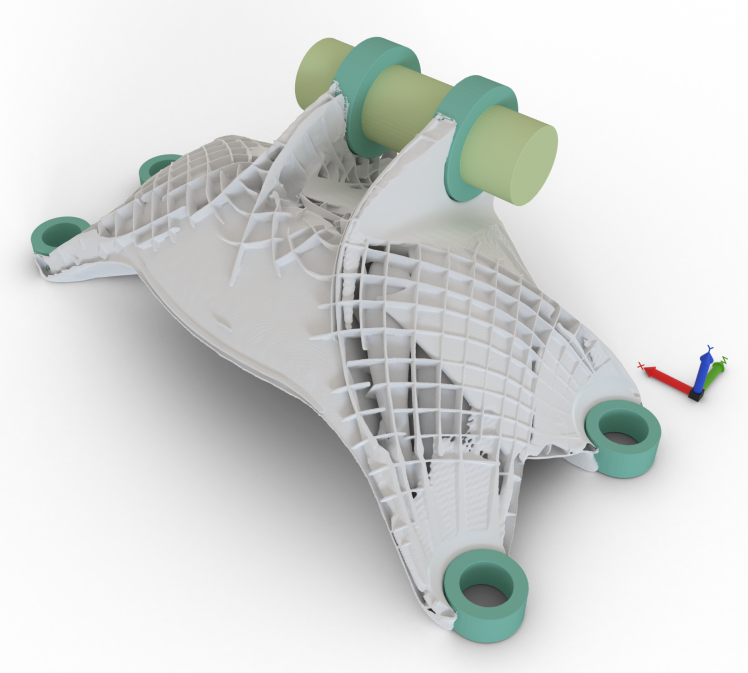}
            \caption{}
            \label{fig:Bracket_LC0_DEHOM_CUT}
        \end{subfigure}%
    }
    \caption{
        Bracket LCA $N_a = 3$.
        (a) Streamline plot of layer normals colored according to the microstructure density.
        (b) The de-homogenized result.
        (c) The outer top surface is removed to expose the internal structure.
    }
    \label{fig:Bracket_LC0_res}

    \makebox[\textwidth][c]{
        \begin{subfigure}{63.333mm}
            \includegraphics[width=\linewidth]{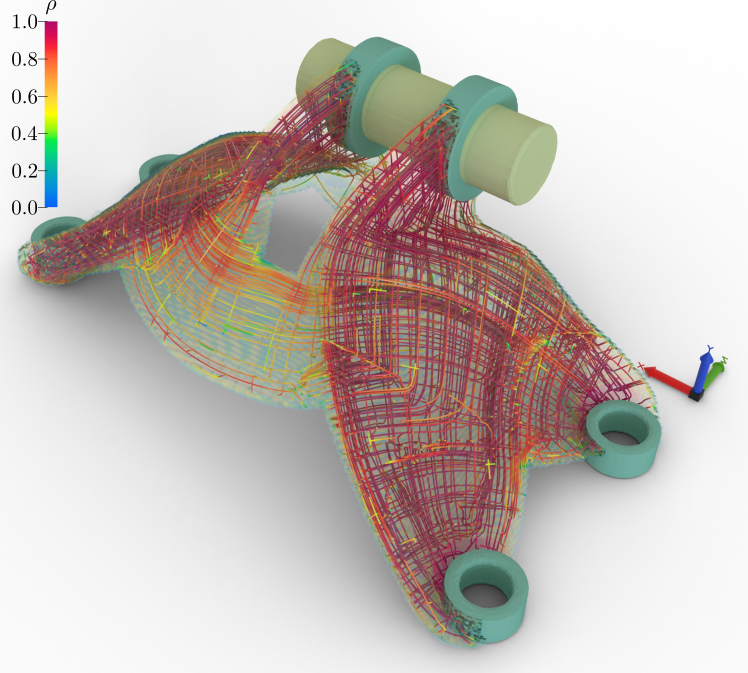}
            \caption{}
            \label{fig:Bracket_LC2_HOM_STREAM}
        \end{subfigure}%
        \begin{subfigure}{63.333mm}
            \includegraphics[width=\linewidth]{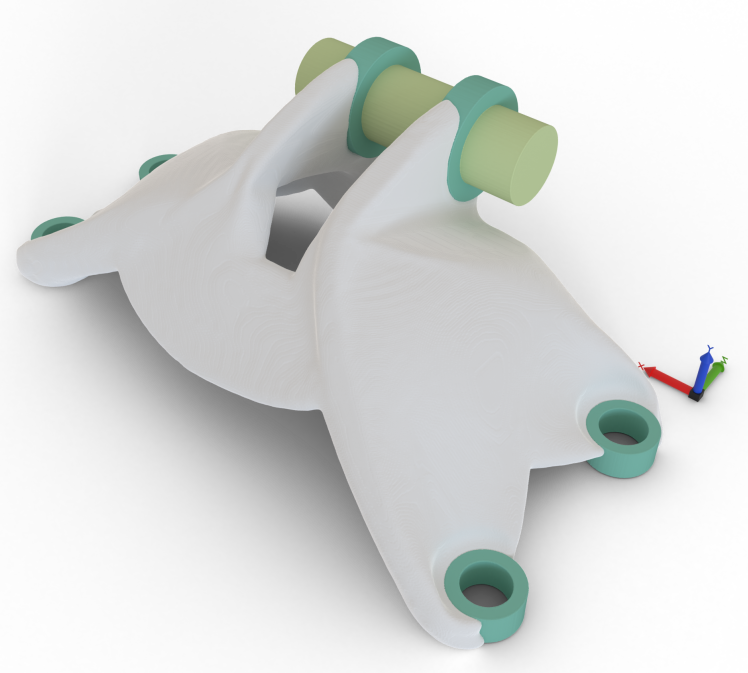}
            \caption{}
            \label{fig:Bracket_LC2_DEHOM}
        \end{subfigure}%
        \begin{subfigure}{63.333mm}
            \includegraphics[width=\linewidth]{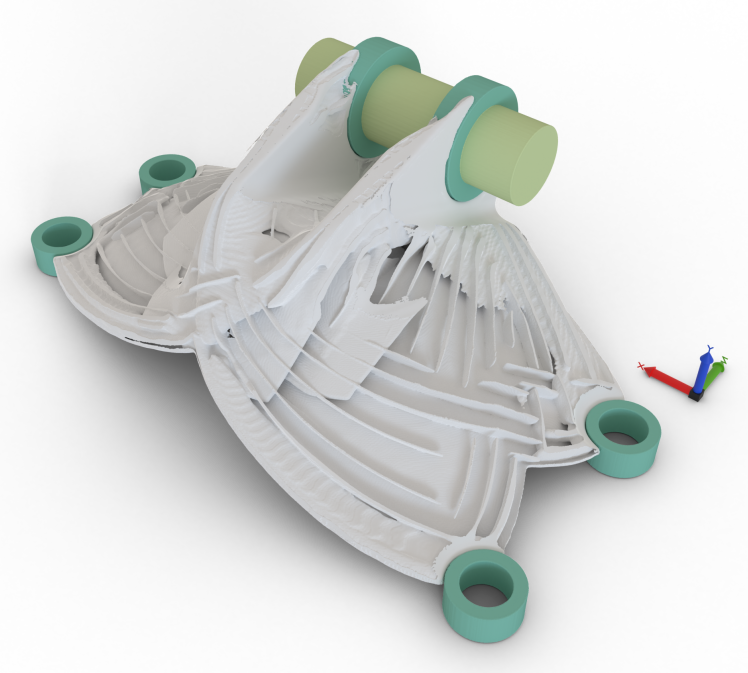}
            \caption{}
            \label{fig:Bracket_LC4_DEHOM_CUT}
        \end{subfigure}%
    }
    \caption{
        Bracket LC2 $N_a = 3$.
        (a) Streamline plot of layer normals colored according to the microstructure density.
        (b) The de-homogenized result.
        (c) The outer top surface is removed to expose the internal structure.
    }
    \label{fig:Bracket_LC2_res}

    \makebox[\textwidth][c]{
        \begin{subfigure}{63.333mm}
            \includegraphics[width=\linewidth]{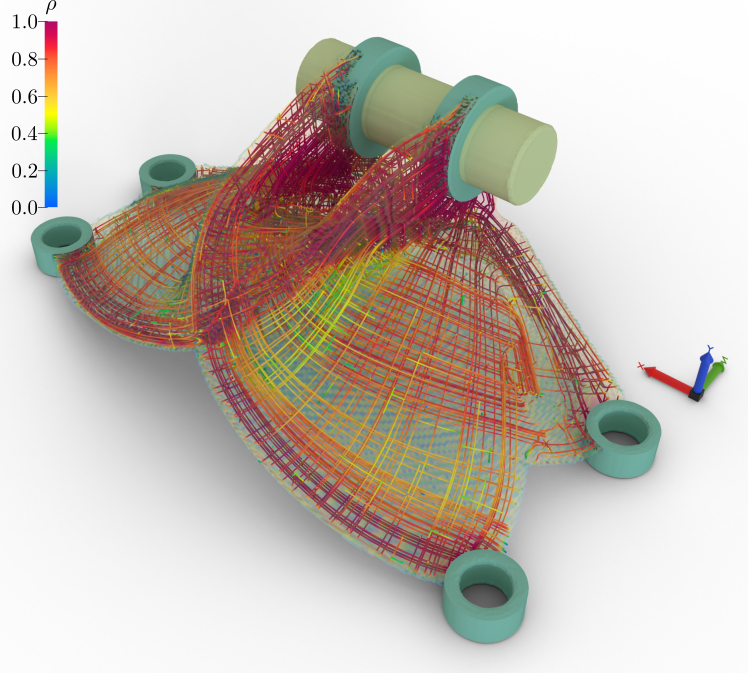}
            \caption{}
            \label{fig:Bracket_LC4_HOM_STREAM}
        \end{subfigure}
        \begin{subfigure}{63.333mm}
            \includegraphics[width=\linewidth]{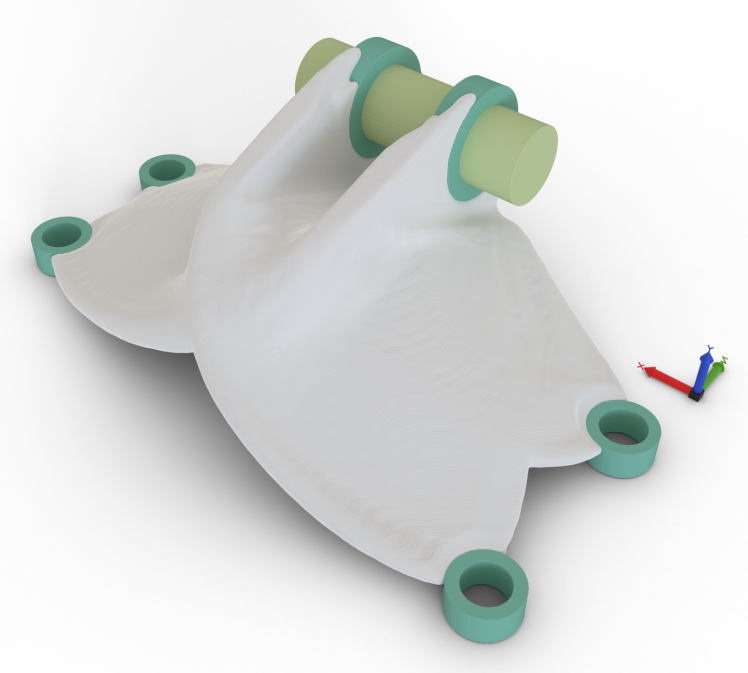}
            \caption{}
            \label{fig:Bracket_LC4_DEHOM}
        \end{subfigure}%
        \begin{subfigure}{63.333mm}
            \includegraphics[width=\linewidth]{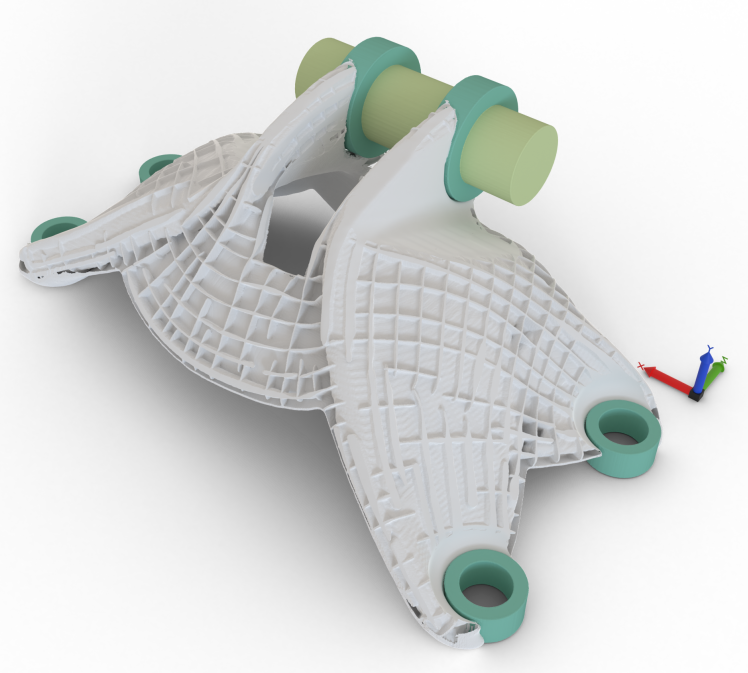}
            \caption{}
            \label{fig:Bracket_LC2_DEHOM_CUT}
        \end{subfigure}%

    }%
    \caption{
        Bracket LC4 $N_a = 2$.
        (a) Streamline plot of layer normals colored according to the microstructure density.
        (b) The de-homogenized result.
        (c) The outer top surface is removed to expose the internal structure.
    }
    \label{fig:Bracket_LC4_res}
\end{figure}
\newpage
\begin{table}[t!]
    \centering
    %\begin{threeparttable}
        \caption{Bracket de-homogenization results.
        For each model, the following is reported;
        Compliance of the design ($\mathcal{J}^S$),
        Volume fraction of design ($f^S$),
        Weighted compliance ($\varsigma^S$),
        Relative difference to optimized design in volume and weighted compliance ($\delta(f^*)$, $\delta(\varsigma^*)$),
        Relative difference to SIMP design in weighted compliance ($\delta(\varsigma^\rho)$),
        Time spent ($T^S_\text{wall}$, $T^S_\text{CPU}$).}
        \begin{tabular}{@{}lllllllllll@{}}
            \toprule
            Model         & $\mathcal{J}^S$            & $f^S$ & $\varsigma^S$             & $\delta(f^*)$ & $\delta(\varsigma^*)$ & $\delta\left(\varsigma^\rho\right)$ & $T_\text{wall}^S$ & $T_\text{CPU}^S$ \\ \midrule
            LCA $N_a = 0$ & $ 628.538 \times 10^{2} $  & 0.118 & $74.021 \times 10^{2} $   & -14.12        & 50.68                 & 45.36                               & 03:25:30          & 39:24:57         \\
            LCA $N_a = 3$ & $ 458.084 \times 10^{2} $  & 0.128 & $ 58.420 \times 10^{2} $  & -7.00         & 18.93                 & 14.73                               & 02:35:54          & 51:21:13         \\ \midrule
            LC2 $N_a = 0$ & $ 1232.162 \times 10^{1} $ & 0.117 & $ 144.725 \times 10^{1} $ & -14.35        & 30.66                 & 24.85                               & 03:13:05          & 51:57:43         \\
            LC2 $N_a = 3$ & $ 909.225 \times 10^{1} $  & 0.135 & $ 122.276 \times 10^{1} $ & -1.93         & 8.68                  & 5.49                                & 05:15:04          & 59:31:55         \\ \midrule
            LC4 $N_a = 0$ & $ 302.207 \times 10^{1} $  & 0.111 & $ 33.518 \times 10^{1} $  & -19.12        & 24.38                 & 20.66                               & 03:26:40          & 54:16:00         \\
            LC4 $N_a = 2$ & $ 226.603 \times 10^{1} $  & 0.126 & $ 28.625 \times 10^{1} $  & -7.88         & 6.03                  & 3.04                                & 04:10:57          & 56:47:26         \\
            \bottomrule
        \end{tabular}
        \label{tab:bracketDeHom}
    %\end{threeparttable}
\end{table}
\begin{multicols}{2}
Density plots of the bracket structures are shown in \cref{fig:Bracket_LC0_HOM_STREAM}, \cref{fig:Bracket_LC2_HOM_STREAM}, and \cref{fig:Bracket_LC4_HOM_STREAM} for LCA with $N_a = 2$, LC2 with $N_a = 3$, and LC4 with $N_a = 2$, respectively. The density values are visualized as streamlines, which also provide information about the surface normals. It can be observed that the densities are relatively higher in LC2 and LC4 compared to LCA. The streamlines represent the individual orientation fields, resulting in patches of smooth fields being visible.

The de-homogenized bracket structures have a minimum length scale of $\delta_{\min} = 0.25 \unit{mm}$. The results of the de-homogenization process are presented in \cref{tab:bracketDeHom} and \cref{fig:Bracket_LC0_res}, \cref{fig:Bracket_LC2_res}, and \cref{fig:Bracket_LC4_res} for $N_a = 2$  and $3$, while for $N_a = 0$  is seen in \cref{fig:Bracket_A0}. The strain energy density distribution is seen in \cref{fig:Bracket_res_E}.

Similar to the Lotte Tower example, the de-homogenized bracket exhibits a behavior influenced by the overall feature size and periodicity, particularly in the case of $N_a = 0$. This behavior arises due to the relatively high density, which adversely affects the accuracy of the volume fraction obtained from the de-homogenization method.

However, considering the additional stability benefits associated with $N_a > 1$, it is more reasonable from an engineering perspective to consider these models. Furthermore, it is evident from the results that the structural components have an infill-like internal structure which is beneficial for additive manufacturing. In these cases, both the volume fraction and weighted compliance are within 10\% of the optimized results for the single load case problems. This implies that compared to the reference large-scale SIMP model, a computational speedup of almost up to 30 can be achieved by using an even finer length scale.

A speedup of 30 is relatively low compared to what has been observed in previous de-homogenization results. The complexity of the grid has led to lower-quality multigrid levels in the homogenization-based model, while the fine discretization of the large-scale SIMP model allows for more and higher-quality multigrid levels. However, the computational complexity is still significantly higher for the large-scale SIMP model. The grid complexity has also affected the surface generation and synthesis of the volumetric solid in the de-homogenization process, as shown in \cref{tab:overview2}.

The larger discrepancy for de-homogenization results of the multiple load case problems can be caused by the fact that the underlying microstructure model is non-optimal for multiple load cases problems, leading to a more complex frame field with the limited orientations and layers, as also seen in 2D~\cite{Jensen2022}. Ideally, a rank-6 (or rank-4) microstructure should be considered for multiple load case problems, however, imposing manufacturability constraints in the sense of minimum wall (lamination) thickness will also be a problem for higher rank microstructure.

\end{multicols}

\begin{multicols}{2}[\section{Discussion and Concluding Remarks}\label{sec:Discussion}]
This work presents a stream surface-based de-homogenization topology optimization procedure tailored for complex geometries on unstructured finite element grids using a rectangular-hole microstructure as the base material. The capabilities of the proposed procedure is demonstrated through three different examples with varying levels of complexity.

The structural performance of the de-homogenized designs aligns well with current state-of-the-art methods, however, the proposed approach provides a significant reduction in computational time, achieving de-homogenized results more than 20 times faster than previous published methods.

The proposed method display minor discrepancies between the mapped results and the underlying multi-scale solution. These discrepancies can be attributed to several factors. First, the grid discretization size and quality are not uniform, as ensuring this is challenging, if not impossible for complex geometries. Second, the relatively large minimal feature size compared to the dimensions of $\Omega$ limits the separation of scales and affects the target periodicity. While using a smaller minimal feature size can enhance the solution, it is not practical for post-evaluation and possible manufacturing in most cases. Remark, that it should be possible to control the length scale for each layer ensuring the target periodicity, howevere this is left for future work.

An additional user-controlled functionality allowing the number of active layers to be varied. It is demonstrated that requiring multiple active layers ensures that the mapped solution becomes more mechanically stable, resulting in a significant increase in the critical buckling load factor without any additional computational expense. The control of active layers also provides direct control over the internal structure, i.e., infill, which demonstrates that the infill is incorporated as a structural component. Hence, enabling direct control over the active layers presents a promising approach for enhancing manufacturability. Future research should focus on expanding the layer control capability to encompass local volume fraction control of the infill, addressing overhang limitations, and incorporating features such as powder and resin evacuation holes. These advancements will further improve the manufacturability of the de-homogenization procedure.

To benchmark the proposed approach, a large-scale SIMP model obtained using supercomputing was used. The computational efficiency of the proposed method ranged from just below 10 times to over 250 times faster for single-load case problems. The multi-scale optimization step is the most computationally expensive, primarily due to the challenges in obtaining a high-quality coarse hexahedral discretization. The use of tetrahedrals could be explored as an alternative, although it has not been attempted due to the inferiority of this element when using a linear interpolation. The variation in speed-up observed is largely influenced by the heuristic nature of the stream surface-based de-homogenization step, which could be mitigated through more problem-specific parameter tuning.

In summary, the proposed approach achieves promising results for complex, engineering-relevant examples, with volume fractions and weighted compliance values within 5\% of the large-scale SIMP model obtained at a fraction of the computational cost.

Finally, the study also identifies the need to extend the multi-scale optimization procedure with either rank-4 or rank-6 base materials to handle multiple load-case problems, which is left for future work.

\end{multicols}

\begin{multicols}{2}

\section*{Acknowledgments}
The authors acknowledge the financial support from the InnoTop VILLUM investigator project through the Villum Foundation and nTopology inc. Furthermore, the authors would like to express their gratitude to the members of the TopOpt group at DTU for valuable discussions during the preparation of this work.

\section*{Declarations}

The authors declare that they have no known competing financial interests or personal relationships that could have appeared to influence the work reported in this paper.

\end{multicols}

\FloatBarrier

\appendix

\renewcommand{\thesection}{\Alph{section}}

\begin{multicols}{2}[\section{Rectangular-hole Microstructures}\label{sec:microMat}]
The rectangular-hole microstructure consists of three orthogonal periodic walls, referred to as laminates for convenience. The walls are defined by a stiff isotropic material represented by $(+)$, while the surrounding void is defined by compliant isotropic material represented by $(-)$, hence mimicking a void material. Hence the local elasticity varies between the two linear-elastic material phases with Young’s moduli $E^+$ and $E^-$. Both phases have an identical Poisson’s ratio $\nu$. The relative thickness for the stiff material phases is given by the relative thickness parameters $w_n$, respectively. With $w_n \in [0,1], \quad n \in \{1,2,3\}$. The elastic properties of the microstructure in the $\bm{y}$ reference frame is only dependent on thickness parameters. Furthermore, the microstructure volume fraction, $\rho$, is found as
\begin{equation}
    \rho = 1 - (1 - w_1)(1 - w_2)(1 - w_3).
\end{equation}

Due to the lack of a closed-form solution to the homogenized elasticity tensor of this microstructure for freely varying thickness parameters, the homogenized elasticity tensor can be computed with numerical homogenization as proposed by~\cite{BendsoeKikuchi1988}. The homogenized elasticity tensor is computed from 6 independent unit strain fields using periodic boundary conditions.

The numerical homogenization is obtained with a PETSc-based homogenization code from~\cite{Wang2021}, much similar to the publicly available MATLAB code by~\cite{Andreassen2014}. A mesh of $40 \times 40 \times 40$ tri-linear hexahedral finite elements is considered to compute the homogenized elasticity matrix $\mathbf{C}^H \in \mathbb{R}^{6\times 6}$ (Voigt notation), as a data set $\{ \mathbf{C}^H \}_{ijk}$ for $(i,j,k) \in \{1,...,21\}^3$ different linear combinations of the relative thickness parameters $\{w_n\}_{ijk}$. Symmetry can be exploited so only a small subset of the combinations needs to be computed.

A continuous function of the homogenized elasticity matrix as a function of the three relative thickness parameters is obtained from a tri-linear interpolation function with 
\begin{equation}
    \mathbf{C}^H(w_n) = \mathcal{L} \left( \{ \mathbf{C}^H \}_{ijk}, \{w_n\}_{ijk},w_n \right),
\end{equation}
where $\mathcal{L}$ is a tri-linear interpolation function. Resulting in a proper data driven constitutive function. 

$\mathbf{C}^H(w_n)$ is aligned with microscopic reference frame $\bm{y}$. As the elastic property is directionally dependent, the design space of the microstructure is expanded to geometric rotations of the microscopic reference frame $\bm{y}$, with respect to $\bm{x}$. Let $\mathbf{R} \in \mathbb{R}^{3\times 3}$ be a proper orthogonal rotation matrix,
\begin{equation}
    \mathbf{R} = \left[\begin{array}{ccc}
        c_{2} c_{3} & c_{3} s_{1} s_{2}-c_{1} s_{3} & c_{1} c_{3} s_{2}+s_{1} s_{3} 
        \\
         c_{2} s_{3} & s_{1} s_{2} s_{3}+c_{1} c_{3} & c_{1} s_{2} s_{3}-c_{3} s_{1} 
        \\
         -s_{2} & s_{1} c_{2} & c_{1} c_{2} 
        \end{array}\right],
\end{equation}
where $c_n = \cos(\theta_n)$ and $s_n = \sin(\theta_n)$, for $n \in \{1,2,3\}$.  With $\theta_n  \in [-4\pi,4 \pi], \quad n \in \{1,2,3\}$ are the frame orientation angles, about ${x}_3$, ${x}_2$, and ${x}_1$, respectively. Then let $\mathbf{T} \in \mathbb{R}^{6\times 6}$ be a transformation matrix, for instance found in~\cite{cook_concepts_2001}, and shown here for convenience,

\begin{equation}
    \mathbf{T} = \left[\begin{array}{cc}
    \mathbf{T}_{11} & \mathbf{T}_{12} \\
    \mathbf{T}_{21} & \mathbf{T}_{22} 
    \end{array}\right],
\end{equation}

\begin{equation}
    \mathbf{T}_{11} = \left[\begin{array}{ccc}
(\mathbf{n}_1)_1^{2} & (\mathbf{n}_2)_1^{2} & (\mathbf{n}_3)_1^{2} \\
 (\mathbf{n}_1)_2^{2} & (\mathbf{n}_2)_2^{2} & (\mathbf{n}_3)_2^{2} \\
 (\mathbf{n}_1)_3^{2} & (\mathbf{n}_2)_3^{2} & (\mathbf{n}_3)_3^{2} 
\end{array}\right],
\end{equation}

\begin{equation}
    \mathbf{T}_{12} = \left[\begin{array}{ccc}
2 (\mathbf{n}_2)_1 (\mathbf{n}_3)_1 & 2 (\mathbf{n}_3)_1 (\mathbf{n}_1)_1 & 2 (\mathbf{n}_1)_1 (\mathbf{n}_2)_1 \\
2 (\mathbf{n}_2)_2 (\mathbf{n}_3)_2 & 2 (\mathbf{n}_3)_2 (\mathbf{n}_1)_2 & 2 (\mathbf{n}_1)_2 (\mathbf{n}_2)_2 \\
2 (\mathbf{n}_2)_3 (\mathbf{n}_3)_3 & 2 (\mathbf{n}_3)_3 (\mathbf{n}_1)_3 & 2 (\mathbf{n}_1)_3 (\mathbf{n}_2)_3 
\end{array}\right],
\end{equation}

\begin{equation}
    \mathbf{T}_{21} = \left[\begin{array}{ccc}
 (\mathbf{n}_1)_2 (\mathbf{n}_1)_3 & (\mathbf{n}_2)_2 (\mathbf{n}_2)_3 & (\mathbf{n}_3)_2 (\mathbf{n}_3)_3 \\
 (\mathbf{n}_1)_3 (\mathbf{n}_1)_1 & (\mathbf{n}_2)_3 (\mathbf{n}_2)_1 & (\mathbf{n}_3)_3 (\mathbf{n}_3)_1 \\
 (\mathbf{n}_1)_1 (\mathbf{n}_1)_2 & (\mathbf{n}_2)_1 (\mathbf{n}_2)_2 & (\mathbf{n}_3)_1 (\mathbf{n}_3)_2 
\end{array}\right],
\end{equation}

\begin{equation}
\begin{split}
    \mathbf{T}_{22} = \left[\begin{array}{cc}
 (\mathbf{n}_2)_2 (\mathbf{n}_3)_3+(\mathbf{n}_2)_3 (\mathbf{n}_3)_2 & (\mathbf{n}_1)_2 (\mathbf{n}_3)_3+(\mathbf{n}_1)_3 (\mathbf{n}_3)_2
\\
 (\mathbf{n}_2)_1 (\mathbf{n}_3)_3+(\mathbf{n}_2)_3 (\mathbf{n}_3)_1 & (\mathbf{n}_1)_1 (\mathbf{n}_3)_3+(\mathbf{n}_1)_3 (\mathbf{n}_3)_1
\\
 (\mathbf{n}_2)_1 (\mathbf{n}_3)_2+(\mathbf{n}_2)_2 (\mathbf{n}_3)_1 & (\mathbf{n}_1)_1 (\mathbf{n}_3)_2+(\mathbf{n}_1)_2 (\mathbf{n}_3)_1
\end{array}\right.\\
\left.\begin{array}{ccc}
  (\mathbf{n}_1)_2 (\mathbf{n}_2)_3+(\mathbf{n}_1)_3 (\mathbf{n}_2)_2 
\\
 (\mathbf{n}_1)_1 (\mathbf{n}_2)_3+(\mathbf{n}_1)_3 (\mathbf{n}_2)_1 
\\
 (\mathbf{n}_1)_1 (\mathbf{n}_2)_2+(\mathbf{n}_1)_2 (\mathbf{n}_2)_1 
\end{array}\right].
\end{split}
\end{equation}

where $\mathbf{n}_n \in \mathbb{R}^3$, for $n \in \{1,2,3\}$, are the transformed normals aligned with $\bm{y}$ to the three orthogonal laminates, defined as $\mathbf{n}_n = \mathbf{R}\mathbf{e}_n $, where $\mathbf{e}_n \in \mathbb{R}^3$, be the $n$th unit vector.
Hence $\mathbf{R}$ is mapped to $\mathbf{T}$, so that $\mathbf{C}^H(\mathbf{w})$ can be freely rotated in 3D space,
\begin{equation}
    \tilde{\mathbf{C}}^H(w_n,\theta_n) = \mathbf{T}(\theta_n) \, \mathbf{C}^H(w_n)\mathbf{T}^\top \, (\theta_n).
\end{equation}
Here $\tilde{\mathbf{C}}^H$ is the rotated homogenized elasticity matrix.

\end{multicols}

\begin{multicols}{2}[\section{Isotropic Material Models}]
\subsection{Two material Isotropic Hashin-Shtrikman bounds}\label{app:hs}
The isotropic Hashin-Shtrikman (HS) upper-bound~\cite{Hashin1963} of a two-phase material. The effective bulk, $K$, and shear, $G$, modulus from a two-material compound is expressed in the following from,
\begin{align}
    K^* &= K_{1}+\frac{1-\varrho_1}{\frac{1}{K_{2}-K_{1}}+\frac{3 \varrho_1}{3 K_{1}+4 G_{1}}},\\
    G^* &= G_{1}+\frac{1-\varrho_1}{\frac{1}{G_{2}-G_{1}}+\frac{6 \left(K_{1}+2 G_{1}\right) \varrho_1}{5 G_{1} \left(3 K_{1}+4 G_{1}\right)}},
\end{align}
where the subscripts indicate the two isotropic linear elastic materials. $\varrho_1$ is the relative material fraction of material 1 in relation to material 2. With the known relationship between Young's, shear, bulk modulus and Poisson's ratio for isotropic materials,
\begin{equation}
    E = 2 G (1 - \nu) = 3 K (1 - 2\nu).
\end{equation}
The effective Young's modulus and Poisson's ratio are expressed as for two materials.
\begin{equation}\label{eq:estar}
    E^* = \frac{9 K^* G^*}{3 K^* + G^*}, \quad \nu^* = \frac{3 K^* - 2 G^*}{2 (3 K^* + G^*)}.
\end{equation}

\subsection{SIMP model}\label{app:simp}
The Solid Isotropic Microstructure with Penalization (SIMP)~\cite{Bendsoe1999} model is defined with the following effective Young's modulus,
\begin{equation}
    E^* = E_{\min}+\left(E_{0}-E_{\min}\right) \rho^{p}
\end{equation}
where $E_0$ is the Young's modulus of the base material and $E_{\min}$ Young's modulus of the void material. The porous material is penalized with $p=3$.
\end{multicols}

\begin{multicols}{2}[\section{Additional results}]
The following appendix contains additional results from the three examples.
\end{multicols}
\FloatBarrier
\begin{table}[htbp]
    \centering
    %\begin{threeparttable}
        \caption{Total overview of the computational performance of each example.
        For each example the number of elements (in
            millions) of the homogenized and de-homogenized design in addition
            to the SIMP design ($N_e^{H}$, $N_e^{D}$ and $N_e^{SIMP}$
            respectively) is reported. Additionally the CPU time for the
            homogenization and de-homogenization ($T_\text{CPU}$,
            $T_\text{CPU}^\text{S}$). Finally the combined CPU time for the
            homogenization and de-homogenization ($T_\text{CPU}^\text{Tot}$), is
            compared to the SIMP solution ($T_\text{CPU}^\text{SIMP}$) and the
            speed-up is reported ($S$). The timings is reported as
                [hh:mm:ss].}
        \label{tab:overview1}

        \begin{tabular}{@{}l|rrr|rr|rr|r@{}}
            \toprule
            Model                                              &
            \multicolumn{1}{c}{$N_e^\text{H}$}                 &
            \multicolumn{1}{c}{$N_e^\text{D}$}                 &
            \multicolumn{1}{c|}{$N_e^\text{SIMP}$}             &
            \multicolumn{1}{c}{$T_\text{CPU}$}                 &
            \multicolumn{1}{c|}{$T_\text{CPU}^\text{S}$}       &
            \multicolumn{1}{c}{$T_\text{CPU}^\text{Tot}$}      &
            \multicolumn{1}{c|}{$T_\text{CPU}^\text{SIMP}$}    &
            \multicolumn{1}{c}{$S$}
            \\
            \midrule
            Cantilever $N_a = 0$                               & $ 0.23 $ & $ 2.69 $  & $ 115.61 $ & $ 13 \mathord{:} 39 \mathord{:} 18 $   & $ 06 \mathord{:} 06 \mathord{:} 11 $ & $ 19 \mathord{:} 45 \mathord{:} 29 $   & $ 4510 \mathord{:} 58 \mathord{:} 55 $  & $ 228 \times $ \\
            Cantilever $N_a = 0$ (No hull)                     & $ 0.23 $ & $ 2.53 $  & $ 115.61 $ & $ 13 \mathord{:} 39 \mathord{:} 18 $   & $ 03 \mathord{:} 59 \mathord{:} 40 $ & $ 17 \mathord{:} 38 \mathord{:} 58 $   & $ 4510 \mathord{:} 58 \mathord{:} 55 $  & $ 256 \times $ \\
            Cantilever $N_a = 0$ ($0.5 \delta_{\mathrm{min}}$) & $ 0.23 $ & $ 19.48 $ & $ 115.61 $ & $ 13 \mathord{:} 39 \mathord{:} 18 $   & $ 13 \mathord{:} 40 \mathord{:} 47 $ & $ 27 \mathord{:} 20 \mathord{:} 05 $   & $ 4510 \mathord{:} 58 \mathord{:} 55 $  & $ 165 \times $ \\
            Cantilever $N_a = 0$ (Sym.)                        & $ 0.23 $ & $ 2.99 $  & $ 115.61 $ & $ 13 \mathord{:} 39 \mathord{:} 18 $   & $ - \mathord{:} - \mathord{:} - $    & $ - \mathord{:} - \mathord{:} - $      & $ 4510 \mathord{:} 58 \mathord{:} 55 $  & $ - $          \\
            Cantilever $N_a = 2$                               & $ 0.23 $ & $ 2.76 $  & $ 115.61 $ & $ 12 \mathord{:} 38 \mathord{:} 20 $   & $ 06 \mathord{:} 41 \mathord{:} 55 $ & $ 19 \mathord{:} 20 \mathord{:} 15 $   & $ 4510 \mathord{:} 58 \mathord{:} 55 $  & $ 233 \times $ \\
            Cantilever $N_a = 2$ (Sym.)                        & $ 0.23 $ & $ 3.08 $  & $ 115.61 $ & $ 12 \mathord{:} 38 \mathord{:} 20 $   & $ - \mathord{:} - \mathord{:} - $    & $ - \mathord{:} - \mathord{:} - $      & $ 4510 \mathord{:} 58 \mathord{:} 55 $  & $ - $          \\
            Cantilever $N_a = 3$                               & $ 0.23 $ & $ 2.71 $  & $ 115.61 $ & $ 13 \mathord{:} 12 \mathord{:} 21 $   & $ 07 \mathord{:} 35 \mathord{:} 46 $ & $ 20 \mathord{:} 48 \mathord{:} 07 $   & $ 4510 \mathord{:} 58 \mathord{:} 55 $  & $ 217 \times $ \\
            Cantilever $N_a = 3$ (Sym.)                        & $ 0.23 $ & $ 3.06 $  & $ 115.61 $ & $ 13 \mathord{:} 12 \mathord{:} 21 $   & $ - \mathord{:} - \mathord{:} - $    & $ - \mathord{:} - \mathord{:} - $      & $ 4510 \mathord{:} 58 \mathord{:} 55 $  & $ - $          \\ \midrule
            Lotte $N_a = 0$                                    & $ 0.42 $ & $ 4.63 $  & $ 26.64 $  & $ 29 \mathord{:} 24 \mathord{:} 33 $   & $ 14 \mathord{:} 24 \mathord{:} 40 $ & $ 43 \mathord{:} 49 \mathord{:} 13 $   & $ 710 \mathord{:} 26 \mathord{:} 04 $   & $ 16 \times $  \\
            Lotte $N_a = 0$ (SG)                               & $ 0.42 $ & $ 4.60 $  & $ 26.64 $  & $ 30 \mathord{:} 09 \mathord{:} 05 $   & $ 14 \mathord{:} 55 \mathord{:} 54 $ & $ 45 \mathord{:} 04 \mathord{:} 59 $   & $ 710 \mathord{:} 26 \mathord{:} 04 $   & $ 16 \times $  \\
            Lotte $N_a = 2$                                    & $ 0.42 $ & $ 4.77 $  & $ 26.64 $  & $ 25 \mathord{:} 32 \mathord{:} 06 $   & $ 24 \mathord{:} 18 \mathord{:} 40 $ & $ 49 \mathord{:} 50 \mathord{:} 46 $   & $ 710 \mathord{:} 26 \mathord{:} 04 $   & $ 14 \times $  \\
            Lotte $N_a = 3$                                    & $ 0.42 $ & $ 5.09 $  & $ 26.64 $  & $ 25 \mathord{:} 18 \mathord{:} 26 $   & $ 61 \mathord{:} 25 \mathord{:} 10 $ & $ 86 \mathord{:} 43 \mathord{:} 36 $   & $ 710 \mathord{:} 26 \mathord{:} 04 $   & $ 8 \times $   \\ \midrule
            Bracket LCA $N_a = 0$                              & $ 1.19 $ & $ 14.64 $ & $ 63.94 $  & $ 5685 \mathord{:} 10 \mathord{:} 53 $ & $ 39 \mathord{:} 24 \mathord{:} 57 $ & $ 5724 \mathord{:} 35 \mathord{:} 50 $ & $ 10260 \mathord{:} 52 \mathord{:} 01 $ & $ 2 \times $   \\
            Bracket LCA $N_a = 3$                              & $ 1.19 $ & $ 16.29 $ & $ 63.94 $  & $ 5634 \mathord{:} 44 \mathord{:} 03 $ & $ 51 \mathord{:} 21 \mathord{:} 13 $ & $ 5686 \mathord{:} 05 \mathord{:} 16 $ & $ 10260 \mathord{:} 52 \mathord{:} 01 $ & $ 2 \times $   \\
            Bracket LC2 $N_a = 0$                              & $ 1.19 $ & $ 13.58 $ & $ 63.94 $  & $ 101 \mathord{:} 29 \mathord{:} 58 $  & $ 51 \mathord{:} 57 \mathord{:} 43 $ & $ 153 \mathord{:} 27 \mathord{:} 41 $  & $ 4941 \mathord{:} 11 \mathord{:} 27 $  & $ 32 \times $  \\
            Bracket LC2 $N_a = 3$                              & $ 1.19 $ & $ 14.64 $ & $ 63.94 $  & $ 98 \mathord{:} 39 \mathord{:} 20 $   & $ 59 \mathord{:} 31 \mathord{:} 55 $ & $ 158 \mathord{:} 11 \mathord{:} 15 $  & $ 4941 \mathord{:} 11 \mathord{:} 27 $  & $ 31 \times $  \\
            Bracket LC4 $N_a = 0$                              & $ 1.19 $ & $ 13.21 $ & $ 63.94 $  & $ 96 \mathord{:} 46 \mathord{:} 08 $   & $ 54 \mathord{:} 16 \mathord{:} 00 $ & $ 151 \mathord{:} 02 \mathord{:} 08 $  & $ 4189 \mathord{:} 58 \mathord{:} 03 $  & $ 28 \times $  \\
            Bracket LC4 $N_a = 2$                              & $ 1.19 $ & $ 14.18 $ & $ 63.94 $  & $ 154 \mathord{:} 33 \mathord{:} 39 $  & $ 56 \mathord{:} 47 \mathord{:} 26 $ & $ 211 \mathord{:} 21 \mathord{:} 05 $  & $ 4189 \mathord{:} 58 \mathord{:} 03 $  & $ 20 \times $  \\
            \bottomrule
        \end{tabular}
    %\end{threeparttable}
\end{table}
\begin{table}[htbp]
    \centering
    %\begin{threeparttable}
        \caption{Total overview of the wall time spent of each example.
        This table reports the wall time spent for the homogenization
            ($T_\text{wall}$), all stages of the de-homogenization algorithm
            ($T_\text{wall}^\text{Gen}$, $T_\text{wall}^\text{Opt}$,
            $T_\text{wall}^\text{SS}$, $T_\text{wall}^\text{Syn}$) and the total
            wall time ($T_\text{wall}^\text{Tot}$). Additionally, the wall time for SIMP is reported. The timings is reported as [hh:mm:ss].}
        \label{tab:overview2}

        \begin{tabular}{@{}l|r|rrrr|rr@{}}
            \toprule
            Model                                              &
            \multicolumn{1}{c|}{$T_\text{wall}$}               &
            \multicolumn{1}{c}{$T_\text{wall}^\text{Gen}$}     &
            \multicolumn{1}{c}{$T_\text{wall}^\text{Opt}$}     &
            \multicolumn{1}{c}{$T_\text{wall}^\text{SS}$}      &
            \multicolumn{1}{c|}{$T_\text{wall}^\text{Syn}$}    &
            \multicolumn{1}{c}{$T_\text{wall}^\text{Tot}$}     &
            \multicolumn{1}{c}{$T_\text{wall}^\text{SIMP}$}
            \\
            \midrule
            Cantilever $N_a = 0$                               & $ 00 \mathord{:} 25 \mathord{:} 43 $ & $ 00 \mathord{:} 08 \mathord{:} 07 $ & $ 00 \mathord{:} 01 \mathord{:} 40 $ & $ 00 \mathord{:} 01 \mathord{:} 17 $ & $ 00 \mathord{:} 15 \mathord{:} 45 $ & $ 00 \mathord{:} 52 \mathord{:} 33 $ & $ 03 \mathord{:} 55 \mathord{:} 10 $ \\
            Cantilever $N_a = 0$ (No hull)                     & $ 00 \mathord{:} 25 \mathord{:} 43 $ & $ 00 \mathord{:} 05 \mathord{:} 48 $ & $ 00 \mathord{:} 00 \mathord{:} 39 $ & $ 00 \mathord{:} 00 \mathord{:} 50 $ & $ 00 \mathord{:} 06 \mathord{:} 26 $ & $ 00 \mathord{:} 39 \mathord{:} 28 $ & $ 03 \mathord{:} 55 \mathord{:} 10 $ \\
            Cantilever $N_a = 0$ ($0.5 \delta_{\mathrm{min}}$) & $ 00 \mathord{:} 25 \mathord{:} 43 $ & $ 00 \mathord{:} 16 \mathord{:} 21 $ & $ 00 \mathord{:} 46 \mathord{:} 05 $ & $ 00 \mathord{:} 01 \mathord{:} 20 $ & $ 00 \mathord{:} 52 \mathord{:} 07 $ & $ 02 \mathord{:} 21 \mathord{:} 37 $ & $ 03 \mathord{:} 55 \mathord{:} 10 $ \\
            Cantilever $N_a = 0$ (Sym.)                        & $ 00 \mathord{:} 25 \mathord{:} 43 $ & $ - \mathord{:} - \mathord{:} - $    & $ - \mathord{:} - \mathord{:} - $    & $ - \mathord{:} - \mathord{:} - $    & $ - \mathord{:} - \mathord{:} - $    & $ - \mathord{:} - \mathord{:} - $    & $ 03 \mathord{:} 55 \mathord{:} 10 $ \\
            Cantilever $N_a = 2$                               & $ 00 \mathord{:} 23 \mathord{:} 48 $ & $ 00 \mathord{:} 10 \mathord{:} 01 $ & $ 00 \mathord{:} 06 \mathord{:} 10 $ & $ 00 \mathord{:} 01 \mathord{:} 10 $ & $ 00 \mathord{:} 09 \mathord{:} 18 $ & $ 00 \mathord{:} 50 \mathord{:} 28 $ & $ 03 \mathord{:} 55 \mathord{:} 10 $ \\
            Cantilever $N_a = 2$ (Sym.)                        & $ 00 \mathord{:} 23 \mathord{:} 48 $ & $ - \mathord{:} - \mathord{:} - $    & $ - \mathord{:} - \mathord{:} - $    & $ - \mathord{:} - \mathord{:} - $    & $ - \mathord{:} - \mathord{:} - $    & $ - \mathord{:} - \mathord{:} - $    & $ 03 \mathord{:} 55 \mathord{:} 10 $ \\
            Cantilever $N_a = 3$                               & $ 00 \mathord{:} 24 \mathord{:} 52 $ & $ 00 \mathord{:} 11 \mathord{:} 19 $ & $ 00 \mathord{:} 01 \mathord{:} 22 $ & $ 00 \mathord{:} 01 \mathord{:} 21 $ & $ 00 \mathord{:} 12 \mathord{:} 11 $ & $ 00 \mathord{:} 51 \mathord{:} 06 $ & $ 03 \mathord{:} 55 \mathord{:} 10 $ \\
            Cantilever $N_a = 3$ (Sym.)                        & $ 00 \mathord{:} 24 \mathord{:} 52 $ & $ - \mathord{:} - \mathord{:} - $    & $ - \mathord{:} - \mathord{:} - $    & $ - \mathord{:} - \mathord{:} - $    & $ - \mathord{:} - \mathord{:} - $    & $ - \mathord{:} - \mathord{:} - $    & $ 03 \mathord{:} 55 \mathord{:} 10 $ \\ \midrule
            Lotte $N_a = 0$                                    & $ 00 \mathord{:} 55 \mathord{:} 18 $ & $ 00 \mathord{:} 21 \mathord{:} 31 $ & $ 00 \mathord{:} 02 \mathord{:} 24 $ & $ 00 \mathord{:} 00 \mathord{:} 28 $ & $ 00 \mathord{:} 39 \mathord{:} 35 $ & $ 01 \mathord{:} 59 \mathord{:} 17 $ & $ 02 \mathord{:} 28 \mathord{:} 09 $ \\
            Lotte $N_a = 0$ (SG)                               & $ 00 \mathord{:} 56 \mathord{:} 38 $ & $ 00 \mathord{:} 24 \mathord{:} 21 $ & $ 00 \mathord{:} 03 \mathord{:} 26 $ & $ 00 \mathord{:} 00 \mathord{:} 31 $ & $ 00 \mathord{:} 24 \mathord{:} 05 $ & $ 01 \mathord{:} 49 \mathord{:} 03 $ & $ 02 \mathord{:} 28 \mathord{:} 09 $ \\
            Lotte $N_a = 2$                                    & $ 00 \mathord{:} 48 \mathord{:} 01 $ & $ 00 \mathord{:} 38 \mathord{:} 56 $ & $ 00 \mathord{:} 00 \mathord{:} 25 $ & $ 00 \mathord{:} 03 \mathord{:} 18 $ & $ 00 \mathord{:} 26 \mathord{:} 38 $ & $ 01 \mathord{:} 57 \mathord{:} 19 $ & $ 02 \mathord{:} 28 \mathord{:} 09 $ \\
            Lotte $N_a = 3$                                    & $ 00 \mathord{:} 47 \mathord{:} 35 $ & $ 01 \mathord{:} 42 \mathord{:} 28 $ & $ 00 \mathord{:} 01 \mathord{:} 27 $ & $ 00 \mathord{:} 09 \mathord{:} 11 $ & $ 00 \mathord{:} 27 \mathord{:} 36 $ & $ 03 \mathord{:} 08 \mathord{:} 19 $ & $ 02 \mathord{:} 28 \mathord{:} 09 $ \\ \midrule
            Bracket LCA $N_a = 0$                              & $ 08 \mathord{:} 04 \mathord{:} 38 $ & $ 00 \mathord{:} 53 \mathord{:} 45 $ & $ 01 \mathord{:} 28 \mathord{:} 50 $ & $ 00 \mathord{:} 10 \mathord{:} 51 $ & $ 00 \mathord{:} 52 \mathord{:} 02 $ & $ 11 \mathord{:} 30 \mathord{:} 08 $ & $ 14 \mathord{:} 34 \mathord{:} 41 $ \\
            Bracket LCA $N_a = 3$                              & $ 08 \mathord{:} 00 \mathord{:} 20 $ & $ 01 \mathord{:} 14 \mathord{:} 27 $ & $ 00 \mathord{:} 06 \mathord{:} 27 $ & $ 00 \mathord{:} 13 \mathord{:} 59 $ & $ 01 \mathord{:} 00 \mathord{:} 59 $ & $ 10 \mathord{:} 36 \mathord{:} 14 $ & $ 14 \mathord{:} 34 \mathord{:} 41 $ \\
            Bracket LC2 $N_a = 0$                              & $ 03 \mathord{:} 10 \mathord{:} 25 $ & $ 01 \mathord{:} 10 \mathord{:} 56 $ & $ 00 \mathord{:} 02 \mathord{:} 05 $ & $ 00 \mathord{:} 13 \mathord{:} 02 $ & $ 01 \mathord{:} 47 \mathord{:} 01 $ & $ 06 \mathord{:} 23 \mathord{:} 30 $ & $ 07 \mathord{:} 02 \mathord{:} 43 $ \\
            Bracket LC2 $N_a = 3$                              & $ 03 \mathord{:} 05 \mathord{:} 05 $ & $ 01 \mathord{:} 24 \mathord{:} 43 $ & $ 00 \mathord{:} 51 \mathord{:} 00 $ & $ 00 \mathord{:} 03 \mathord{:} 17 $ & $ 02 \mathord{:} 56 \mathord{:} 03 $ & $ 08 \mathord{:} 20 \mathord{:} 09 $ & $ 07 \mathord{:} 02 \mathord{:} 43 $ \\
            Bracket LC4 $N_a = 0$                              & $ 03 \mathord{:} 01 \mathord{:} 32 $ & $ 01 \mathord{:} 12 \mathord{:} 21 $ & $ 00 \mathord{:} 07 \mathord{:} 39 $ & $ 00 \mathord{:} 15 \mathord{:} 13 $ & $ 01 \mathord{:} 51 \mathord{:} 26 $ & $ 06 \mathord{:} 28 \mathord{:} 12 $ & $ 05 \mathord{:} 57 \mathord{:} 16 $ \\
            Bracket LC4 $N_a = 2$                              & $ 04 \mathord{:} 49 \mathord{:} 54 $ & $ 01 \mathord{:} 13 \mathord{:} 43 $ & $ 00 \mathord{:} 09 \mathord{:} 00 $ & $ 00 \mathord{:} 13 \mathord{:} 05 $ & $ 02 \mathord{:} 35 \mathord{:} 07 $ & $ 09 \mathord{:} 00 \mathord{:} 51 $ & $ 05 \mathord{:} 57 \mathord{:} 16 $ \\
            \bottomrule
        \end{tabular}

    %\end{threeparttable}
\end{table}
\begin{figure}[tbp]
    \centering
    \makebox[\textwidth][c]{
        \includegraphics{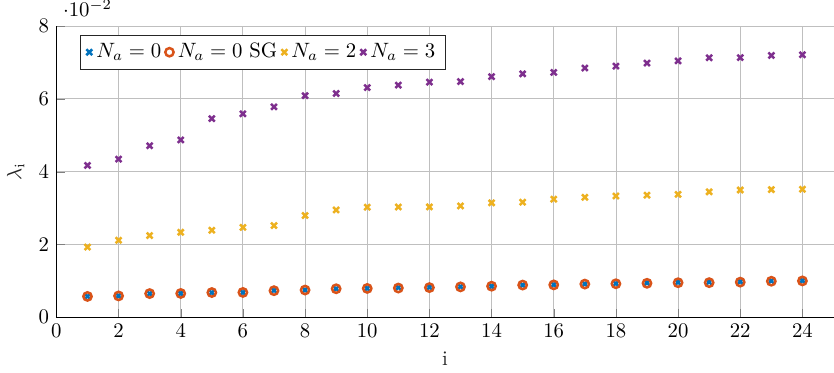}
    }
    \caption{Plot of the 24 buckling load factors (BLFs) of the linear buckling analysis of the Lotte tower with different layer configurations.}
    \label{fig:lotteTowerBuckling}
\end{figure}

\begin{figure}[tbp]
    \centering
    \makebox[\textwidth][c]{
        \begin{subfigure}{71mm}
            \includegraphics[width=\linewidth]{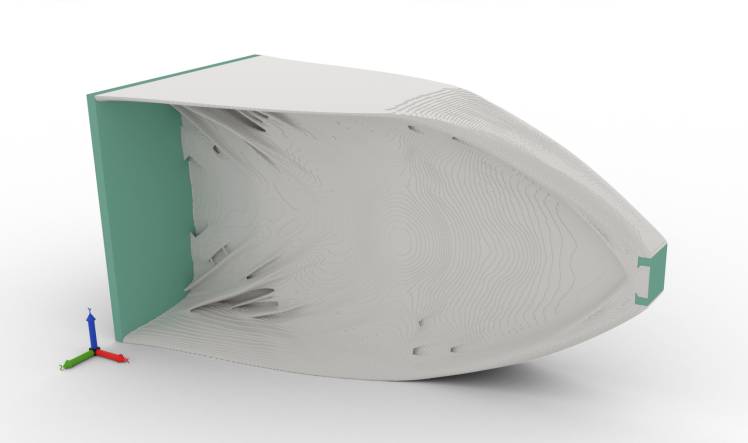}
            \caption{}
            \label{fig:michell_SIMP}
        \end{subfigure}%
        \begin{subfigure}{71mm}
            \includegraphics[width=\linewidth]{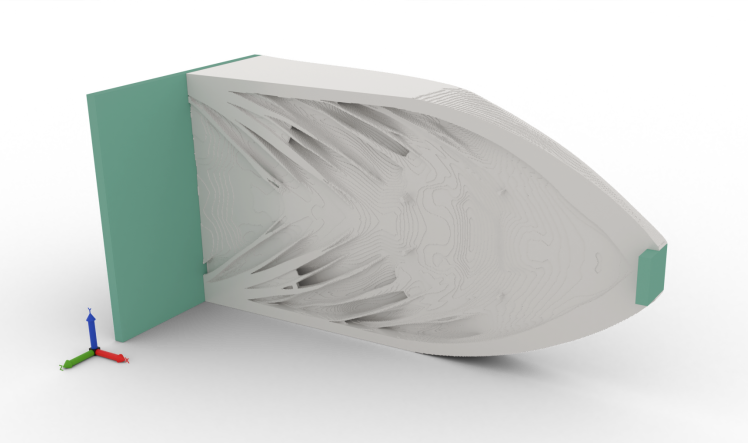}
            \caption{}
            \label{fig:michell_SIMP_CUT}
        \end{subfigure}%
        \begin{subfigure}{47.5mm}
            \includegraphics[width=\linewidth]{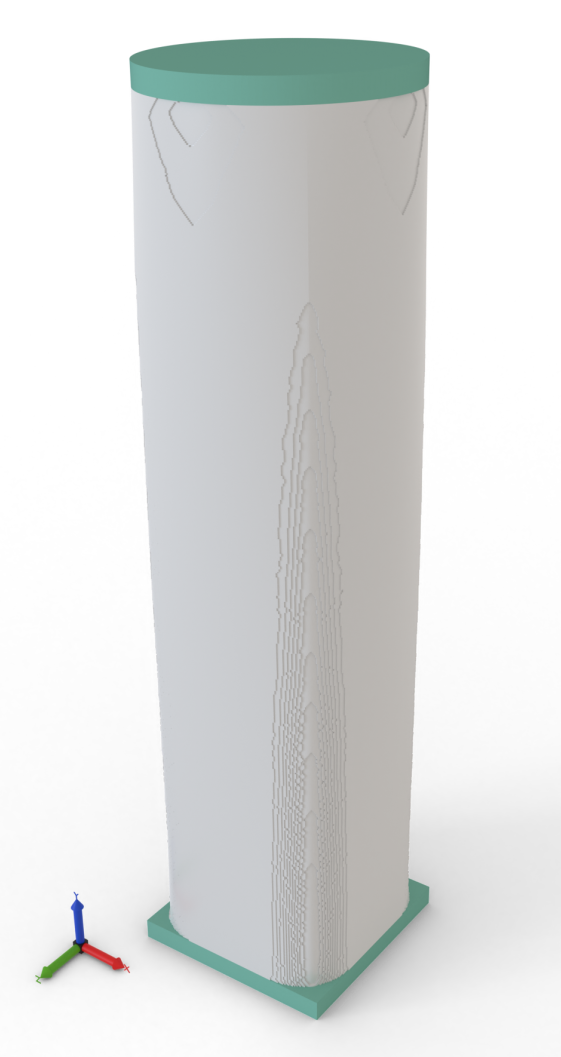}
            \caption{}
            \label{fig:lotteTower_SIMP}
        \end{subfigure}%
    }
    \makebox[\textwidth][c]{
        \begin{subfigure}{63.333mm}
            \includegraphics[width=\linewidth]{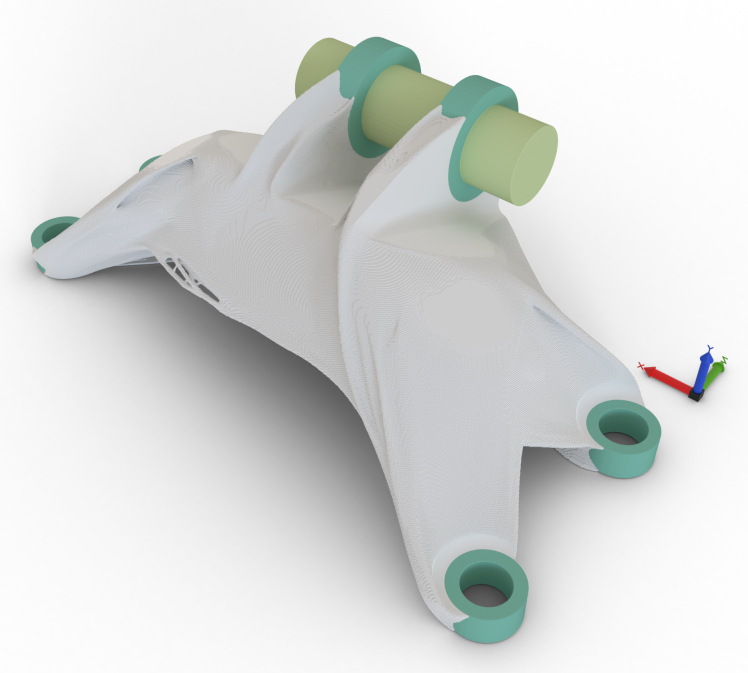}
            \caption{}
            \label{fig:Bracket_LC0_SIMP}
        \end{subfigure}%
        \begin{subfigure}{63.333mm}
            \includegraphics[width=\linewidth]{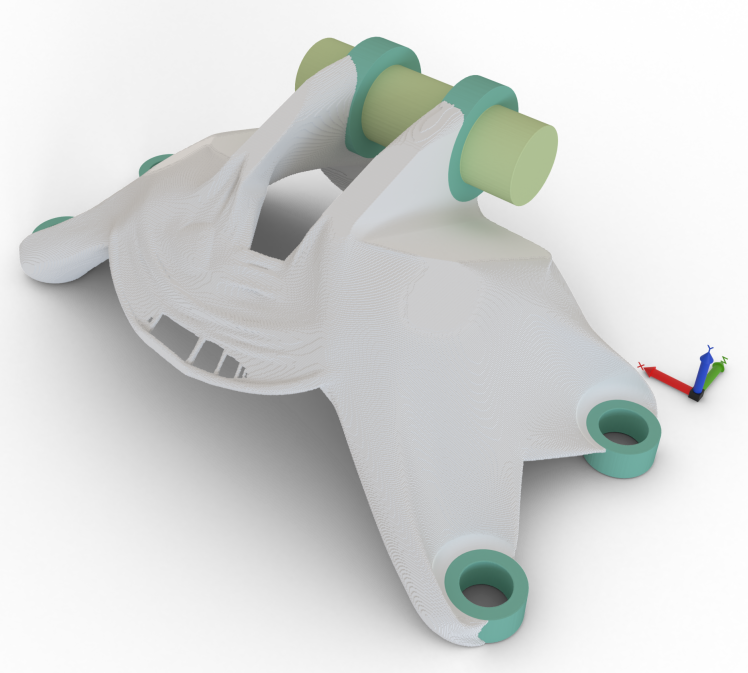}
            \caption{}
            \label{fig:Bracket_LC2_SIMP}
        \end{subfigure}%
        \begin{subfigure}{63.333mm}
            \includegraphics[width=\linewidth]{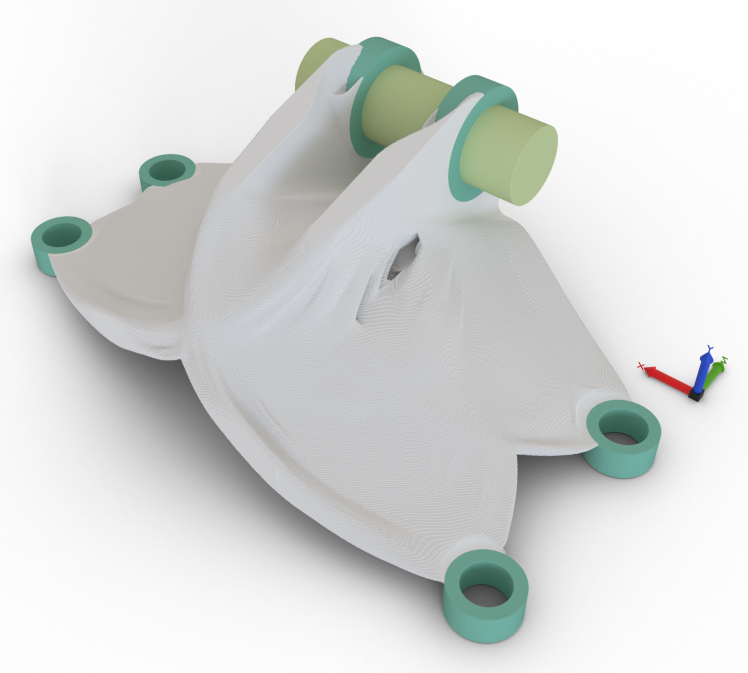}
            \caption{}
            \label{fig:Bracket_LC4_SIMP}
        \end{subfigure}%
    }%
    \caption{Large-scale SIMP optimization result.
    (a)-(b) Michell cantilever (with slice cut reviling the internal structure). (c) Lotte tower. (d) Bracket LCA. (e) Bracket LC2. (f) Bracket LC4.}
    \label{fig:SIMPres}
\end{figure}

\begin{figure}[tbp!]
    \centering
    \makebox[\textwidth][c]{
        \begin{subfigure}{63.333mm}
            \includegraphics[width=\linewidth]{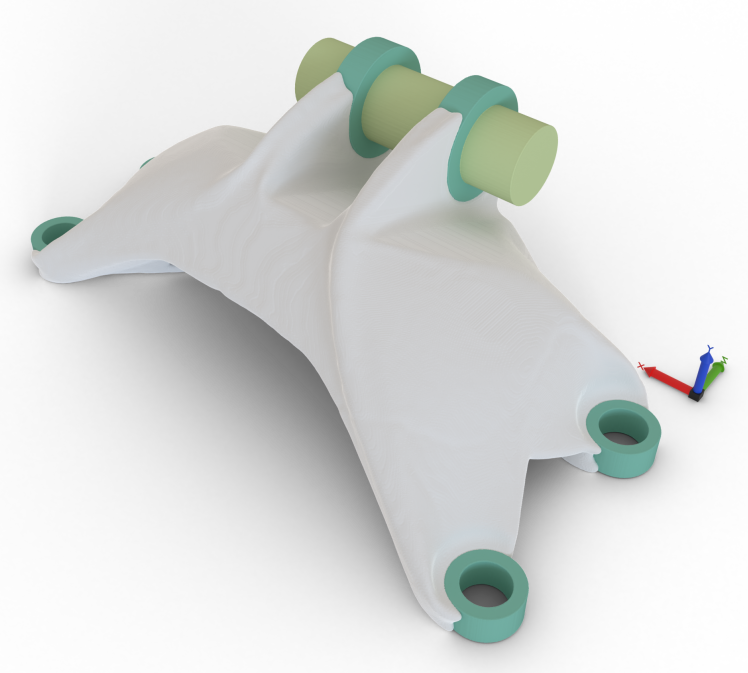}
            \caption{}
            \label{fig:Bracket_LC0_A0_DEHOM}
        \end{subfigure}%
        \begin{subfigure}{63.333mm}
            \includegraphics[width=\linewidth]{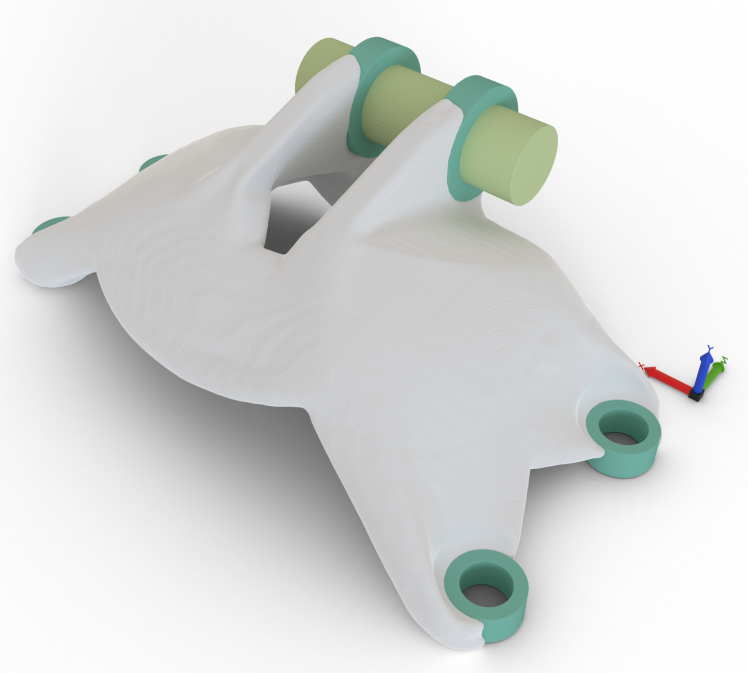}
            \caption{}
            \label{fig:Bracket_LC2_A0_DEHOM}
        \end{subfigure}%
        \begin{subfigure}{63.333mm}
            \includegraphics[width=\linewidth]{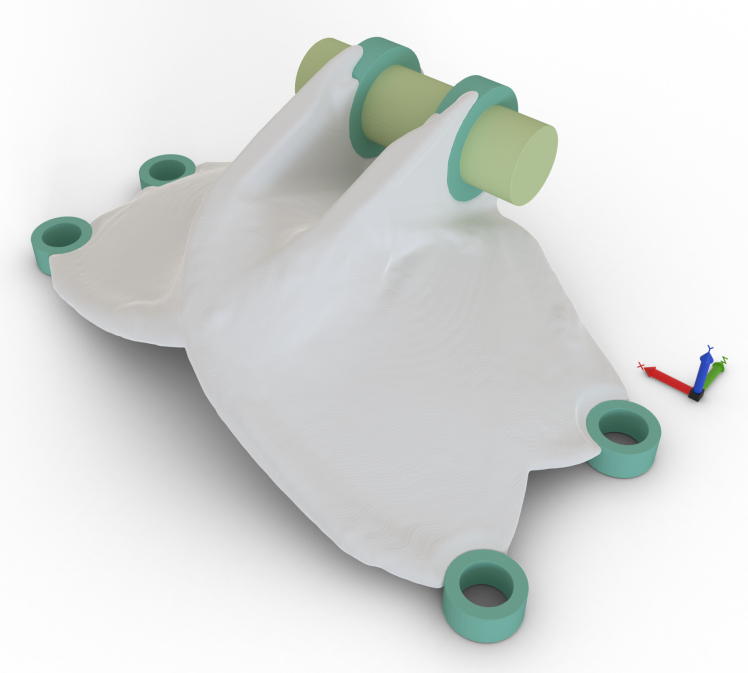}
            \caption{}
            \label{fig:Bracket_LC4_A0_DEHOM}
        \end{subfigure}
    }
    \makebox[\textwidth][c]{
        \begin{subfigure}{63.333mm}
            \includegraphics[width=\linewidth]{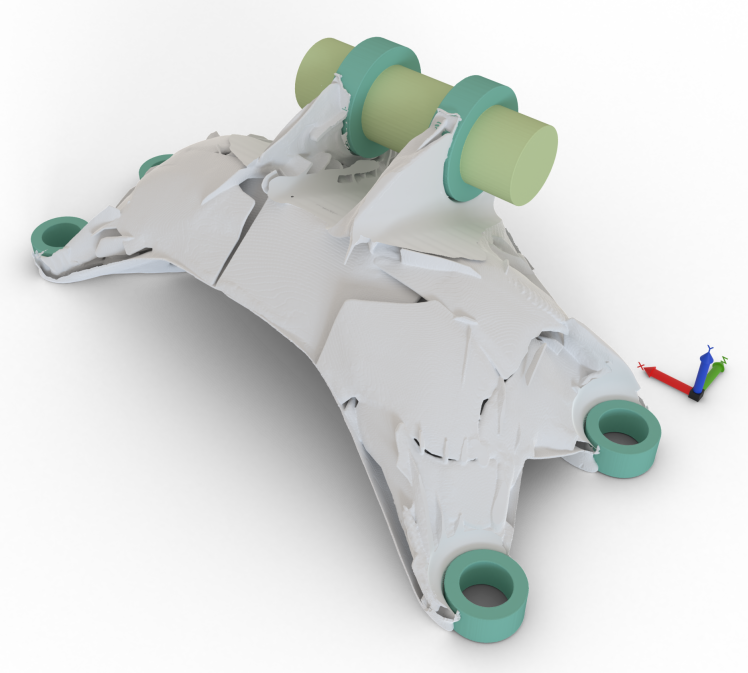}
            \caption{}
            \label{fig:Bracket_LC0_A0_DEHOM_cut}
        \end{subfigure}%
        \begin{subfigure}{63.333mm}
            \includegraphics[width=\linewidth]{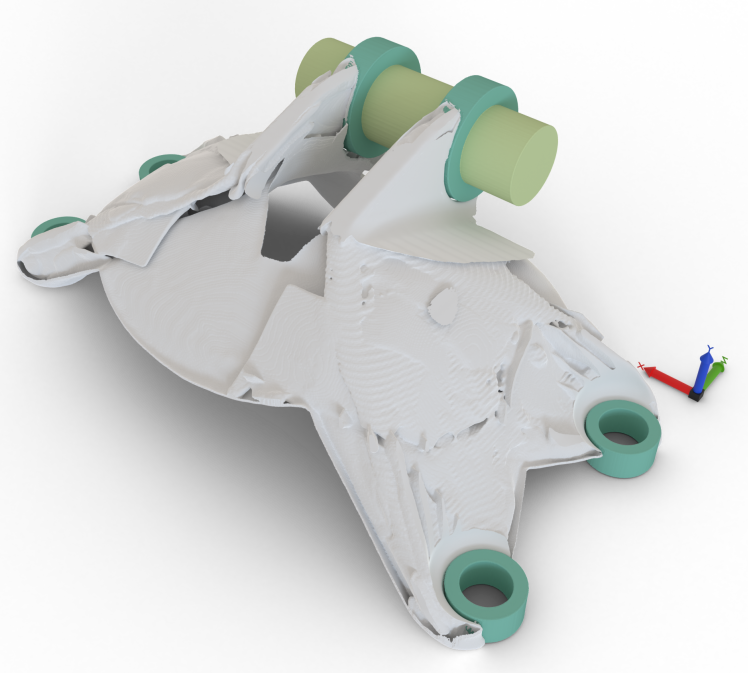}
            \caption{}
            \label{fig:Bracket_LC2_A0_DEHOM_CUT}
        \end{subfigure}%
        \begin{subfigure}{63.333mm}
            \includegraphics[width=\linewidth]{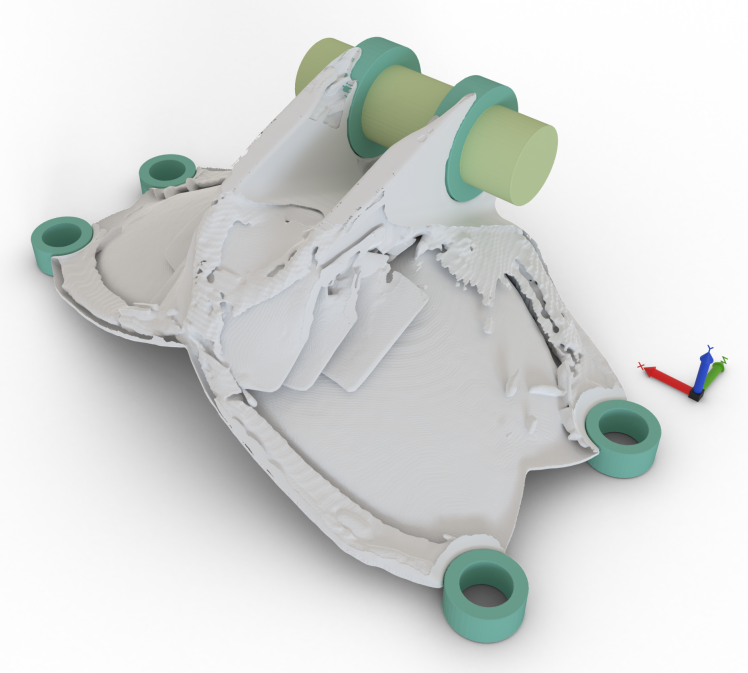}
            \caption{}
            \label{fig:Bracket_LC4_A0_DEHOM_CUT}
        \end{subfigure}%
    }
    \caption{
    De-homogenized result of 
    (a),(d) Bracket LCA $N_a = 0$.
    (b),(e) Bracket LC2 $N_a = 0$.
    (c),(f) Bracket LC4 $N_a = 0$.
    (d)-(f) The outer top surface is removed to expose the internal structure.}
    \label{fig:Bracket_A0}
\end{figure}

\begin{figure}[tb!]
    \centering
    \makebox[\textwidth][c]{
        \begin{subfigure}{63.333mm}
            \includegraphics[width=\linewidth]{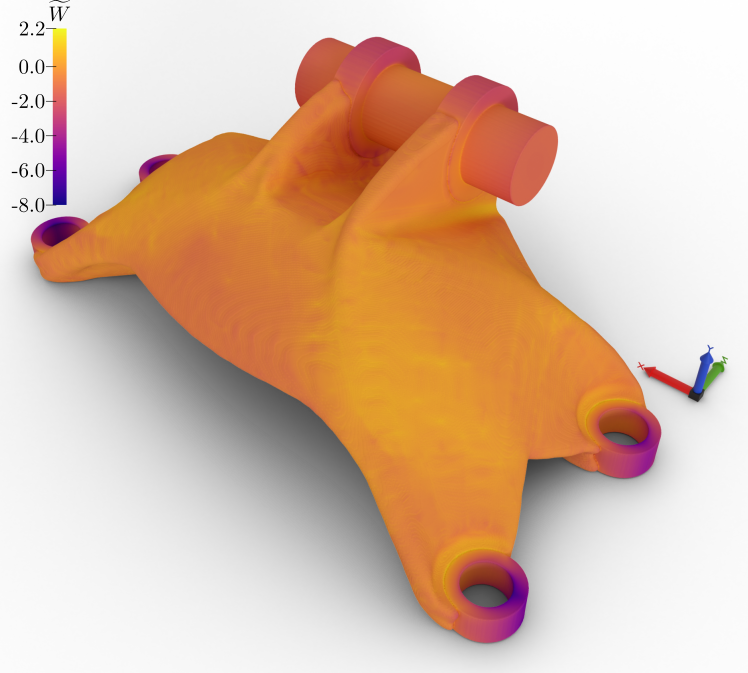}
            \caption{}
            \label{fig:Bracket_LC0_A13_DEHOM_E}
        \end{subfigure}%
        \begin{subfigure}{63.333mm}
            \includegraphics[width=\linewidth]{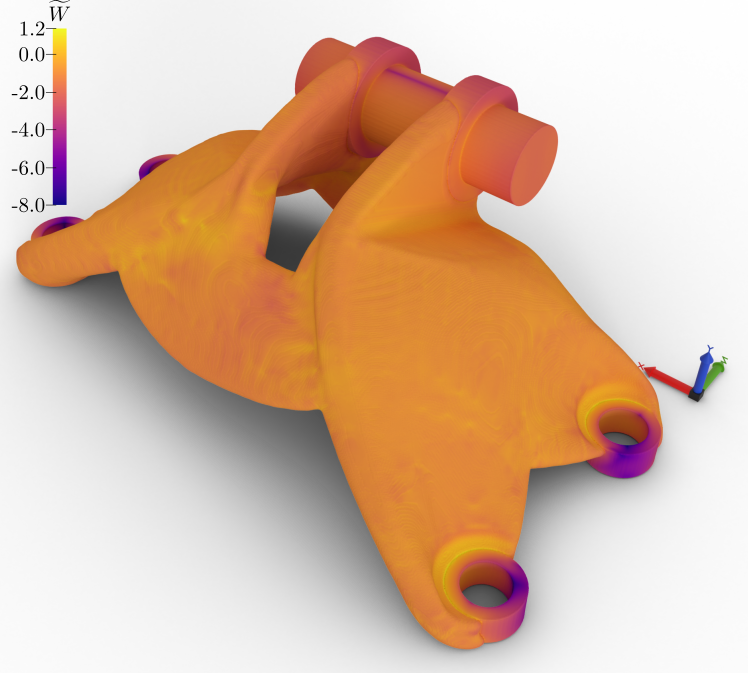}
            \caption{}
            \label{fig:Bracket_LC2_DEHOM_E}
        \end{subfigure}%
        \begin{subfigure}{63.333mm}
            \includegraphics[width=\linewidth]{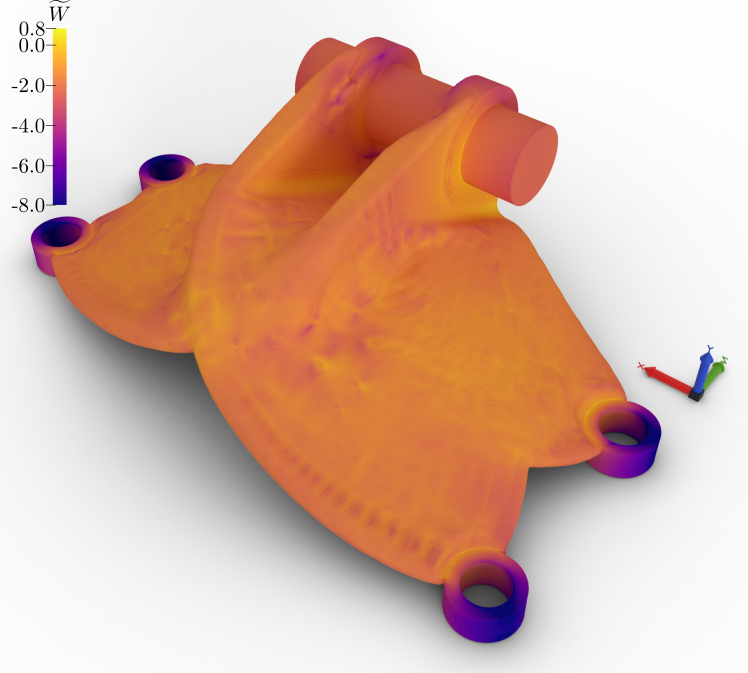}
            \caption{}
            \label{fig:Bracket_LC4_DEHOM_E}
        \end{subfigure}
    }
    \makebox[\textwidth][c]{
        \begin{subfigure}{63.333mm}
            \includegraphics[width=\linewidth]{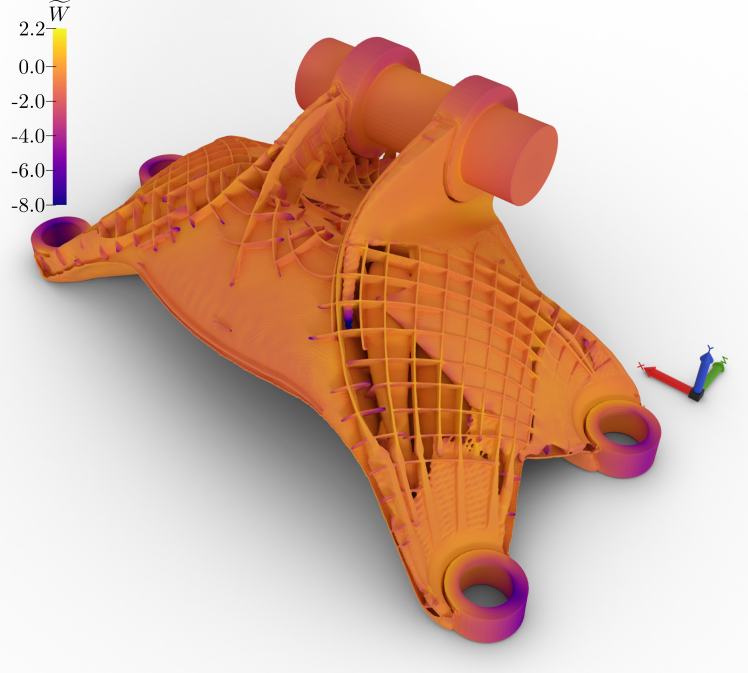}
            \caption{}
            \label{fig:Bracket_LC0_A13_DEHOM_cut_E}
        \end{subfigure}%
        \begin{subfigure}{63.333mm}
            \includegraphics[width=\linewidth]{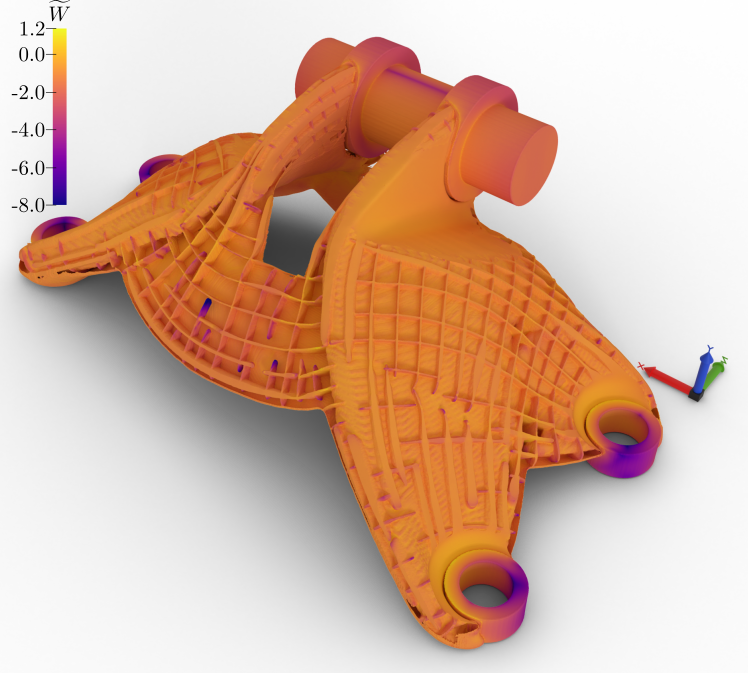}
            \caption{}
            \label{fig:Bracket_LC2_DEHOM_CUT_E}
        \end{subfigure}%
        \begin{subfigure}{63.333mm}
            \includegraphics[width=\linewidth]{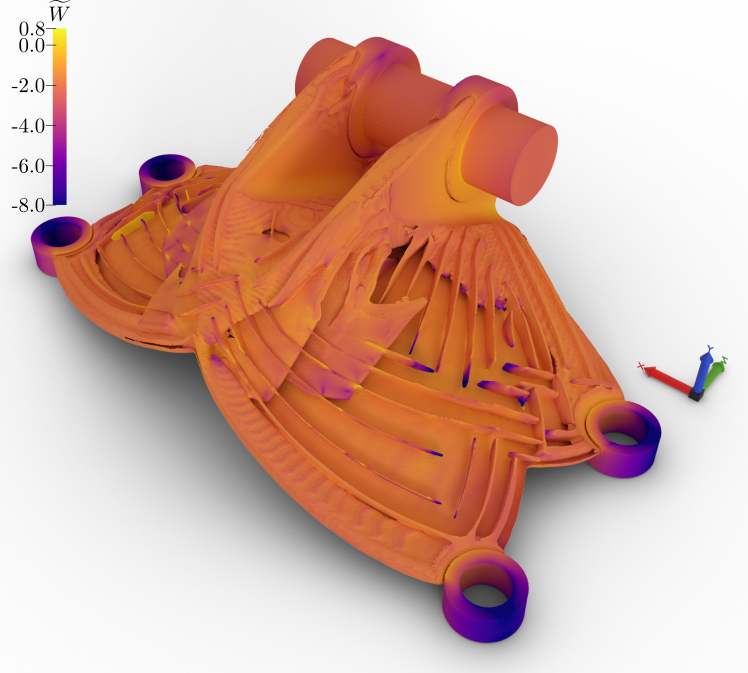}
            \caption{}
            \label{fig:Bracket_LC4_DEHOM_CUT_E}
        \end{subfigure}%
    }
    \caption{
    Strain energy density distribution of the de-homogenized result of 
    (a),(d) Bracket LCA $N_a = 3$.
    (b),(e) Bracket LC2 $N_a = 3$.
    (c),(f) Bracket LC4 $N_a = 2$.
    (d)-(f) The outer top surface is removed to expose the internal structure.}
    \label{fig:Bracket_res_E}
\end{figure}

\FloatBarrier

\begin{multicols}{2}

%%%%%%%%%%%% Supplementary Methods %%%%%%%%%%%%
%\footnotesize
%\section*{Methods}

%%%%%%%%%%%%% Acknowledgements %%%%%%%%%%%%%
%\footnotesize
%\section*{Acknowledgements}

%%%%%%%%%%%%%%   Bibliography   %%%%%%%%%%%%%%
% \normalsize
% \begingroup
% \raggedright

% \bibliography{references}

\bibliographystyle{apalike}
\bibliography{references.bib}

\begin{thebibliography}{}

\bibitem[Aage et~al., 2015]{Aage2015}
Aage, N., Andreassen, E., and Lazarov, B.~S. (2015).
\newblock {Topology optimization using PETSc: An easy-to-use, fully parallel,
  open source topology optimization framework}.
\newblock {\em Struct. Multidiscip. Optim.}, 51(3):565--572.

\bibitem[Aage et~al., 2017]{Aage2017}
Aage, N., Andreassen, E., Lazarov, B.~S., and Sigmund, O. (2017).
\newblock {Giga-voxel computational morphogenesis for structural design}.
\newblock {\em Nature}, 550(7674):84--86.

\bibitem[Aage and Lazarov, 2013]{Aage2013}
Aage, N. and Lazarov, B.~S. (2013).
\newblock {Parallel framework for topology optimization using the method of
  moving asymptotes}.
\newblock {\em Structural and Multidisciplinary Optimization}, 47(4):493--505.

\bibitem[Allaire, 2002]{Allaire2002}
Allaire, G. (2002).
\newblock {\em {Shape Optimization by the Homogenization Method}}, volume 146
  of {\em Applied Mathematical Sciences}.
\newblock Springer New York, New York, NY.

\bibitem[Amir et~al., 2014]{Amir2013}
Amir, O., Aage, N., and Lazarov, B.~S. (2014).
\newblock {On multigrid-CG for efficient topology optimization}.
\newblock {\em Struct. Multidiscip. Optim.}, 49(5):815--829.

\bibitem[Andreassen and Andreasen, 2014]{Andreassen2014}
Andreassen, E. and Andreasen, C.~S. (2014).
\newblock {How to determine composite material properties using numerical
  homogenization}.
\newblock {\em Computational Materials Science}, 83:488--495.

\bibitem[Avellaneda, 1987]{Avellaneda1987}
Avellaneda, M. (1987).
\newblock {Optimal Bounds and Microgeometries for Elastic Two-Phase
  Composites}.
\newblock {\em SIAM Journal on Applied Mathematics}, 47(6):1216--1228.

\bibitem[Baandrup et~al., 2020]{Baandrup2020}
Baandrup, M., Sigmund, O., Polk, H., and Aage, N. (2020).
\newblock {Closing the gap towards super-long suspension bridges using
  computational morphogenesis}.
\newblock {\em Nature Communications}, 11(1):2735.

\bibitem[Balay et~al., 2022a]{petsc-user-ref}
Balay, S., Abhyankar, S., Adams, M.~F., Benson, S., Brown, J., Brune, P.,
  Buschelman, K., Constantinescu, E., Dalcin, L., Dener, A., Eijkhout, V.,
  Faibussowitsch, J., Gropp, W.~D., Hapla, V., Isaac, T., Jolivet, P., Karpeev,
  D., Kaushik, D., Knepley, M.~G., Kong, F., Kruger, S., May, D.~A., McInnes,
  L.~C., Mills, R.~T., Mitchell, L., Munson, T., Roman, J.~E., Rupp, K., Sanan,
  P., Sarich, J., Smith, B.~F., Zampini, S., Zhang, H., Zhang, H., and Zhang,
  J. (2022a).
\newblock {PETSc/TAO} users manual.
\newblock Technical Report ANL-21/39 - Revision 3.18, Argonne National
  Laboratory.

\bibitem[Balay et~al., 2022b]{petsc-web-page}
Balay, S., Abhyankar, S., Adams, M.~F., Benson, S., Brown, J., Brune, P.,
  Buschelman, K., Constantinescu, E.~M., Dalcin, L., Dener, A., Eijkhout, V.,
  Faibussowitsch, J., Gropp, W.~D., Hapla, V., Isaac, T., Jolivet, P., Karpeev,
  D., Kaushik, D., Knepley, M.~G., Kong, F., Kruger, S., May, D.~A., McInnes,
  L.~C., Mills, R.~T., Mitchell, L., Munson, T., Roman, J.~E., Rupp, K., Sanan,
  P., Sarich, J., Smith, B.~F., Zampini, S., Zhang, H., Zhang, H., and Zhang,
  J. (2022b).
\newblock {PETS}c {W}eb page.
\newblock \url{https://petsc.org/}.

\bibitem[Balay et~al., 1997]{petsc-efficient}
Balay, S., Gropp, W.~D., McInnes, L.~C., and Smith, B.~F. (1997).
\newblock Efficient management of parallelism in object oriented numerical
  software libraries.
\newblock In Arge, E., Bruaset, A.~M., and Langtangen, H.~P., editors, {\em
  Modern Software Tools in Scientific Computing}, pages 163--202.
  Birkh{\"{a}}user Press.

\bibitem[Bends{\o}e and Kikuchi, 1988]{BendsoeKikuchi1988}
Bends{\o}e, M.~P. and Kikuchi, N. (1988).
\newblock {Generating optimal topologies in structural design using a
  homogenization method}.
\newblock {\em Computer Methods in Applied Mechanics and Engineering},
  71(2):197--224.

\bibitem[Bends{\o}e and Sigmund, 1999]{Bendsoe1999}
Bends{\o}e, M.~P. and Sigmund, O. (1999).
\newblock {Material interpolation schemes in topology optimization}.
\newblock {\em Archive of Applied Mechanics}, 69(9-10):635--654.

\bibitem[Bends{\o}e and Sigmund, 2004]{Bendsoe2004}
Bends{\o}e, M.~P. and Sigmund, O. (2004).
\newblock {\em {Topology Optimization}}.
\newblock Springer Berlin Heidelberg, Berlin, Heidelberg.

\bibitem[Carter et~al., 2014]{Carter2014}
Carter, W., Erno, D., Abbott, D., Bruck, C., Wilson, G., Wolfe, J., Finkhousen,
  G., Tepper, A., and Stevens, R. (2014).
\newblock The ge aircraft engine bracket challenge: An experiment in
  crowdsourcing for mechanical design concepts.
\newblock 2014 International Solid Freeform Fabrication Symposium.

\bibitem[Chapra, 2011]{Chapra2011Applied}
Chapra, S. (2011).
\newblock {\em Applied Numerical Methods {W/MATLAB}: for Engineers \&
  Scientists}.
\newblock McGraw-Hill Science/Engineering/Math, 3 edition.

\bibitem[Clausen et~al., 2016]{Clausen2016}
Clausen, A., Aage, N., and Sigmund, O. (2016).
\newblock {Exploiting Additive Manufacturing Infill in Topology Optimization
  for Improved Buckling Load}.
\newblock {\em Engineering}, 2(2):250--257.

\bibitem[Cook et~al., 2001]{cook_concepts_2001}
Cook, R.~D., Malkus, D.~S., Plesha, M.~E., and Witt, R.~J. (2001).
\newblock {\em Concepts and Applications of Finite Element Analysis, 4th
  Edition}.
\newblock Wiley, 4 edition.

\bibitem[{Coreform LLC}, 2023]{cubit}
{Coreform LLC} (2023).
\newblock Coreform trelis pro.
\newblock \url{http://coreform.com}.

\bibitem[Dormand and Prince, 1980]{Dormand1980}
Dormand, J. and Prince, P. (1980).
\newblock A family of embedded runge-kutta formulae.
\newblock {\em Journal of Computational and Applied Mathematics}, 6(1):19--26.

\bibitem[Evgrafov et~al., 2008]{Evgrafov2008}
Evgrafov, A., Rupp, C.~J., Maute, K., and Dunn, M.~L. (2008).
\newblock {Large-scale parallel topology optimization using a dual-primal
  substructuring solver}.
\newblock {\em Struct. Multidiscip. Optim.}, 36:329--345.

\bibitem[Ferrari et~al., 2021]{Ferrari2021}
Ferrari, F., Sigmund, O., and Guest, J.~K. (2021).
\newblock {Topology optimization with linearized buckling criteria in 250 lines
  of Matlab}.
\newblock {\em Structural and Multidisciplinary Optimization},
  63(6):3045--3066.

\bibitem[Francfort et~al., 1995]{Francfort1995}
Francfort, G., Murat, F., and Tartar, L. (1995).
\newblock {Fourth-order moments of nonnegative measures on S2 and
  applications}.
\newblock {\em Archive for Rational Mechanics and Analysis}, 131(4):305--333.

\bibitem[Francfort and Murat, 1986]{Francfort1986}
Francfort, G.~A. and Murat, F. (1986).
\newblock {Homogenization and optimal bounds in linear elasticity}.
\newblock {\em Archive for Rational Mechanics and Analysis}, 94(4):307--334.

\bibitem[Garnier et~al., 2022]{Garnier2022}
Garnier, D.-H., Schmidt, M.-P., and Rohmer, D. (2022).
\newblock {Growth of oriented orthotropic structures with reaction/diffusion}.
\newblock {\em Structural and Multidisciplinary Optimization}, 65(11):327.

\bibitem[Geoffroy-Donders et~al., 2020]{Geoffroy-Donders2020}
Geoffroy-Donders, P., Allaire, G., and Pantz, O. (2020).
\newblock 3-d topology optimization of modulated and oriented periodic
  microstructures by the homogenization method.
\newblock {\em Journal of Computational Physics}, 401:108994.

\bibitem[Giele et~al., 2021]{Giele2021a}
Giele, R., Groen, J., Aage, N., Andreasen, C.~S., and Sigmund, O. (2021).
\newblock {On approaches for avoiding low-stiffness regions in variable
  thickness sheet and homogenization-based topology optimization}.
\newblock {\em Structural and Multidisciplinary Optimization}, 64(1):39--52.

\bibitem[GrabCAD, 2013]{GEJETBRACKET}
GrabCAD (2013).
\newblock Ge jet engine bracket challenge webpage.
\newblock \url{https://grabcad.com/challenges/ge-jet-engine-bracket-challenge}.
\newblock Accessed on 11/06/2023.

\bibitem[Grant and Boyd, 2014]{cvx}
Grant, M. and Boyd, S. (2014).
\newblock {CVX}: Matlab software for disciplined convex programming, version
  2.1.
\newblock \url{http://cvxr.com/cvx}.

\bibitem[Groen and Sigmund, 2018]{Groen2018}
Groen, J.~P. and Sigmund, O. (2018).
\newblock {Homogenization-based topology optimization for high-resolution
  manufacturable microstructures}.
\newblock {\em International Journal for Numerical Methods in Engineering},
  113(8):1148--1163.

\bibitem[Groen et~al., 2020]{Groen2020}
Groen, J.~P., Stutz, F.~C., Aage, N., B{\ae}rentzen, J.~A., and Sigmund, O.
  (2020).
\newblock {De-homogenization of optimal multi-scale 3D topologies}.
\newblock {\em Computer Methods in Applied Mechanics and Engineering},
  364:112979.

\bibitem[Groen et~al., 2021]{Groen2021}
Groen, J.~P., Thomsen, C.~R., and Sigmund, O. (2021).
\newblock {Multi-scale topology optimization for stiffness and
  de-homogenization using implicit geometry modeling}.
\newblock {\em Structural and Multidisciplinary Optimization},
  63(6):2919--2934.

\bibitem[Hashin and Shtrikman, 1963]{Hashin1963}
Hashin, Z. and Shtrikman, S. (1963).
\newblock {A variational approach to the theory of the elastic behaviour of
  multiphase materials}.
\newblock {\em Journal of the Mechanics and Physics of Solids}, 11(2):127--140.

\bibitem[Hernandez et~al., 2005]{Roman2010}
Hernandez, V., Roman, J.~E., and Vidal, V. (2005).
\newblock {SLEPc}.
\newblock {\em ACM Transactions on Mathematical Software}, 31(3):351--362.

\bibitem[H{\o}gh{\o}j and Tr{\"{a}}ff, 2022]{Hoeghoej2022}
H{\o}gh{\o}j, L.~C. and Tr{\"{a}}ff, E.~A. (2022).
\newblock {An advection-diffusion based filter for machinable designs in
  topology optimization}.
\newblock {\em Computer Methods in Applied Mechanics and Engineering},
  391:114488.

\bibitem[Jensen et~al., 2023]{uTopDeHom}
Jensen, P. D.~L., Olsen, T.~F., Bærentzen, J.~A., Aage, N., and Sigmund, O.
  (2023).
\newblock utopdehom.
\newblock \url{https://doi.org/10.11583/DTU.c.6738699}.

\bibitem[Jensen et~al., 2022]{Jensen2022}
Jensen, P. D.~L., Sigmund, O., and Groen, J.~P. (2022).
\newblock {De-homogenization of optimal 2D topologies for multiple loading
  cases}.
\newblock {\em Computer Methods in Applied Mechanics and Engineering},
  399:115426.

\bibitem[Jensen et~al., 2021]{Jensen2021}
Jensen, P. D.~L., Wang, F., Dimino, I., and Sigmund, O. (2021).
\newblock {Topology Optimization of Large-Scale 3D Morphing Wing Structures}.
\newblock {\em Actuators}, 10(9):217.

\bibitem[Lazarov and Sigmund, 2011]{Lazarov2011}
Lazarov, B.~S. and Sigmund, O. (2011).
\newblock {Filters in topology optimization based on Helmholtz-type
  differential equations}.
\newblock {\em International Journal for Numerical Methods in Engineering},
  86(6):765--781.

\bibitem[Lurie and Cherkaev, 1984]{Lurie1984}
Lurie, K.~A. and Cherkaev, A.~V. (1984).
\newblock {G-closure of a set of anisotropically conducting media in the
  two-dimensional case}.
\newblock {\em Journal of Optimization Theory and Applications},
  42(2):283--304.

\bibitem[Mahdavi et~al., 2006]{Mahdavi2006}
Mahdavi, A., Balaji, R., Frecker, M., and Mockensturm, E.~M. (2006).
\newblock {Topology optimization of 2D continua for minimum compliance using
  parallel computing}.
\newblock {\em Struct. Multidiscip. Optim.}

\bibitem[Milton, 1986]{Milton1986}
Milton, G.~W. (1986).
\newblock {Modelling the Properties of Composites by Laminates}.
\newblock In {\em Homogenization and Effective Moduli of Materials and Media},
  pages 150--174. Springer New York.

\bibitem[{MOSEK ApS}, 2021]{Mosek}
{MOSEK ApS} (2021).
\newblock {\em MOSEK Fusion API for C++ manual, Version 9.2}.

\bibitem[Norris, 1985]{Norris1985}
Norris, A.~N. (1985).
\newblock {A differential scheme for the effective moduli of composites}.
\newblock {\em Mechanics of Materials}, 4(1):1--16.

\bibitem[Pantz and Trabelsi, 2008]{Pantz2008}
Pantz, O. and Trabelsi, K. (2008).
\newblock {A Post-treatment of the homogenization method for shape
  optimization}.
\newblock {\em SIAM Journal on Control and Optimization}, 47(3):1380--1398.

\bibitem[Sigmund et~al., 2016]{Sigmund2016}
Sigmund, O., Aage, N., and Andreassen, E. (2016).
\newblock {On the (non-)optimality of Michell structures}.
\newblock {\em Structural and Multidisciplinary Optimization}, 54(2):361--373.

\bibitem[Stromberg et~al., 2011]{Stromberg2011}
Stromberg, L.~L., Beghini, A., Baker, W.~F., and Paulino, G.~H. (2011).
\newblock {Application of layout and topology optimization using pattern
  gradation for the conceptual design of buildings}.
\newblock pages 165--180.

\bibitem[Stutz et~al., 2020]{Stutz2020}
Stutz, F.~C., Groen, J.~P., Sigmund, O., and B{\ae}rentzen, J.~A. (2020).
\newblock {Singularity aware de-homogenization for high-resolution topology
  optimized structures}.
\newblock {\em Structural and Multidisciplinary Optimization},
  62(5):2279--2295.

\bibitem[Stutz et~al., 2022]{Stutz2022}
Stutz, F.~C., Olsen, T.~F., Groen, J.~P., Trung, T.~N., Aage, N., Sigmund, O.,
  Solomon, J., and B\ae{}rentzen, J.~A. (2022).
\newblock Synthesis of frame field-aligned multi-laminar structures.
\newblock {\em ACM Trans. Graph.}, 41(5).

\bibitem[Sundstr\"{o}m, 2010]{Sundstroem2010}
Sundstr\"{o}m, B. (2010).
\newblock {\em {Handbook of Solid Mechanics}}.
\newblock Department of Solid Mechanics, KTH.

\bibitem[Svanberg, 1987]{Svanberg1987}
Svanberg, K. (1987).
\newblock {The method of moving asymptotes—a new method for structural
  optimization}.
\newblock {\em International Journal for Numerical Methods in Engineering},
  24(2):359--373.

\bibitem[Tr{\"{a}}ff et~al., 2021]{Traff2021}
Tr{\"{a}}ff, E.~A., Sigmund, O., and Aage, N. (2021).
\newblock {Topology optimization of ultra high resolution shell structures}.
\newblock {\em Thin-Walled Structures}, 160(August 2020):107349.

\bibitem[Wallin et~al., 2020]{Wallin2020}
Wallin, M., Ivarsson, N., Amir, O., and Tortorelli, D. (2020).
\newblock {Consistent boundary conditions for PDE filter regularization in
  topology optimization}.
\newblock {\em Structural and Multidisciplinary Optimization},
  62(3):1299--1311.

\bibitem[Wang et~al., 2011]{Wang2011}
Wang, F., Lazarov, B.~S., and Sigmund, O. (2011).
\newblock {On projection methods, convergence and robust formulations in
  topology optimization}.
\newblock {\em Structural and Multidisciplinary Optimization}, 43(6):767--784.

\bibitem[Wang and Sigmund, 2021]{Wang2021}
Wang, F. and Sigmund, O. (2021).
\newblock {3D architected isotropic materials with tunable stiffness and
  buckling strength}.
\newblock {\em Journal of the Mechanics and Physics of Solids}, 152(November
  2020):104415.

\bibitem[Wang and Tamijani, 2022]{Wang2022}
Wang, Z. and Tamijani, A.~Y. (2022).
\newblock {Computational synthesis of large-scale three-dimensional
  heterogeneous lattice structures}.
\newblock {\em Aerospace Science and Technology}, 120:107258.

\bibitem[Wu et~al., 2018]{Sigmund2018}
Wu, J., Aage, N., Westermann, R., and Sigmund, O. (2018).
\newblock {Infill Optimization for Additive Manufacturing—Approaching
  Bone-Like Porous Structures}.
\newblock {\em IEEE Transactions on Visualization and Computer Graphics},
  24(2):1127--1140.

\bibitem[Wu et~al., 2021]{Wu2021b}
Wu, J., Wang, W., and Gao, X. (2021).
\newblock {Design and Optimization of Conforming Lattice Structures}.
\newblock {\em IEEE Transactions on Visualization and Computer Graphics},
  27(1):43--56.

\end{thebibliography}
% \endgroup

\end{multicols}
%%%%%%%%%%%%  Supplementary Figures  %%%%%%%%%%%%
%\clearpage

%%%%%%%%%%%%%%%%   End   %%%%%%%%%%%%%%%%
%\end{multicols}  % Method B for two-column formatting (doesn't play well with line numbers), comment out if using method A
\end{document}